\newif\ifMNRAS
\MNRAStrue 

\PassOptionsToPackage{pdfpagelabels=false}{hyperref} 

\ifMNRAS
    \documentclass[fleqn,usenatbib,useAMS]{mnras}
    \usepackage{graphicx}
\else
    \documentclass[twocolumn]{aastex63}
\fi

\usepackage{acronym}
\usepackage{hyperref}
\usepackage{amsmath,amsfonts,amssymb}
\usepackage{mathrsfs}
\usepackage{mathtools}
\usepackage[utf8]{inputenc}

\usepackage[normalem]{ulem}
\usepackage[dvipsnames]{xcolor}
\usepackage{xspace}

\newcommand{\muller}{M\"uller et al.\xspace}
\newcommand{\fc}{Fields-Couch\xspace}
\newcommand{\ddd}{3D-to-3D\xspace}
\newcommand{\oned}{1D-to-3D\xspace}
\usepackage{pifont}
\newcommand{\xmark}{\ding{55}}

\ifMNRAS
\title[3D Progenitors of CCSNe]{The Collapse and Three-Dimensional Explosion of Three-Dimensional Massive-star Supernova Progenitor Models}

\author[Vartanyan et al.]{
 \href{https://orcid.org/0000-0003-1938-9282}{David Vartanyan$^{1}$}\thanks{E-mail: dvartany@berkeley.edu},
\href{https://orcid.org/0000-0001-5939-5957}{Matthew S. B. Coleman$^{2}$},
\href{https://orcid.org/0000-0002-3099-5024}{Adam Burrows$^{2}$}
\\
$^{1}$Department of Physics and Astronomy, University of California, Berkeley, CA  94720, USA\\
$^{2}$Department of Astrophysical Sciences, 4 Ivy Lane, Princeton University, Princeton, NJ 08544, USA
}

\date{Accepted XXX. Received YYY; in original form ZZZ}

\pubyear{2021}

\else
\shorttitle{3D Progenitor}
\shortauthors{Vartanyan et al.}
\fi

\begin{document}
\label{firstpage}
\pagerange{\pageref{firstpage}--\pageref{lastpage}}
\maketitle

\begin{abstract}
The explosion outcome and diagnostics of core-collapse supernovae depend sensitively on the nature of the stellar progenitor, but most studies to date have focused exclusively on one-dimensional, spherically-symmetric massive star progenitors. We present some of the first core-collapse supernovae simulations of three-dimensional massive star supernovae progenitors, a 12.5- and a 15-M$_{\odot}$ model, evolved in three-dimensions from collapse to bounce through explosion with the radiation-hydrodynamic code F{\sc{ornax}}. We compare the results using those starting from three-dimensional progenitors to three-dimensional simulations of spherically-symmetric, one-dimensional progenitors of the same mass. We find that the models evolved in three dimensions during the final stages of massive star evolution are more prone to explosion. 
The turbulence arising in these multi-dimensional initial models serve as seed turbulence that promotes shock revival. Detection of gravitational waves and neutrinos signals could reveal signatures of pre-bounce turbulence. 
\end{abstract} 

\ifMNRAS
\fi

\section{Introduction}
\label{sec:int}


The successful explosion of a core-collapse supernova (CCSN) has posed a theoretical challenge to the astrophysical community for over half a century \citep{1966ApJ...143..626C,janka2012,burrows2013}. The advent of a growing number of detailed multi-dimensional simulations \citep{jms2016, roberts2016,oconnor_couch2018b,vartanyan2018a,vartanyan2018b,hiroki_2019,nagakura2019,ott2018_rel,summa2018,glas2019,burrows2018,burrows_2019,burrows2020,2020ApJ...896..102K,2021ApJ...916L...5V}, many of which produce robust explosions, has recently transformed the simulation landscape \citep{2021Natur.589...29B} with important implications for future gravitational wave and neutrino observations \citep{2015MNRAS.450..414F,seadrow2018,vsg2018,radice2019,andresen2019, sedda,srivastava2019,pajkos2019,pajkos2020,vartanyan2019,vartanyan2020,kuroda2017,2021MNRAS.tmp.1539N,nagakura2021}. 

Despite these remarkable advances, most studies to date are fundamentally limited in their use of spherically-symmetric progenitors (such as those derived by MESA, e.g., \citealt{2013ApJS..208....4P}, and KEPLER, e.g., \citealt{wh07, swbj16, sukhbold2018}) as initial conditions. The explosion outcome depends sensitively on the details of the progenitor structure. An important next step for progress on CCSNe theory is the study of the role of multi-dimensional progenitors pre-collapse. 

The exploration of stellar evolution in multiple dimensions has progressed through various phases over the last several decades. Early multi-dimensional studies focused on the sun and solar-mass stars \citep{sofia89,sofia2012}, as well as more massive stars \citep{sofia1978}, often including rotation. These studies, focused on stars no more massive than 10-M$_{\odot}$, generally evolved the pre-collapse evolution during phases no further than core carbon-ignition and focused on the validity of the mixing-length approach in approximating multi-dimensional convection \citep{1986ApJ...307..222C}. 

Subsequent studies focused on asphericities in more massive progenitors  pre-collapse
(e.g. \citealt{1989ApJ...341L..63A,1991ApJ...367..619F,1994ApJ...427..932A,1994ApJ...433L..41B,2000ApJ...545..435A,2007ApJ...667..448M,2017IAUS..331....1H, 2017MNRAS.465.2991J})
through two-, and later three-dimensional simulations of oxygen and silicon shell burning. These models did not generally proceed to core collapse, but rather focused again on the viability of mixing length theory in modeling convective overshoot and entrainment to ultimately improve hydrodynamic modeling in spherically-symmetric one-dimensional simulations (e.g., \citealt{2015ApJ...809...30A}). 
Massive stars are unique in that their dynamical timescales are sufficiently short to require no more than a few convective turnover timescales to generate saturated turbulence, making them well suited for the current capabilities of three-dimensional simulations.

Some attention was given to the growth of turbulence on collapse \citep{10.2307/j.ctv301fr4.17,2000ApJ...535..402L,2021MNRAS.503.3617A}
and to the shock-turbulence interaction \citep{2016MNRAS.461.3864A,2018JPhG...45e3003R}. These various explorations evolved in the last decade into efforts focused on the role of perturbations as seeds for turbulence behind the shock \citep{2013ApJ...778L...7C, 2014ApJ...785...82S, 2015ApJ...799....5C,muller_janka_pert,burrows2018,2019MNRAS.483..208N}. Initial turbulence seeds post-shock turbulence, thereby providing a greater stress behind the shock, which is favorable for shock revival into explosion \citep{bhf1995}.

It was not until recently, however, that various groups have evolved in three-dimensions the convective silicon- and oxygen-burning regions of a massive star during the final minutes before core collapse \citep{2013ApJ...778L...7C,2015ApJ...808L..21c,muller2016,2020ApJ...890...94Y,muller_lowmass,takahashi2019,fields2020,2021arXiv210704617F,2021ApJ...908...44Y}, with several models including rotation \citep{2021MNRAS.506L..20Y,2021arXiv210700173M}, and with implications for both stellar evolution \citep{2020MNRAS.497.4644M,varma21} and explosion outcome \citep{muller2017,muller2019,bollig2021}. However, detailed multi-dimensional core-collapse simulations of multi-dimensional progenitors are still lacking. To that end, we here employ our radiation-hydrodynamics code F{\sc{ornax}} to follow the core collapse, bounce, and explosion of two three-dimensional progenitors. 
We find that the three-dimensional pre-collapse progenitors explode more readily than their spherically-symmetric counterparts, emphasizing the role of preexisting progenitor turbulence in the early growth of turbulence that promotes shock revival to explosion.

The paper is organized as follows. In \S\ref{sec:setup}, we summarize the simulation setup and introduce the 3D progenitors studied here. In \S\ref{sec:results}, we discuss the morphology of the stellar models and the results of our simulations, focusing on the dependence of explosion outcome and diagnostics on the structure and dimensionality of the progenitor models. In \S\ref{sec:discussion} we discuss the role of progenitor structure on explosion and comment on both the emergent gravitational waves and the presence of the lepton emission self-sustained asymmetry (LESA). We summarize our conclusions in \S\ref{sec:conc}.

\section{Numerical Methods and Computational Setup} \label{sec:setup}

F{\sc{ornax}} \citep{skinner2019} is a multi-dimensional, multi-group radiation hydrodynamics code constructed to study core-collapse supernovae. It features an M1 solver \citep{2011JQSRT.112.1323V} for neutrino transport with detailed neutrino microphysics and a monopole approximation to general relativity \citep{marek2006}. 

For all iterations here, the models are evolved out to 20,000 km with a resolution of 678$\times$128$\times$256 in ($r$,$\mathrm{\theta}$,$\mathrm{\phi}$) using twelve fully-coupled energy bins. We use the SFHo EOS of \cite{2013ApJ...774...17S}, consistent with most known laboratory nuclear physics constraints \citep{2017ApJ...848..105T}.
Additionally, we note that F{\sc{ornax}} treats three species of neutrinos: electron-neutrinos ($\nu_e$), electron anti-neutrinos ($\bar{\nu}_e$), and $\mu$,$\tau$ neutrinos and their antiparticles bundled together as heavy-neutrinos, ``$\nu_\mu$."
The setup here is otherwise identical to those discussed in \cite{vartanyan2018b}. 
We introduce no perturbations of our own into the models.

We study the 12.5-M$_{\odot}$ progenitor presented in \cite{muller_lowmass} and the 15-M$_{\odot}$ progenitor from \cite{fields2020} (labeled there as 3D32kmPert).
\cite{muller_lowmass} simulate convective burning in the active oxygen shell in 3D for the last few minutes prior to collapse following \cite{2016ApJ...833..124M}.
\cite{fields2020} simulate the last $\sim$424 seconds before core collapse, evolving the Si and O burning shells in 3D. To isolate the role of multi-dimensional effects in the progenitor, we also study the outcomes of otherwise equivalent, but 1D spherically-symmetric, 12.5-M$_{\odot}$ and 15-M${_\odot}$ initial models evolved with MESA (provided by B. M\"uller and C. Fields, respectively; private correspondence).
In total, we perform four 3D simulations, encompassing two stellar progenitors each evolved until core collapse in both 1D, labeled subsequently, M12.5-1D3D and FC15-1D3D, and in 3D, labeled M12.5-3D3D and FC15-3D3D, for the \cite{muller_lowmass} and \cite{fields2020} models, respectively. The 1D3D models are mapped to 3D 10 ms after core-bounce such that, after core-bounce, all models are evolved in 3D.

\begin{figure}
    \centering
    \includegraphics[width=0.47\textwidth]{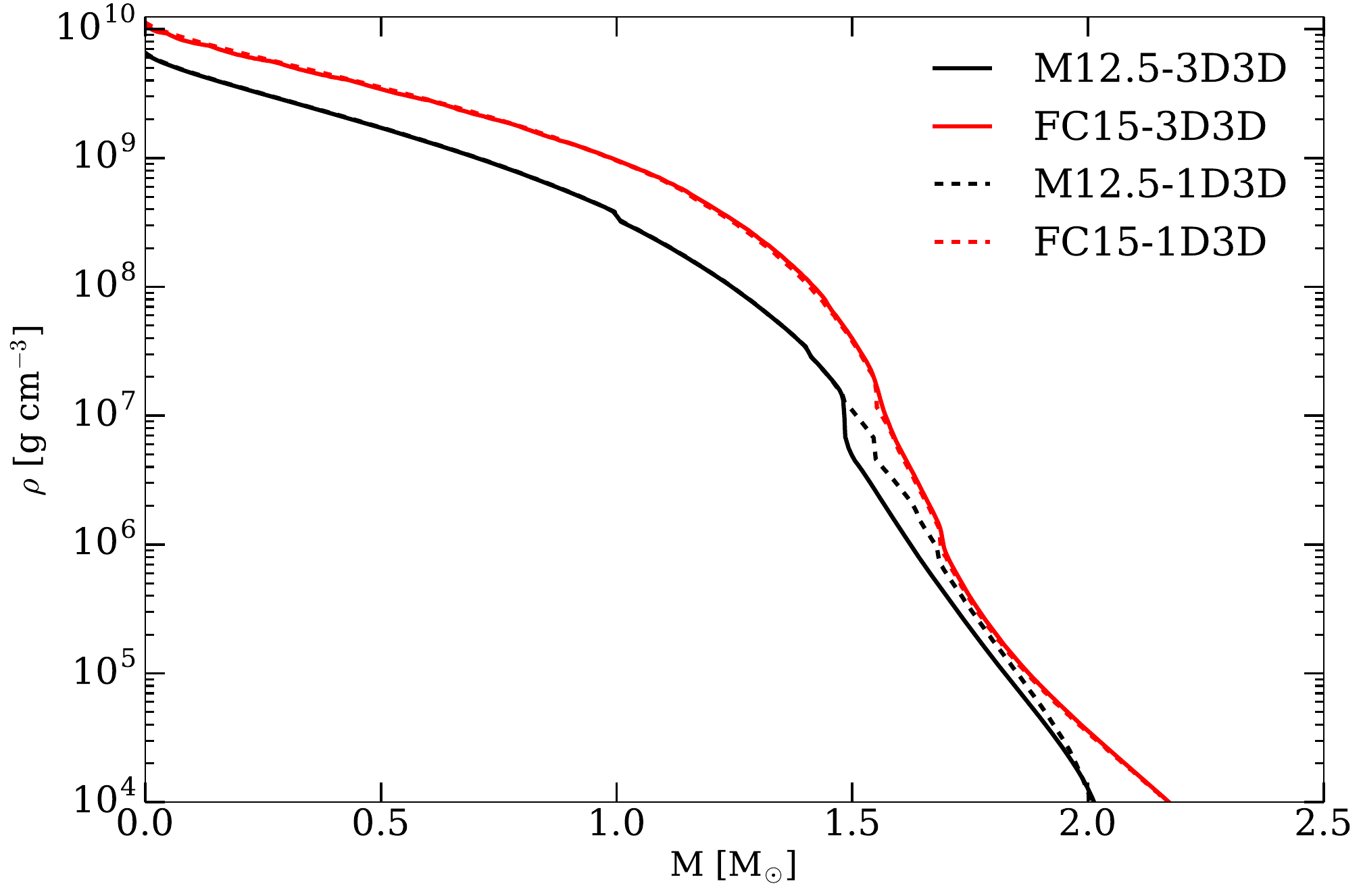}
    \caption{The angle-averaged initial density (in g cm$^{-3}$) profiles as a function of enclosed mass (in M$_{\odot}$) for the four models studied here.}
    \label{fig:profile}
\end{figure}

\begin{figure}
    \centering
     \includegraphics[width=0.47\textwidth]{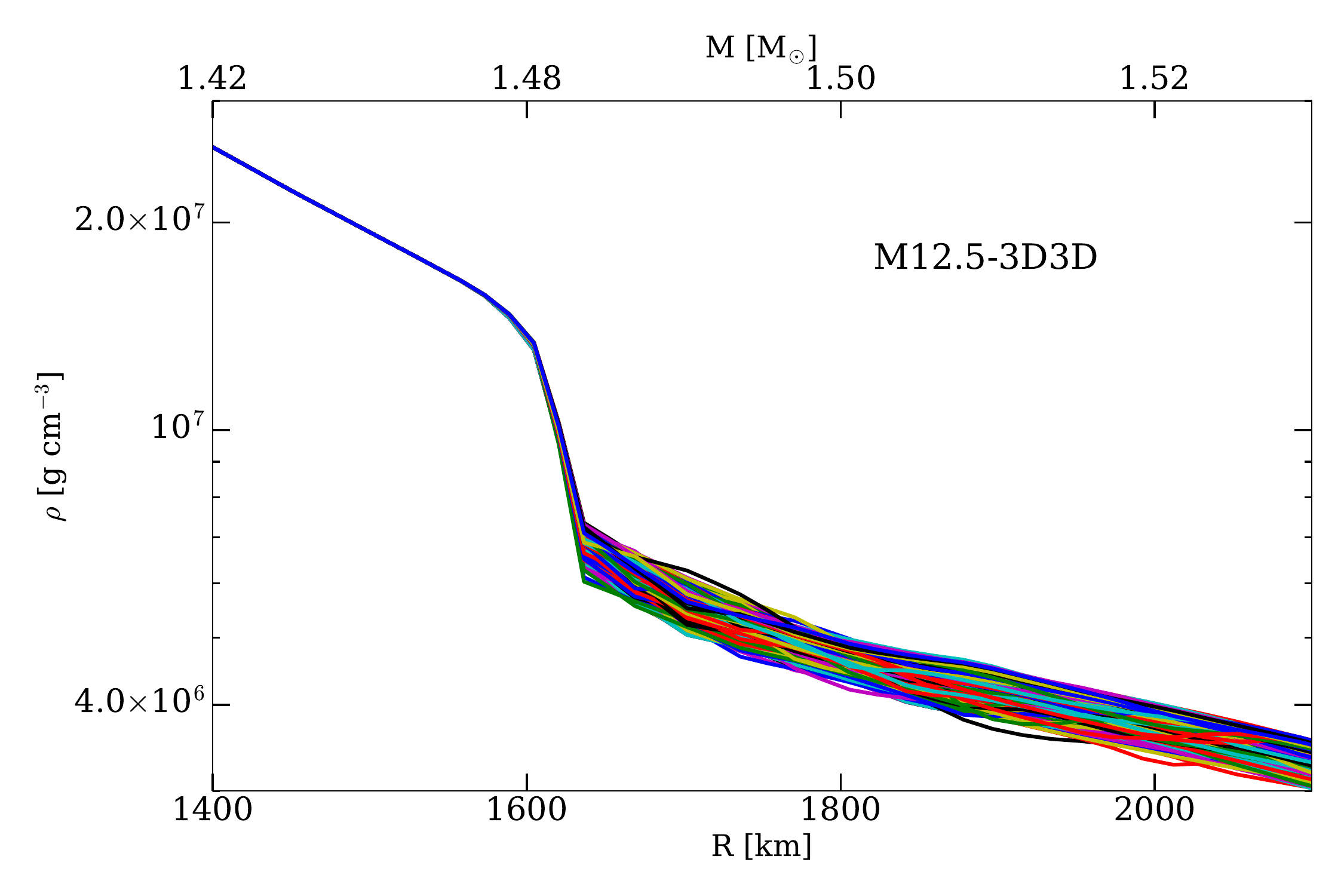}
    \caption{The pre-collapse Si/O interface density (in g cm$^{-3}$) as a function of radius for the stellar interior  (in km) for the M12.5 model as a function of various viewing angles  in $\theta$, $\phi$. Note that the interface location is isotropic (independent of angle) at its inner boundary, i.e. the Si/O interface is at the same radius for all angles for this 3D progenitor. The interface has a width of $\sim$30 km, or $\sim$0.02 M$_{\odot}$. The density drops by a factor of $\sim$3 across the interface, and varies by $\sim$50\% with viewing angle at the outer boundary. } 
    \label{fig:rho_R}
\end{figure}

Many existing studies of multi-dimensional progenitors use simplified neutrino physics and radiation transport (e.g., the ray-by-ray or fast-multigroup approximations to neutrino radiation transport).
Our work here has the advantage of using detailed neutrino microphysics with the robust M1 closure scheme.
We show in Fig.\,\ref{fig:profile} the angle-averaged density profiles of the four models studied here. 
The FC15 models show only slight variations between the 1D and 3D progenitor models when density is angle-averaged. 
In contrast, the M12.5 3D progenitor has a sharper density dropoff at the Si/O interface than the 1D progenitor, which we find promotes explosion (see \S\ref{sec:results}). 
In Fig.\,\ref{fig:rho_R}, we illustrate the nature of the Si/O interface for the M12.5-3D3D progenitor for a variety of viewing angles as a function of enclosed mass and radius. We see a much smaller variation by angle in the pre-collapse FC15-3D3D model around the Si/O interface.
The interface itself is quite narrow $-$ roughly 30 km and $\sim$0.02 M$_{\odot}$ wide. 
With early infall speeds of $\sim$1000 km s$^{-1}$ and accretion rates of 1 M$_{\odot}$ s$^{-1}$, the corresponding time to accrete across the interface is $\sim$10-20 ms. 

\begin{figure*}
    \centering
    \includegraphics[width=0.47\textwidth]{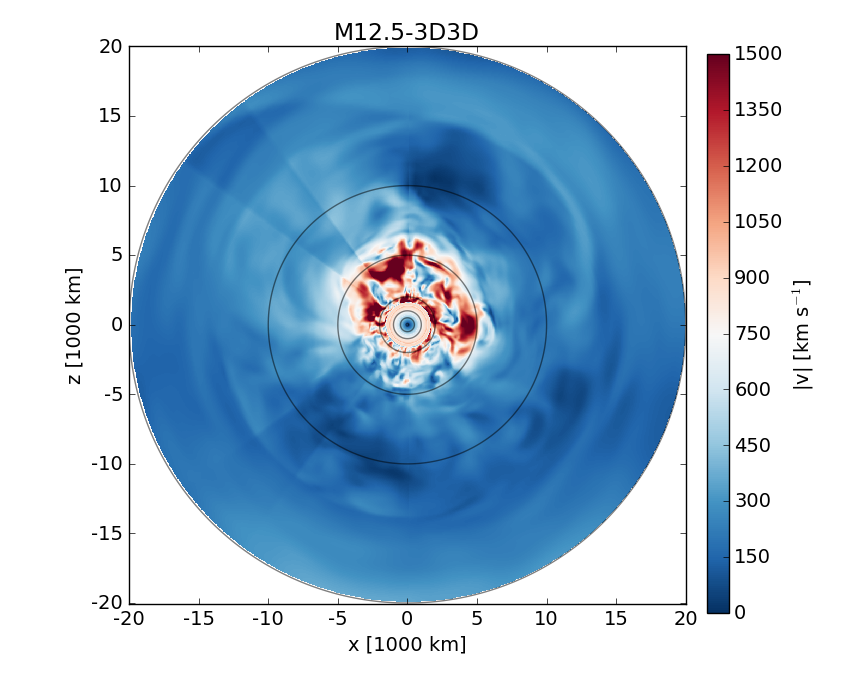}
    \includegraphics[width=0.47\textwidth]{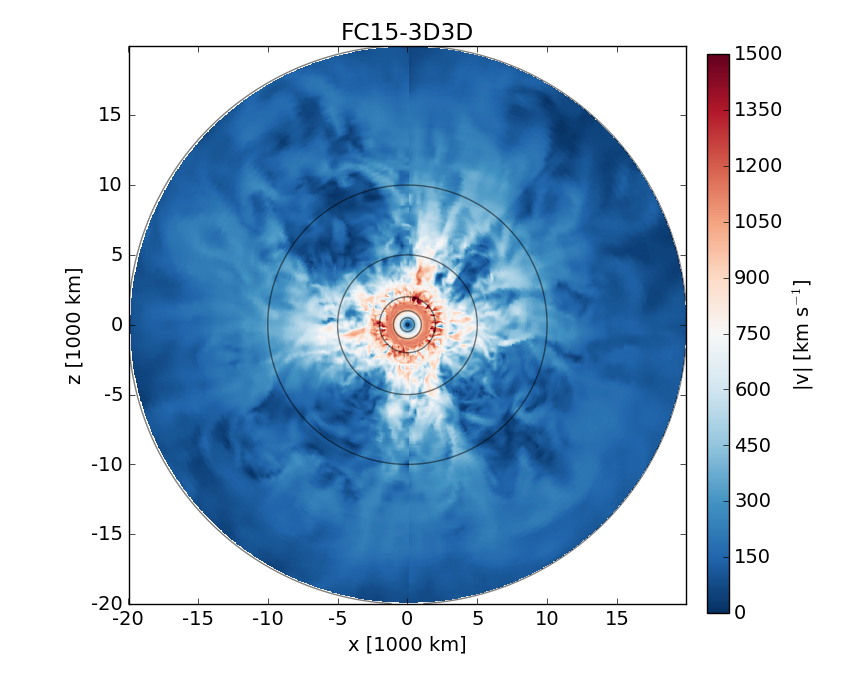}
    \caption{The initial velocity profiles along the xz-plane for the M12.5 3D model (\textbf{left}) and the FC15 3D model (\textbf{right}). Mean radial speeds are $\sim$-1000 km s$^{-1}$ as expected at infall, with turbulent velocities $\sim$500 km s$^{-1}$. We identify a weak axial artifact and the remnants of the quadrupolar perturbations from the Cartesian grid in the FC15 model. The M12.5 model, which uses a yin-yang grid, does not manifest axial artifacts but shows signs of the yin-yang grid stitching at large scales.}
    \label{fig:vel} 
\end{figure*}

The M12.5-3D3D and FC15-3D3D models differ significantly in the extent of turbulence, with the former showing no evidence of turbulence below the sharp Si/O interface.
In Fig.\ref{fig:vel}, we illustrate the initial velocity slices for the M12.5-3D3D model (left) and the FC15-3D3D model (right). 
In both instances, the 3D evolution around the oxygen-burning shells is evident. 
We note that both models feature unique simulation artifacts: the M12.5 model shows yin-yang grid stitching at large scales, and the FC15 model shows a large-scale quadrupolar symmetry. 
In both cases, the artifacts lie outside one of the main region of interest, the Si/O interface. The artifacts at large scales are either not accreted at all, or are accreted so late that they have no influence on the explosion evolution.
The artifacts at small scales are accreted so rapidly, becoming part of the PNS, that they also have no noticeable impact on the explosion evolution.


To further describe our results, we find it useful to define some averaging procedures: for an arbitrary fluid/radiation variable $f$ we define the shell-average and density-weighted shell-average as
\begin{align}
    \label{eqn:shell-mean}
    \left<f\right>_{\Omega}\left(t,r\right) &\equiv \int f\left(t,r,\theta,\phi\right)\;{\rm d}\Omega,\\
    \shortintertext{and}
    \label{eqn:rho_shell-mean}
    \left<f\right>_{\rho}\left(t,r\right)&=\dfrac{\left<\rho f\right>_{\Omega}}{\left<\rho\right>_{\Omega}},
\end{align}
respectively.

\section{Results} \label{sec:results}

\begin{figure}
    \includegraphics[width=\linewidth]{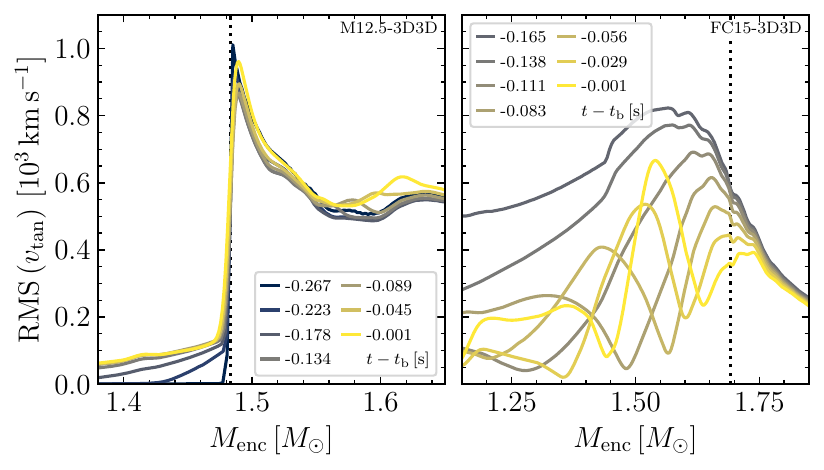}
    \caption{RMS $v_{\rm tan}$ ($\sqrt{\left<v_{\rm tan}^2\right>_\rho}$) profiles at several times before bounce as indicated in the legends (in seconds). The vertical dotted black line is the initial mass coordinate of the Si/O interface. For the \muller \ddd model (left), we can clearly see the Si/O interface in $v_{\rm tan}$, while there is no such discernible feature in the FC15-3D3D model (right). The evolution of $v_{\rm tan}$ before bounce in the M12.5-3D3D model is minimal, while for the FC15-3D3D model $v_{\rm tan}$ changes significantly.
    }
    \label{fig:rms_vtan_init_prof}
\end{figure}

\begin{figure*}
    \centering
    \includegraphics[width=0.47\textwidth]{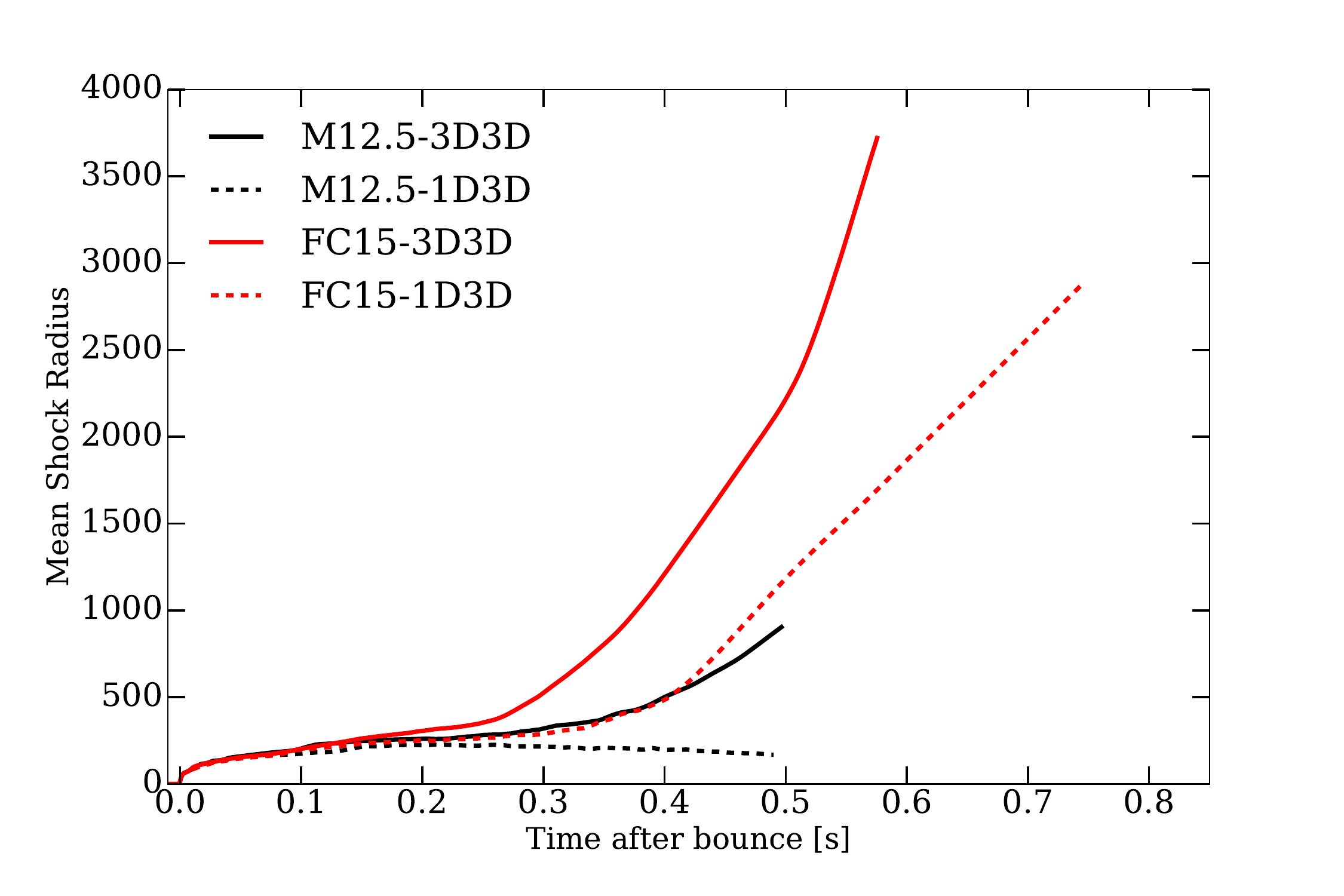}
    \includegraphics[width=0.47\textwidth]{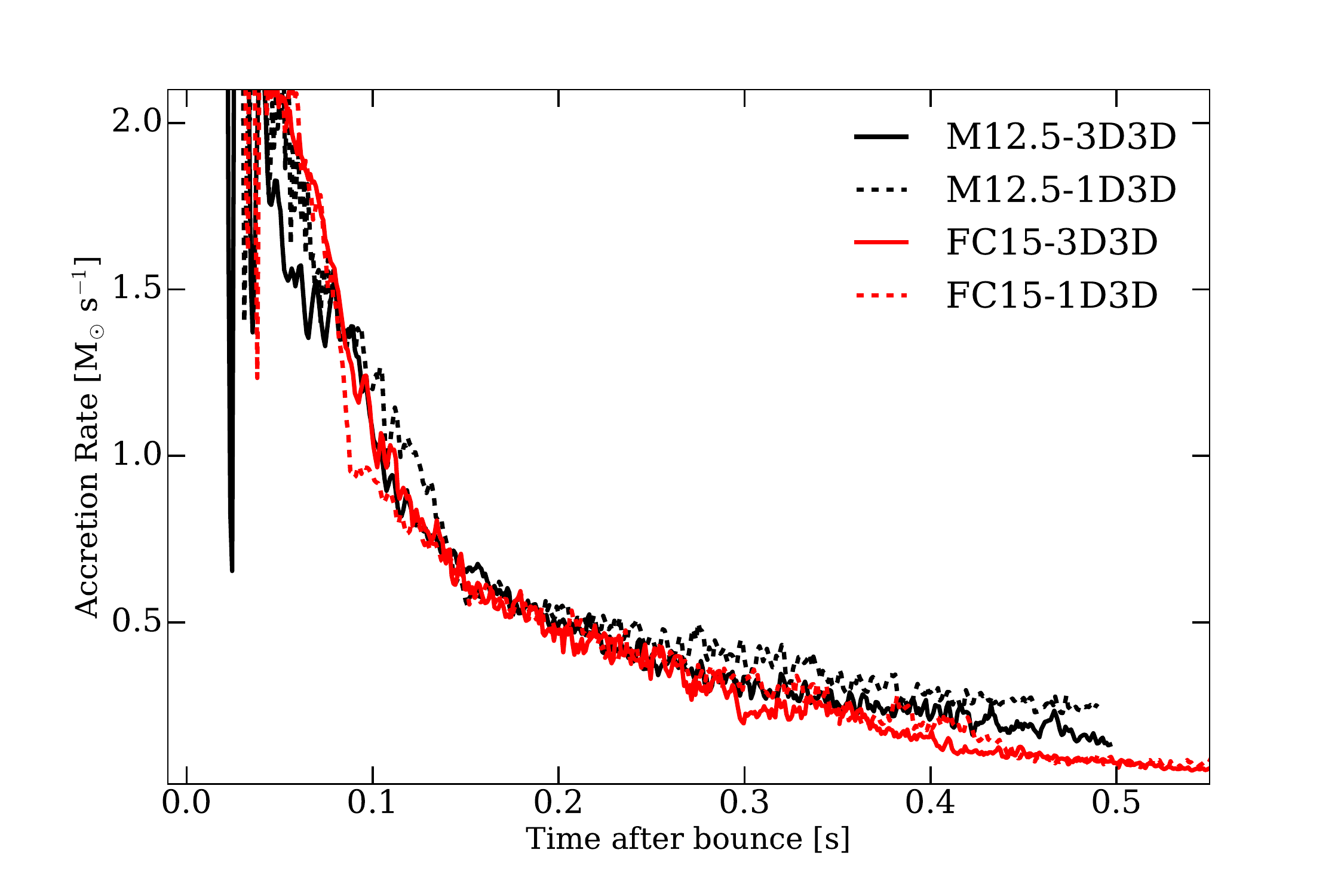}
    \caption{\textbf{Left}: Mean shock radius (in km) as a function of time after bounce (in seconds) for simulations of the 12.5- and 15-M$_{\odot}$ models studied here for both the 1D MESA model mapped to 3D, and the 3D progenitors evolved entirely in 3D. For both models, the fully 3D progenitor is more explodable. For the 12.5-M$_{\odot}$ set, only the fully 3D model (blue) explodes. The 1D MESA progenitor, whether mapped to 3D or 2D, both 10 ms after bounce, fails to explode. Explosion time of the 3D progenitor just after 100 ms corresponds with the accretion of the Si/O interface. \textbf{Right}: Accretion rate (in M$_{\odot}$ s$^{-1}$) as a function of time after bounce (in seconds).}
    \label{fig:rs}
\end{figure*}

The four CCSNe models were evolved in 3D for roughly half a second to one second post bounce to determine their early fates. 
In the early initial pre-bounce phases we plot in Fig.~\ref{fig:rms_vtan_init_prof} the behavior of the RMS tangential velocity profiles versus mass coordinate for both 3D progenitor models, which show nearly no evolution for the M12.5-3D3D model and significant evolution below the Si/O interface for the FC15-3D3D model.
In Fig.\,\ref{fig:rs}, we illustrate the angle-averaged shock radii (left) and accretion rates at 100 km (right). All models except for the M12.5-1D3D explode, with shocks reaching runaway growth within $\sim$300 ms. We find that the 3D progenitor models explode preferentially $-$ the FC15-3D3D model explodes $\sim$100 ms earlier than its FC15-1D3D counterpart, whereas the M12.5-3D3D explodes, while the M12.5-1D3D fails to explode. 

\begin{figure*}
    \centering
    \includegraphics[width=0.47\textwidth]{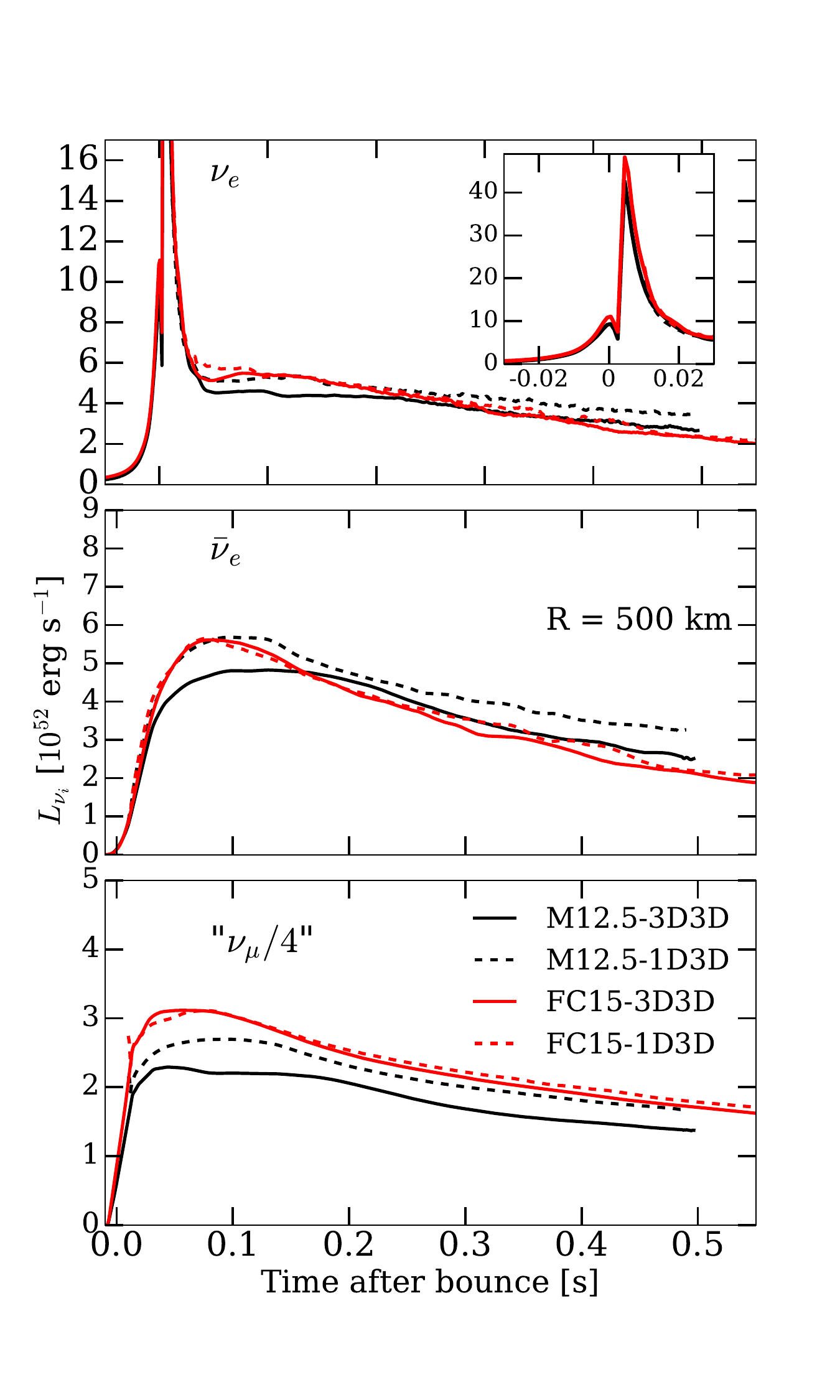}
    \includegraphics[width=0.47\textwidth]{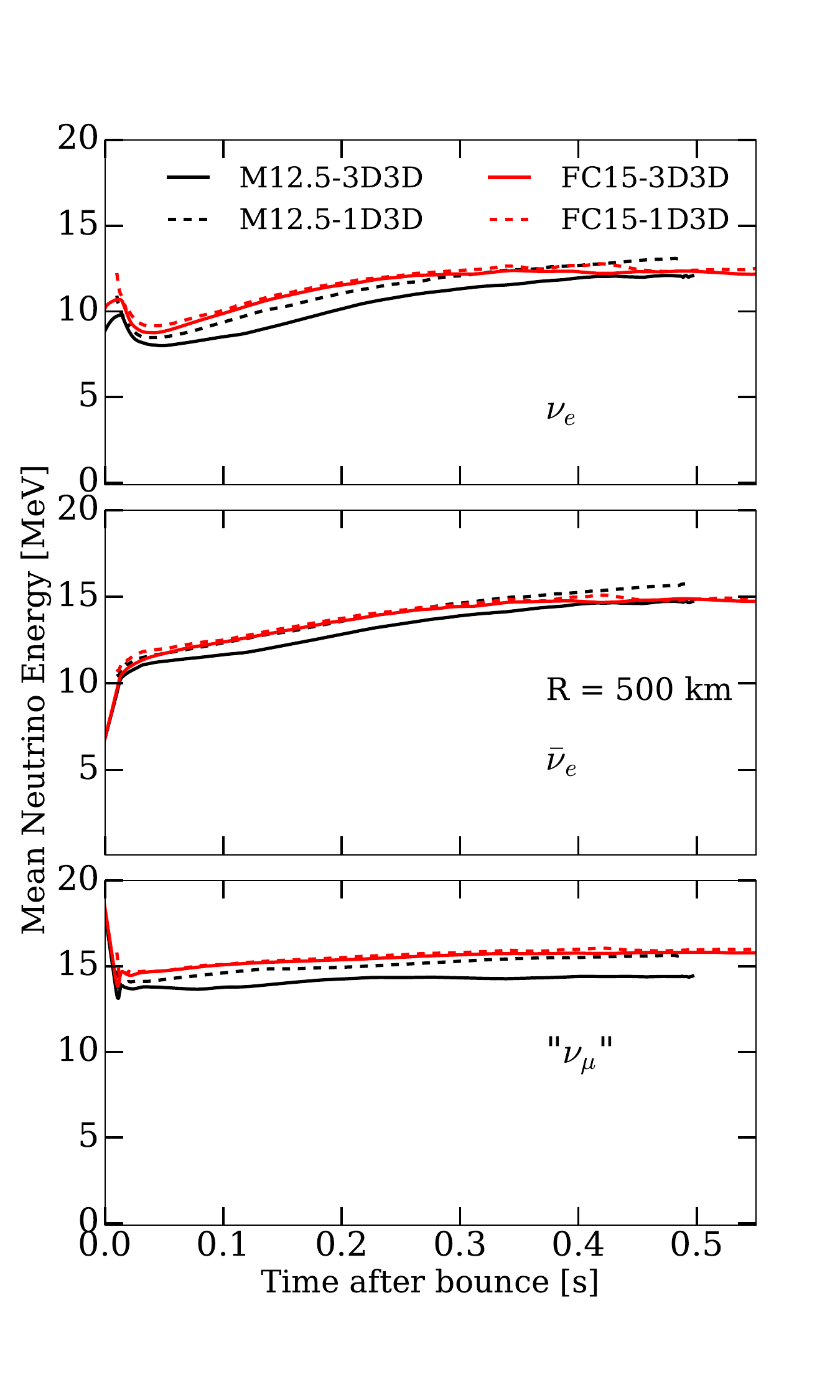}
    \caption{Neutrino luminosity (in 10$^{51}$ erg s$^{-1}$, \textbf{left}) and neutrino average energy (in MeV, \textbf{right}) as a function of time after bounce (in seconds). We include an inset for the electron-neutrino luminosity in the top left panel with a different scale showing the early breakout peak.}
    \label{fig:lum}
\end{figure*}

In Fig.\,\ref{fig:lum}, we illustrate the angle-averaged neutrino luminosities and mean energies for the four models studied here. The FC15 models have harder spectra for all species until $\sim$400 ms, at which point the FC15-3D3D model explodes, accretion is reversed, and the electron neutrino and anti-neutrino energies decrease relative to those of the non-exploding model FC15-1D3D. The FC15-1D3D explodes, and the mean energies respond $\sim$100 ms later. Otherwise, the neutrino luminosity and energy of these two models behave similarly. Note that the explosion barely affects the heavy `$\mu$' neutrino  energies, which remain harder throughout, suggesting a deeper origin in the proto-neutron star. We see similar behavior for the neutrino luminosities, with explosion and accretion reversal truncating the electron-neutrino and anti-neutrino luminosities, but leaving the `heavy' neutrino luminosities unaffected. The non-exploding M12.5-1D3D model has the highest electron-neutrino and anti-neutrino energies at late times due to sustained accretion. Also interesting is that model M12.5-1D3D shows markedly different neutrino luminosities from the M12.5-3D3D model in the first 100 ms, even though the latter has not yet exploded. This is due to progenitor profile differences that manifest in a higher accretion rate for the 1D model than the 3D model. Additionally, model M12.5-1D3D sustains a higher `heavy' neutrino luminosity than the M12.5-3D3D model for the entirety of the simulation, even before the latter explodes.

\begin{figure*}
    \centering
    \includegraphics[width=0.47\textwidth]{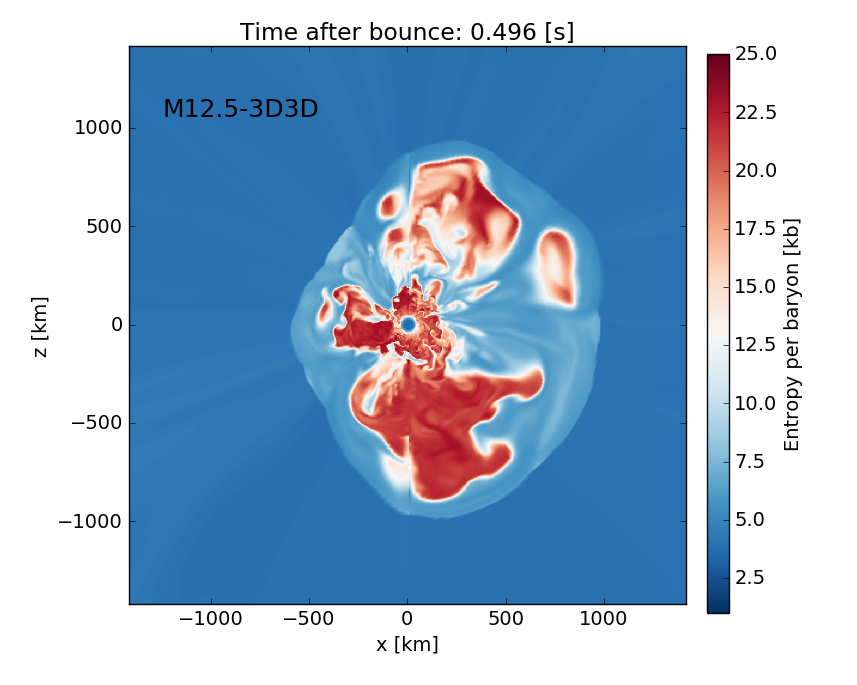}
    \includegraphics[width=0.47\textwidth]{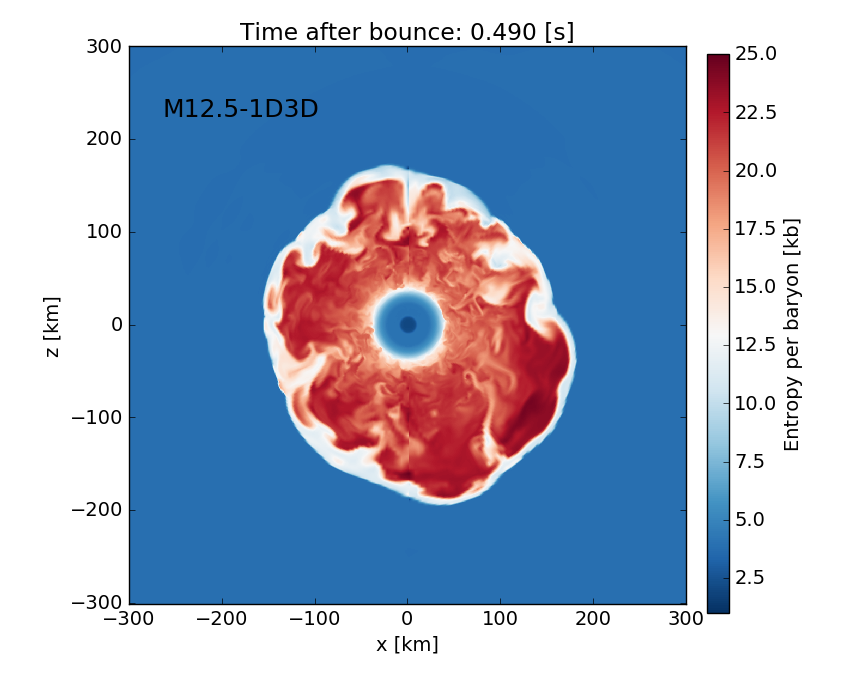}
    \includegraphics[width=0.47\textwidth]{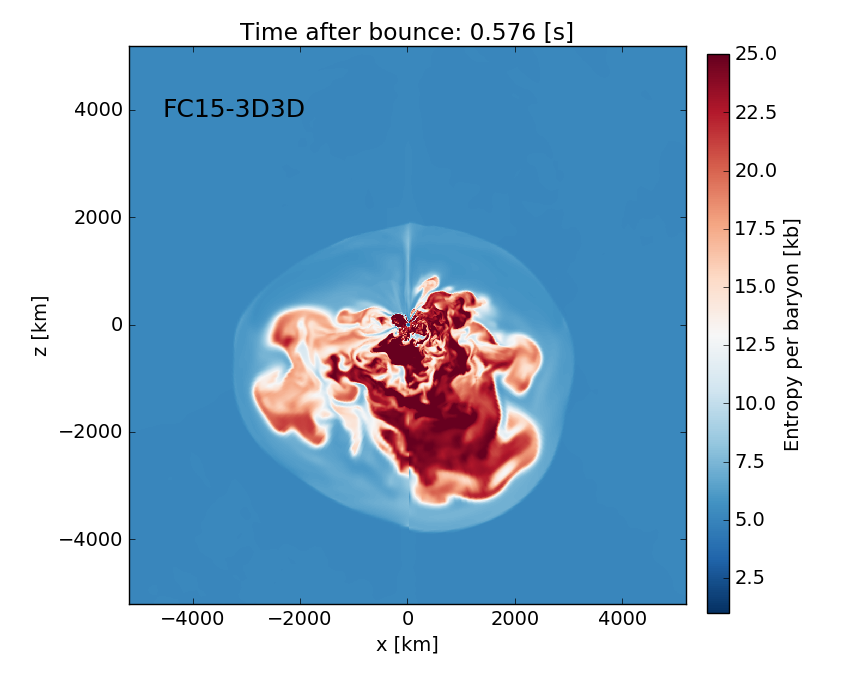}
    \includegraphics[width=0.47\textwidth]{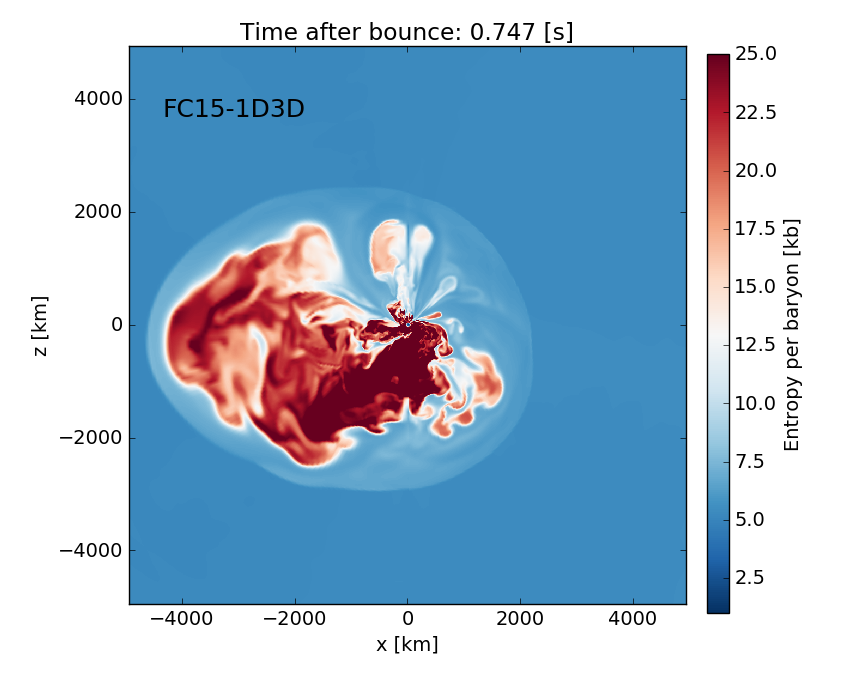}
    \caption{\textbf{Top}: The entropy profiles along the xz-plane of the M12.5-3D3D (\textbf{left}) and M12.5-1D3D (\textbf{right}) models at the last time step of our simulations. The latter fails to explode (note the  different scales). The former explodes asymmetrically. \textbf{Bottom}: Same as above, but for the FC15-3D3D (\textbf{left}) and the FC15-1D3D (\textbf{right}) models. Both models here explode. Note that the more massive FC15 progenitors explosions are focused entirely in one hemisphere, whereas the less massive M12.5-3D3D model explodes with plumes in both hemispheres.}
    \label{fig:ent} 
\end{figure*}

\begin{figure*}
    \includegraphics[width=0.47\textwidth]{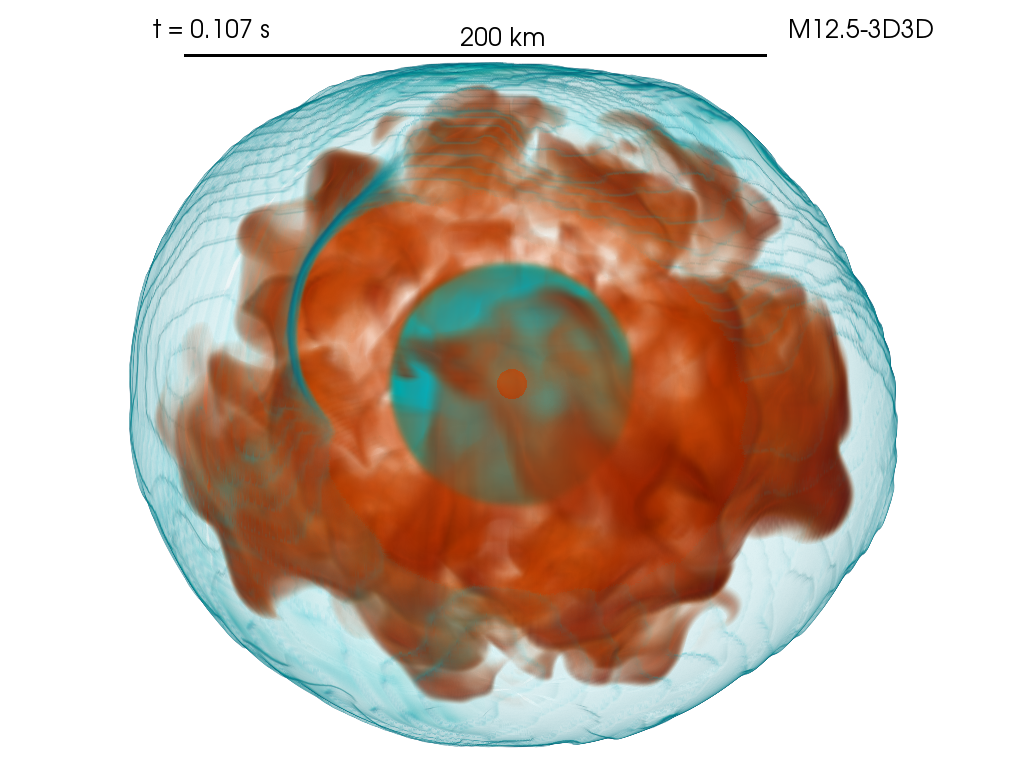}
    \includegraphics[width=0.47\textwidth]{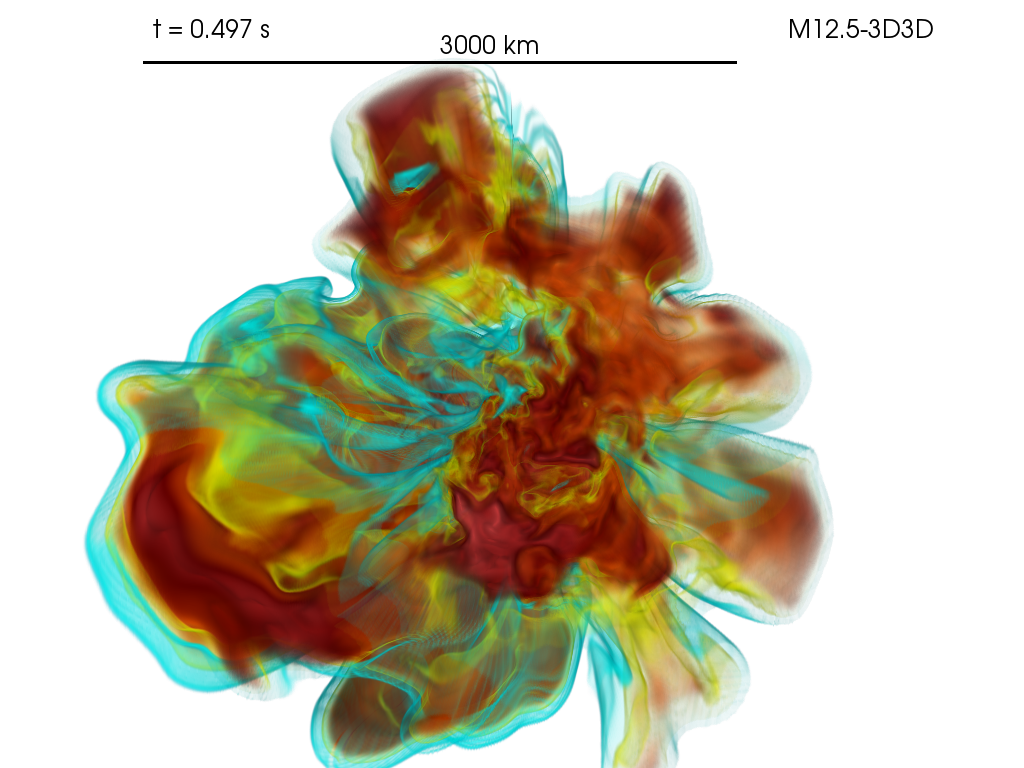}
    \caption{Entropy visualization at various times for model M12.5-3D3D. The cyan veil indicates the shock surface. The length scale changes with time.}
    \label{fig:vis_M3} 
\end{figure*}

\begin{figure*}
    \includegraphics[width=0.47\textwidth]{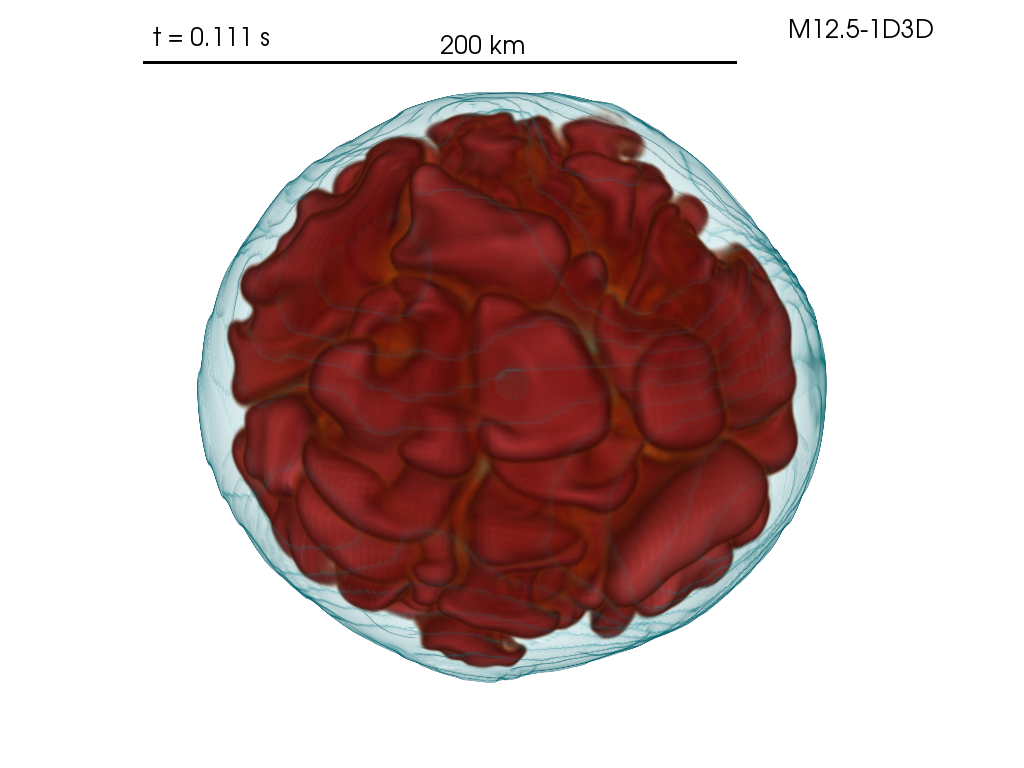}
    \includegraphics[width=0.47\textwidth]{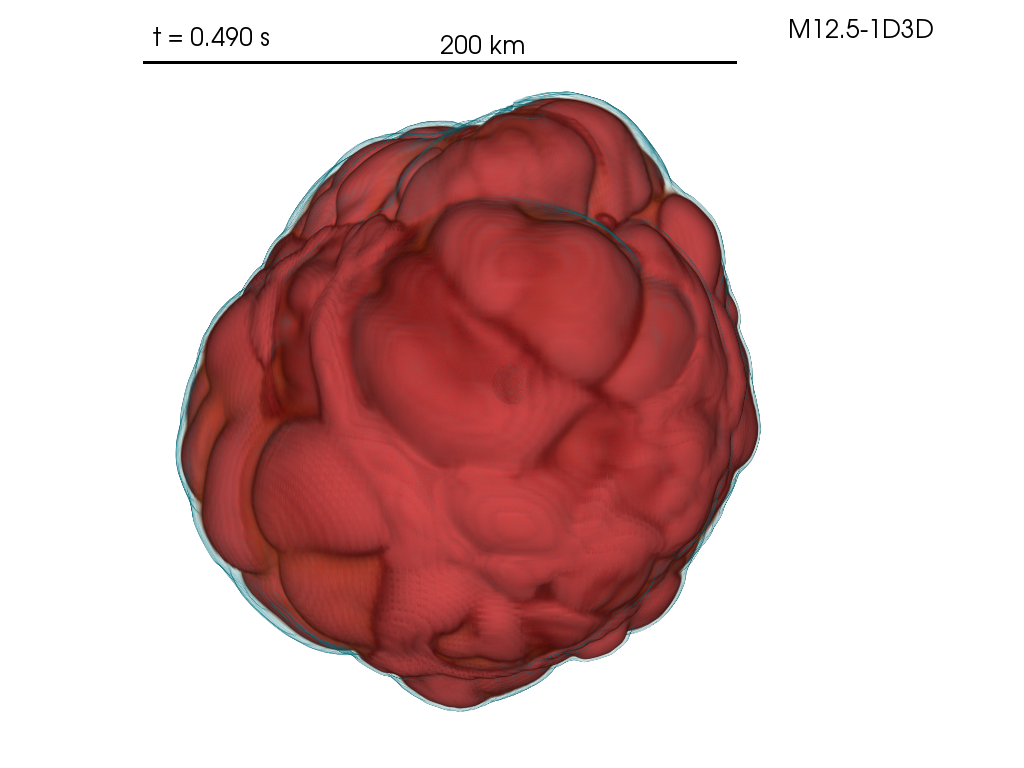}
    \caption{Same as above, but for model M12.5-1D3D, with times chosen analogous to the prior model to indicate the model's time evolution. This model fails to explode. Note the smaller scale turbulence in the leftmost figure compared with model M12.5-3D3D, indicating the significance of pre-bounce perturbations in driving large-scale turbulence.}
    \label{fig:vis_M1} 
\end{figure*}

\begin{figure*}
    \includegraphics[width=0.47\textwidth]{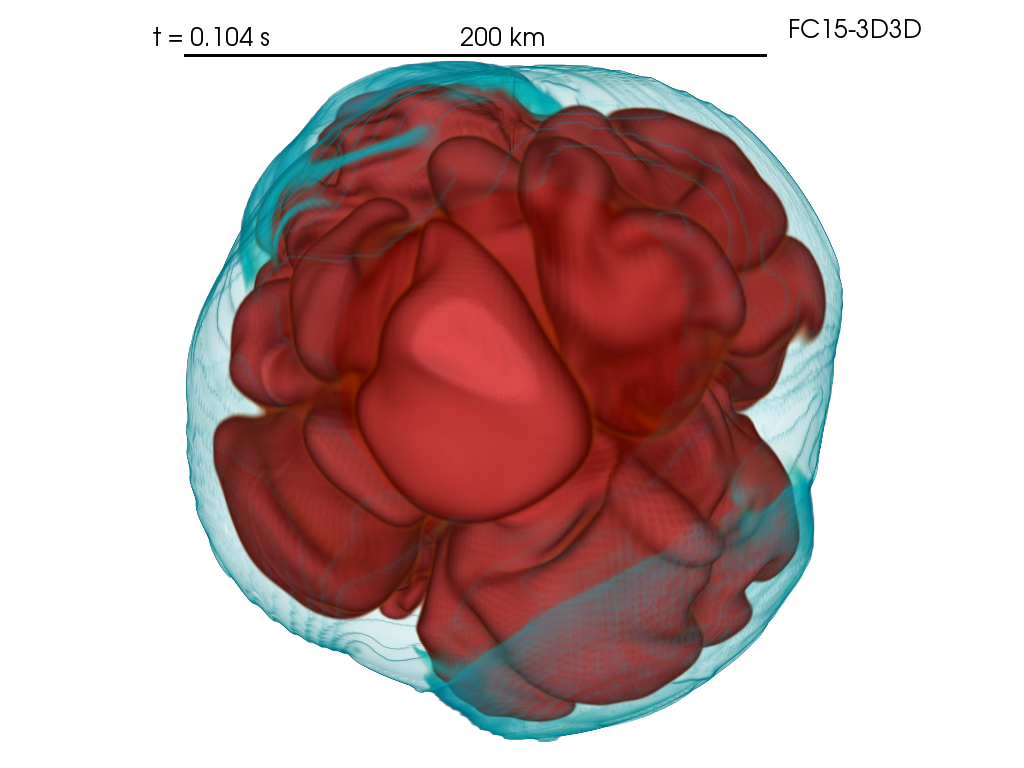}
    \includegraphics[width=0.47\textwidth]{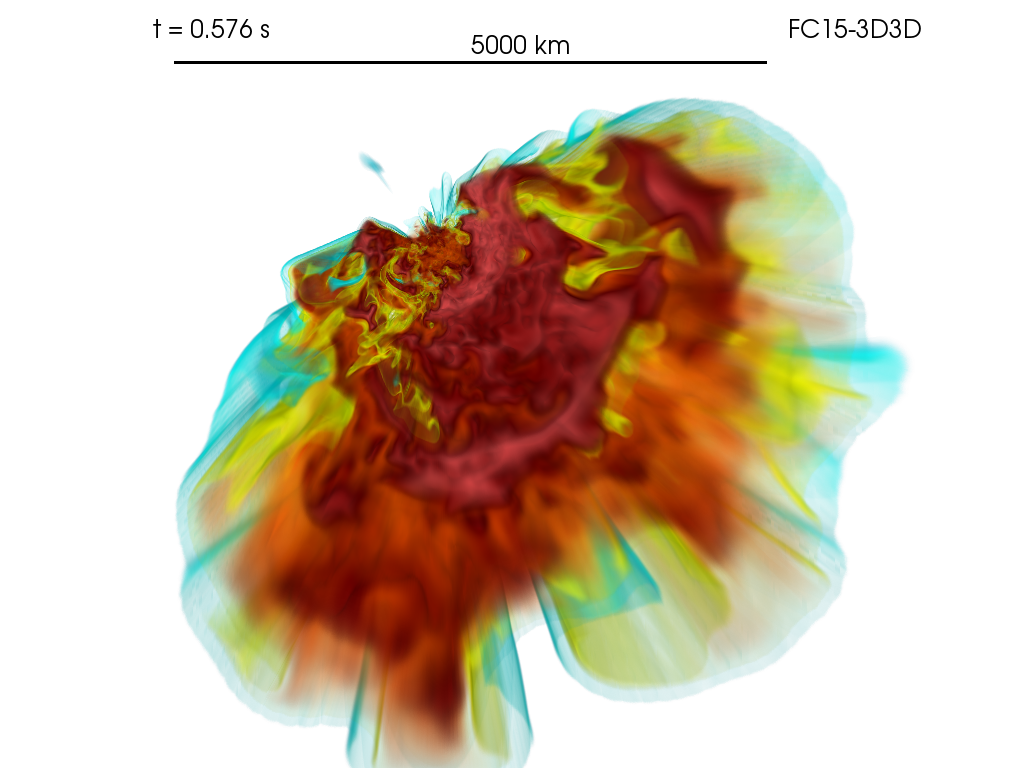}
    \caption{Same as above, but for model FC15-3D3D.}
    \label{fig:vis_F3} 
\end{figure*}

\begin{figure*}
    \includegraphics[width=0.47\textwidth]{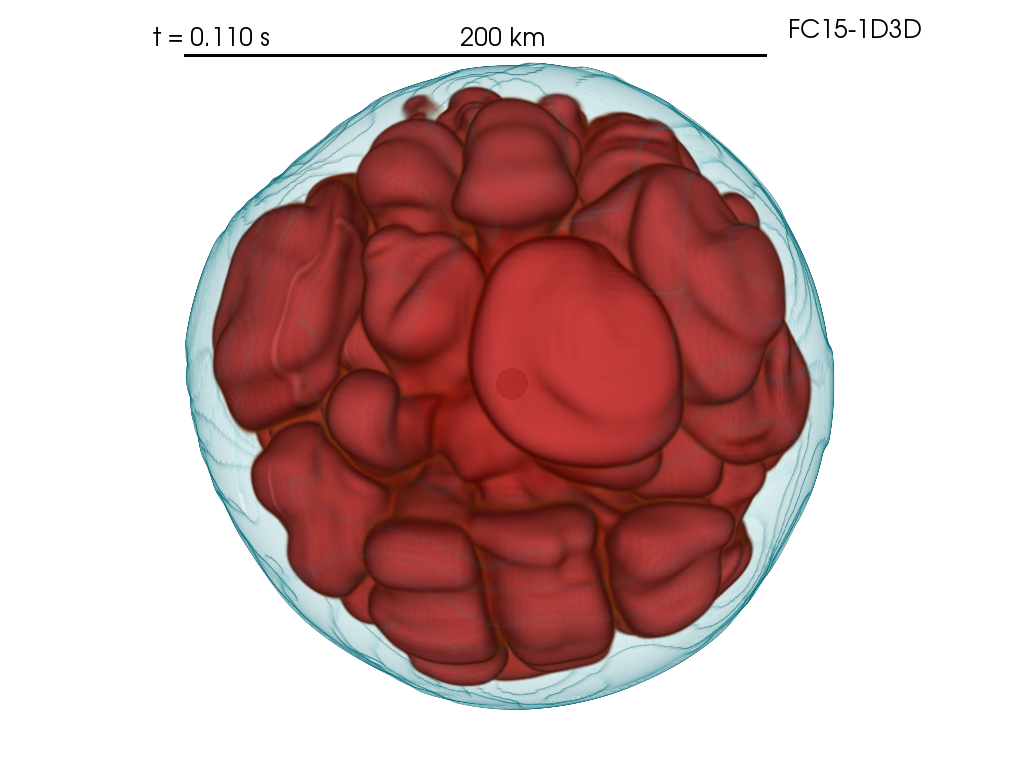}
    \includegraphics[width=0.47\textwidth]{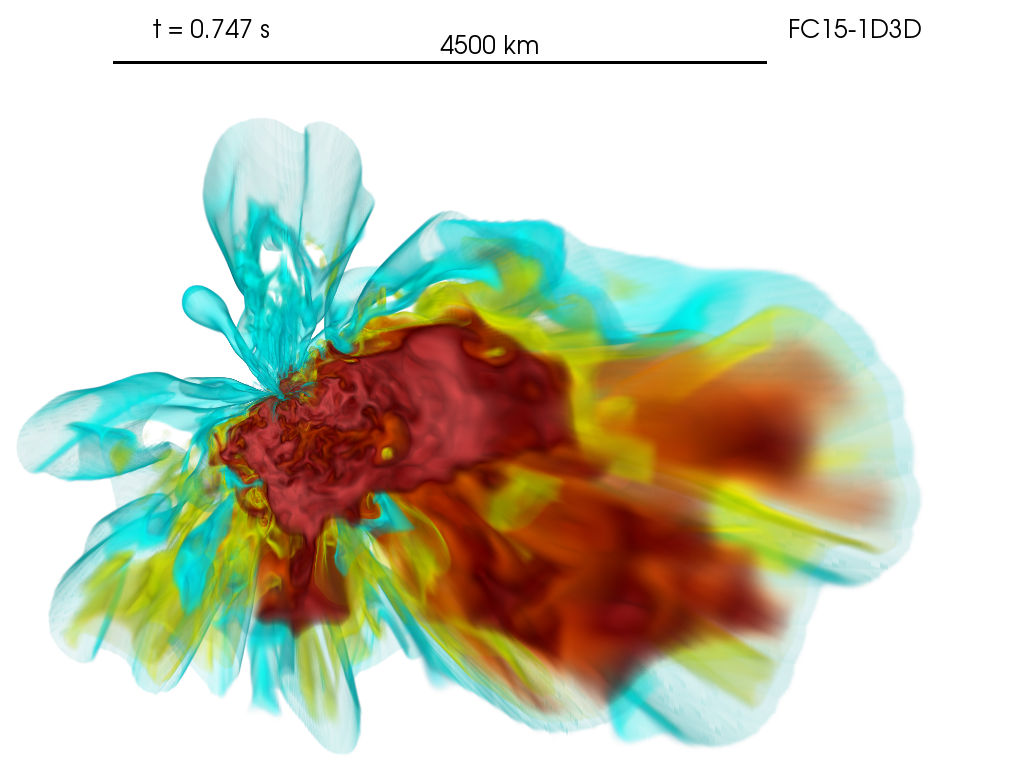}
    \caption{Same as above, but for model FC15-1D3D. Note that the plume size is again slight smaller at similar times post bounce compared to FC15-3D3D}
    \label{fig:vis_F1} 
\end{figure*}

In Fig.\,\ref{fig:ent}, we show late-time 2D slices of the entropy for the four models studied here. All the models that do explode do so asymmetrically, with multiple plumes in various directions. Model M12.5-3D3D maintains a dipolar outflow at late times near the north-south axis, while models FC15-3D3D and FC15-1D3D develop unipolar ejecta. We show 3D renderings of all four models at early and late times in Figs.\,\ref{fig:vis_M3}-\ref{fig:vis_F1}. The more massive models FC15-3D3D and FC15-1D3D explode more asymmetrically than model M12.5-3D3D and with higher entropies at late times. Some salient metrics for all the models are summarized in Table\,\ref{sn_tab}. We note that the compactness parameter does not prove a valuable metric of explodability for these models (see also e.g., \citealt{burrows2020}).\footnote{Model FC15-3D3D explodes preferentially to, but with a larger compactness parameter than, model FC15-1D3D. Model M12.5-3D3D has a smaller compactness parameter, but explodes, whereas model M12.5-1D3D does not.}

\begin{table*}
\center\caption{\Large{Explosion Properties}}
\begin{tabular}{*{7}{p{15mm}}}
    \hline\hline
  Model        & \multicolumn{2}{c}{Explosion?} & \multicolumn{2}{c}{Compactness Parameter} & \multicolumn{2}{c}{Gravitational Overburden [erg]} \\
            (M$_{\odot}$)    & 3D3D &  1D3D & 3D3D &  1D3D  & 3D3D &  1D3D \\\hline
M12.5    &    \checkmark\checkmark     &    \xmark    & 0.312 & 0.368 & -2.80$\times$10$^{49}$ & -4.31$\times$10$^{49}$\\
FC15  &  \checkmark\checkmark       &     \checkmark  & 0.424 & 0.414 & -4.66$\times$10$^{49}$ & -8.24$\times$10$^{49}$\\
\hline
  \end{tabular}

  \begin{flushleft}\large{Table of our 3D simulation results: models with a checkmark explode, and models with an \xmark\, do not explode. Models with two checkmarks explode first for a given row, if the the counterpart explodes at all. We also show the compactness parameter (at 1.75 M$_{\odot}$) and gravitational binding energy, or an overburden, exterior to our grids(in erg). The 1D3D models had larger gravitational overburdens than the 3D3D models. Note that models with a smaller compactness parameter are not necessarily more explodable, as model FC15-3D3D has a higher compactness, but explodes earlier than model FC-1D3D.}  \end{flushleft}
  \label{sn_tab}
\end{table*}\section{Discussion}\label{sec:discussion}

\begin{figure*}
    \centering
    \includegraphics[width=0.47\textwidth]{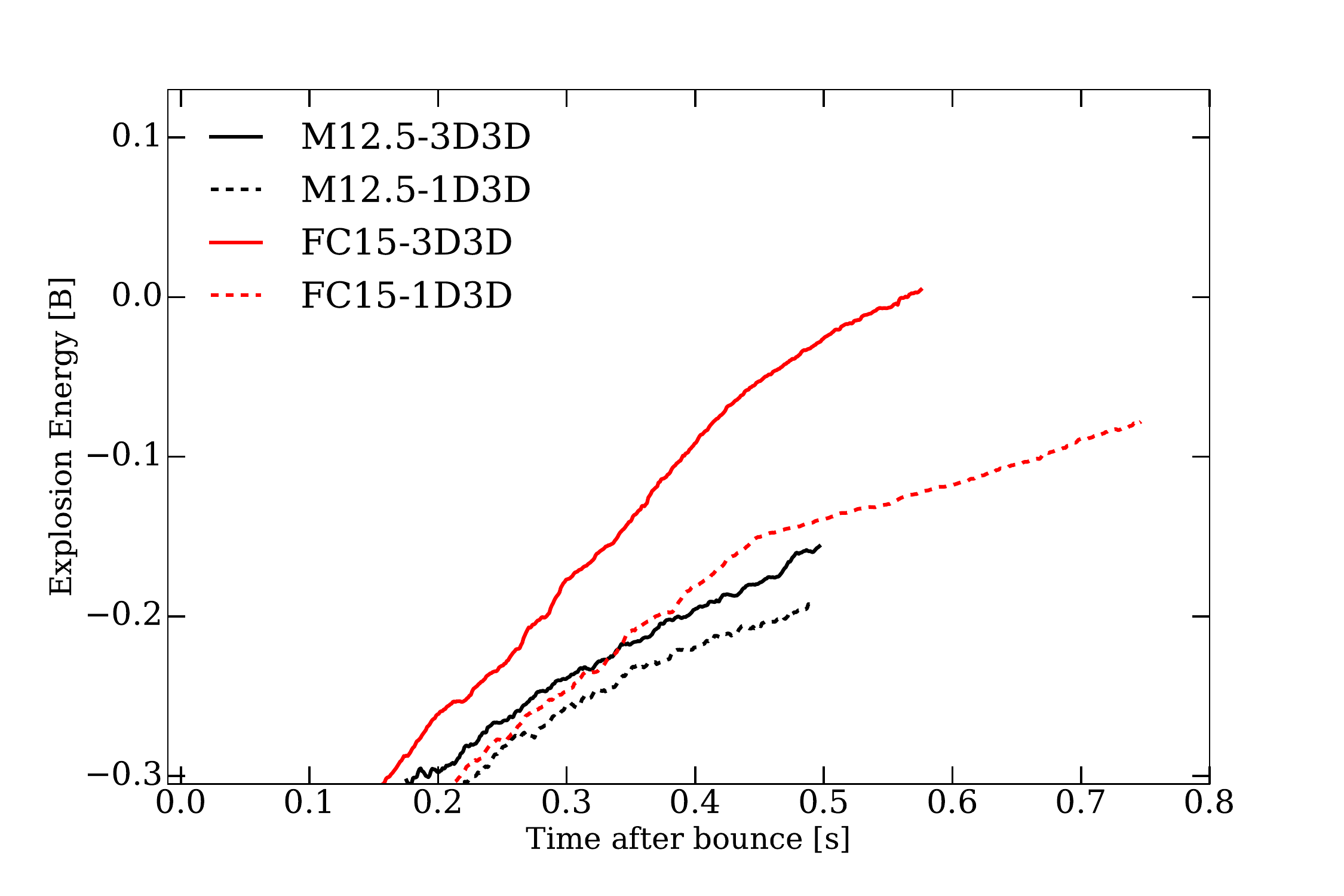}
   \includegraphics[width=0.47\textwidth]{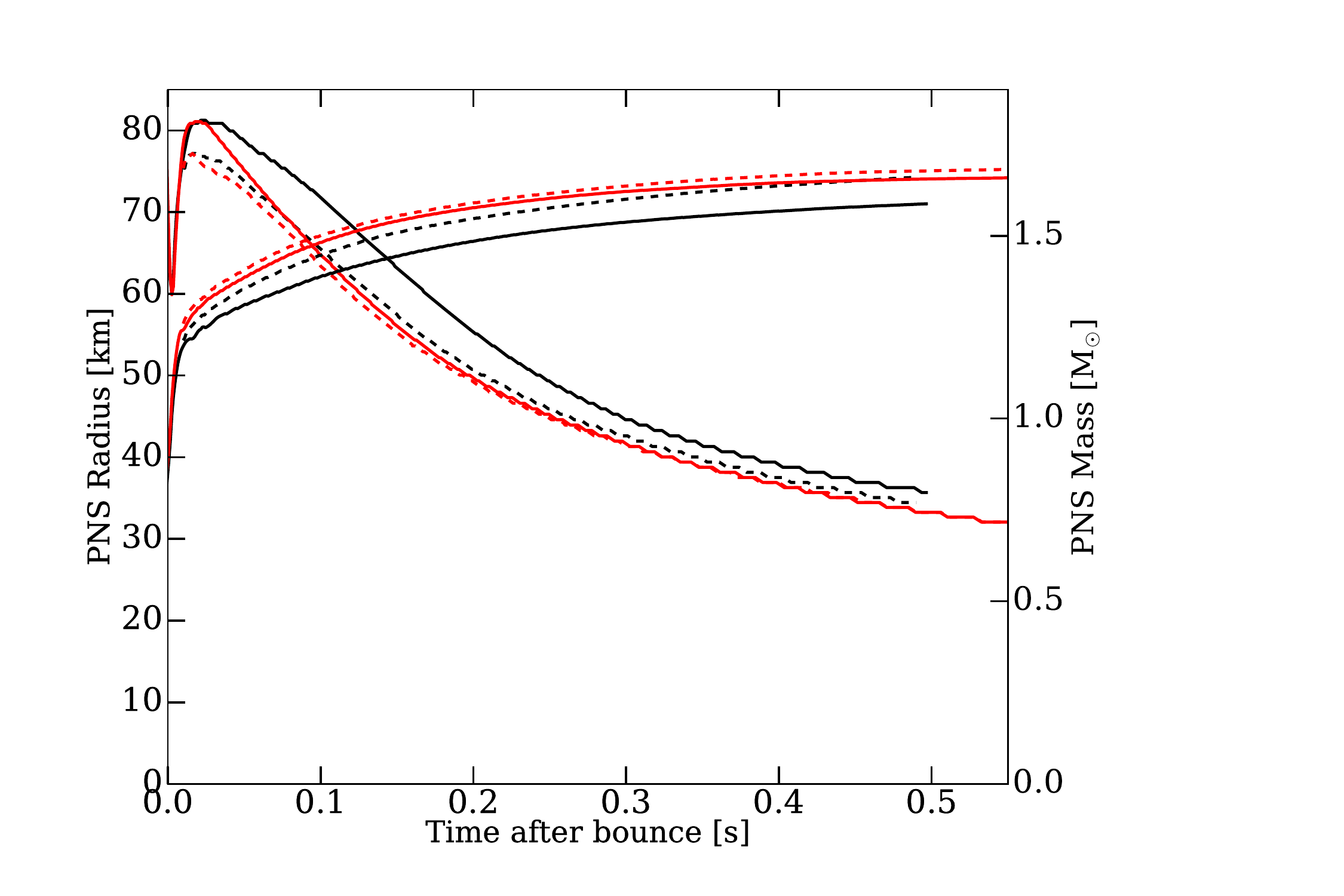}   
    \caption{Net explosion energy (in Bethes) as a function of time after bounce (in s) for the  models studied here. The energy calculated accounts for the gravitational overburden exterior to our grid. At the end of our simulations, the explosion energy is just breaking even, unbinding the star. \textbf{The net explosion energy has a growth rate of $\sim$0.3 Bethes per second for the FC15-M$_{\odot}$ 3D progenitor, and $\sim$0.15 Bethes per second for the 12.5-M$_{\odot}$ 3D progenitor. The explosion energy growth rate for the FC15 3D model is higher than for the 1D MESA progenitor by $\sim$0.1 Bethes per second $\sim$0.5 seconds after bounce. Model M12.5-1D3D does not explode.}  \textbf{Right}: Baryonic PNS mass (in M$_{\odot}$, \textit{right side}) and PNS radius, identified where the density is 10$^{11}$ g cm$^{-3}$ (in km, \textit{left side}) as a function of time after bounce (in seconds).}
    \label{fig:explene} 
\end{figure*}

Here, we investigate the preferential explosion of the 3D progenitors relative to their 1D progenitor counterparts. 
In the left panel of Fig.\,\ref{fig:explene}, we illustrate the net explosion energy (in Bethes, or 10$^{51}$ ergs), including the gravitational overburden of material exterior to our grid, as a function of time after bounce for all models. 
As suggested by the explosion morphology of Fig.\,\ref{fig:ent}, and the discussion in \S\ref{sec:results}, models with unipolar ejecta maintain higher explosion energies than those with dipolar ejecta (such as M12.5-3D3D, see also \citealt{2021Natur.589...29B}), perhaps due to the preferential neutrino heating into one hemisphere. Note that, within the first half second to second $-$ the temporal extent of our simulations $-$ only model FC15-3D3D reaches positive explosion energies. This is not concerning. Rather, simulations require multiple seconds to reach their final asymptotic explosion energies \citep{muller2017,burrows2020,2021Natur.589...29B,bollig2021}. Indeed, the energy growth rates for the M12.5-3D and FC15-3D3D models within the first half second are $\sim$0.35 and $\sim$0.7 Bethes per second, both larger than their 1D3D counterparts by $\sim$0.1 Bethes per second. The explosion energy of the M12.5-3D3D progenitor becomes positive at $\sim$500 ms post bounce (the extent of our simulation), similar to \cite{muller_lowmass} (their Fig.\,3), which studied the same progenitor in 3D. Similarly, in both studies, the shock radius reaches 1000 km approximately 500 ms after core-bounce (theirs is delayed by $\sim$60 ms as it stalls at $\sim$200 km), with shock revival corresponding to the accretion of the Si/O interface. Furthermore, we find that the 3D progenitors typically have smaller gravitational binding energies exterior to our grid. In the right panel of Fig.\,\ref{fig:explene}, we illustrate the baryonic proto-neutron star (PNS) masses and radii, with the models reaching final baryonic PNS masses between 1.6 and 1.7 M$_{\odot}$. 

\begin{figure*}
    \centering
    \includegraphics[width=0.47\textwidth]{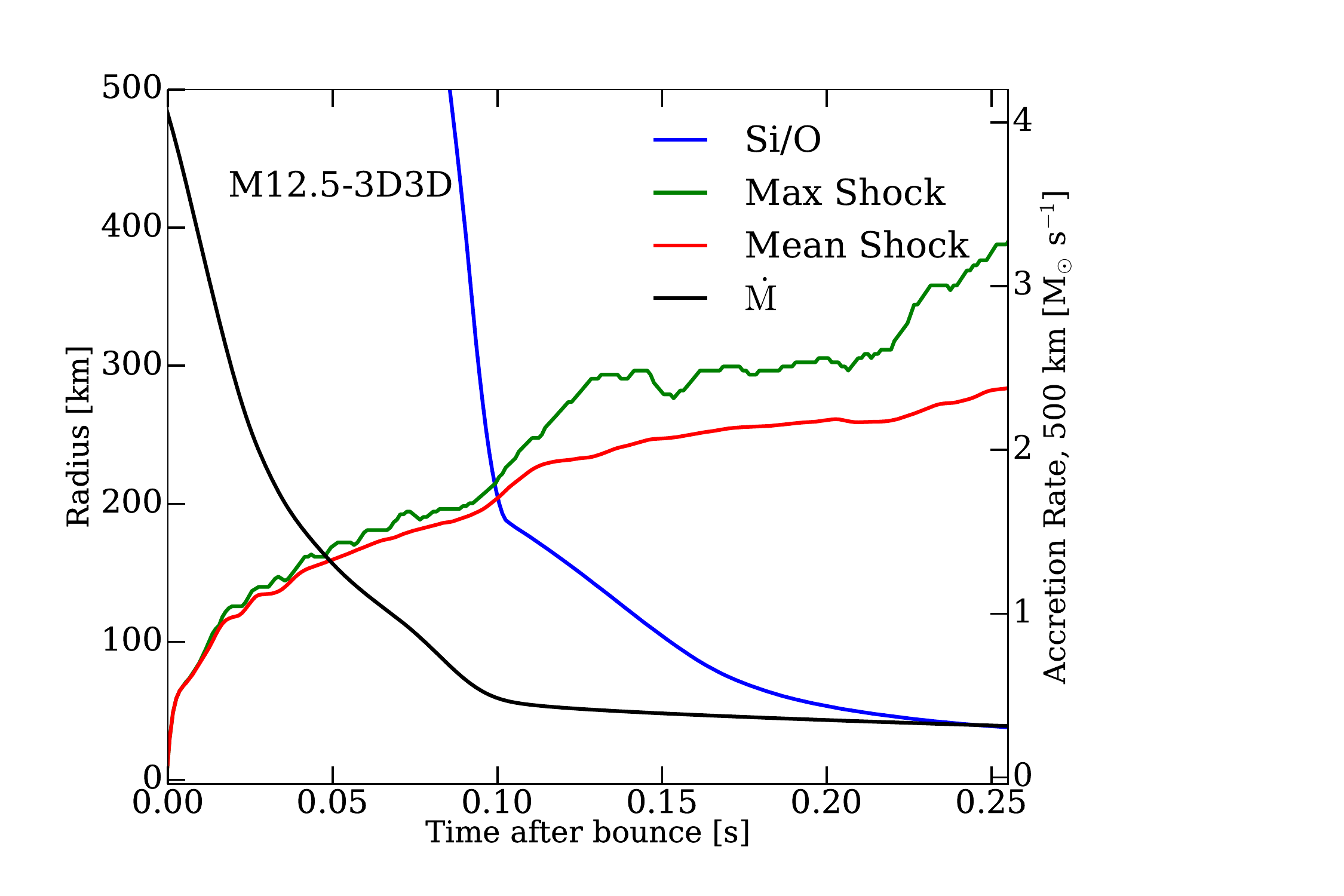}\hfill
    \includegraphics[width=0.47\textwidth]{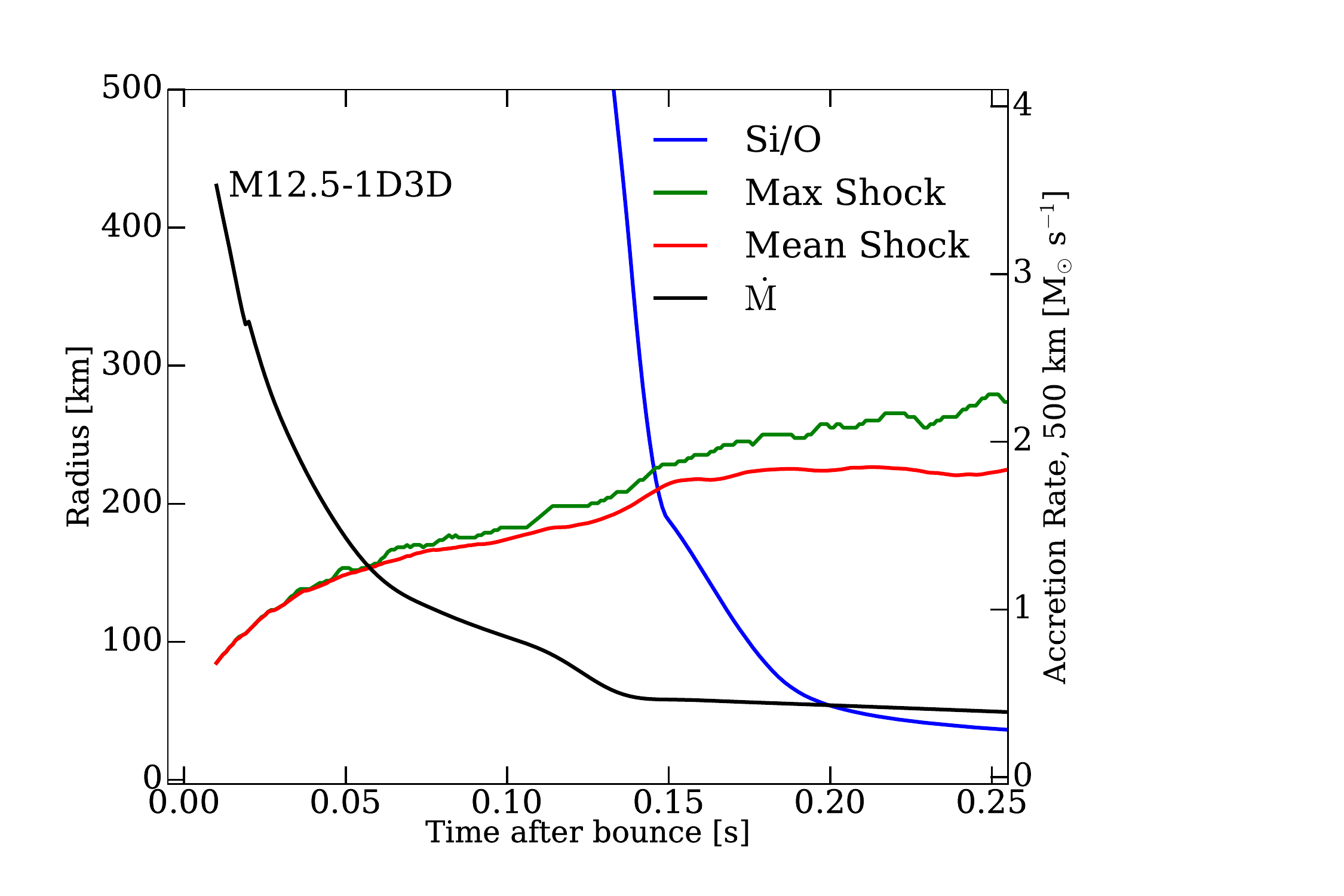}\\
    \vspace{-0.2cm}
    \includegraphics[width=0.47\textwidth]{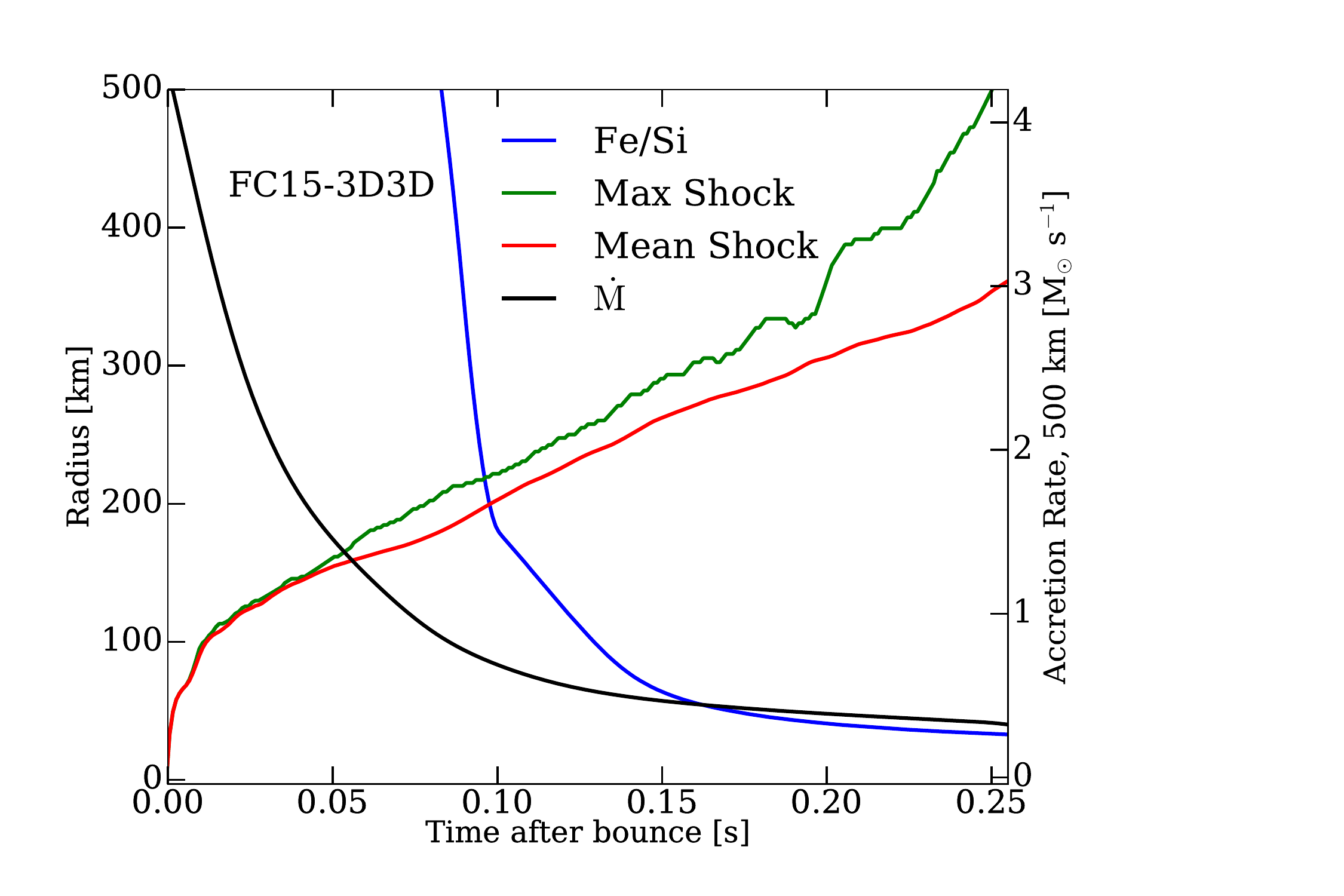}\hfill
    \includegraphics[width=0.47\textwidth]{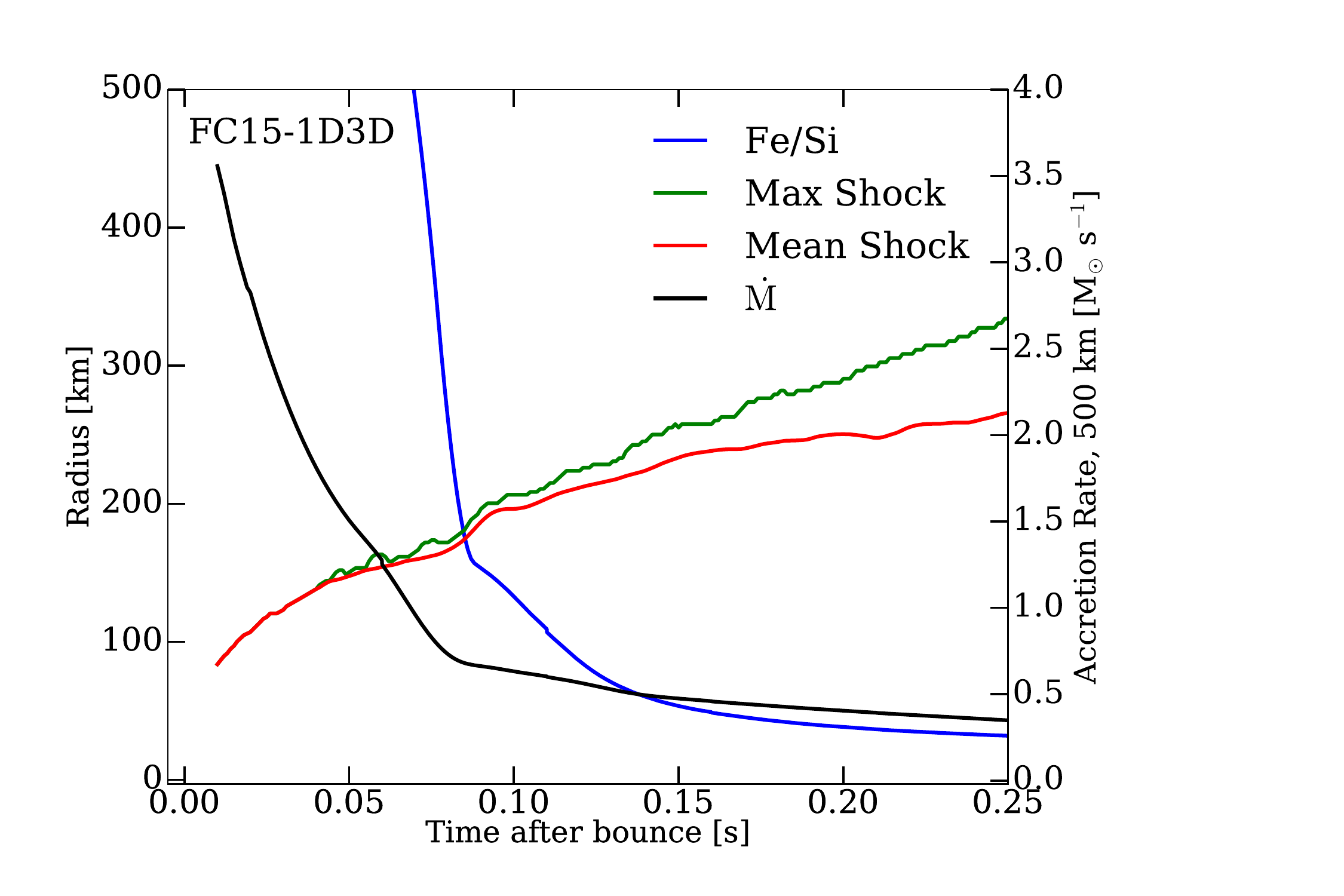}
    \caption{For each of the four models (\textbf{top: M12.5, bottom: FC15, left: 3D3D, right: 1D3D}), we show maximum and minimum shock radii, the position of the infalling Si/O (or Fe/Si for the FC15 models) interface (in km), and the accretion rate (in M$_{\odot}$ s$^{-1}$) as a function of time after bounce (in s) for the first 250 ms post bounce. Accretion of the sharp density interface onto the stalled shock results in a drop in the accretion ram pressure exterior to it, which oftimes leads to the rejuvenation of the shock wave. This is the case for the M12.5 models: the 3D model accretes a sharper, deeper Si/O interface earlier and revives $\sim$50 ms earlier than the MESA model, which ultimately fails to explode. However, even the non-exploding M12.5-1D3D model experiences a modest shock revival shortly after $\sim$150 ms, coincident with accretion of the Si/O interface. For the FC15-3D3D model, however, the model has already begun to explode, and the shock surface reaches the Fe/Si interface before the latter accretes onto it. For the FC15-1D3D model, Fe/Si interface accretion does correspond with shock revival. For both FC15 models, subsequent accretion of the Si/O interface (not shown) occurs after shock revival but still contributes to an accelerating shock front. See \S\ref{sec:discussion} for more details.}
    \label{fig:rSiO} 
\end{figure*}

\begin{figure}
    \centering
    \includegraphics[width=1\linewidth]{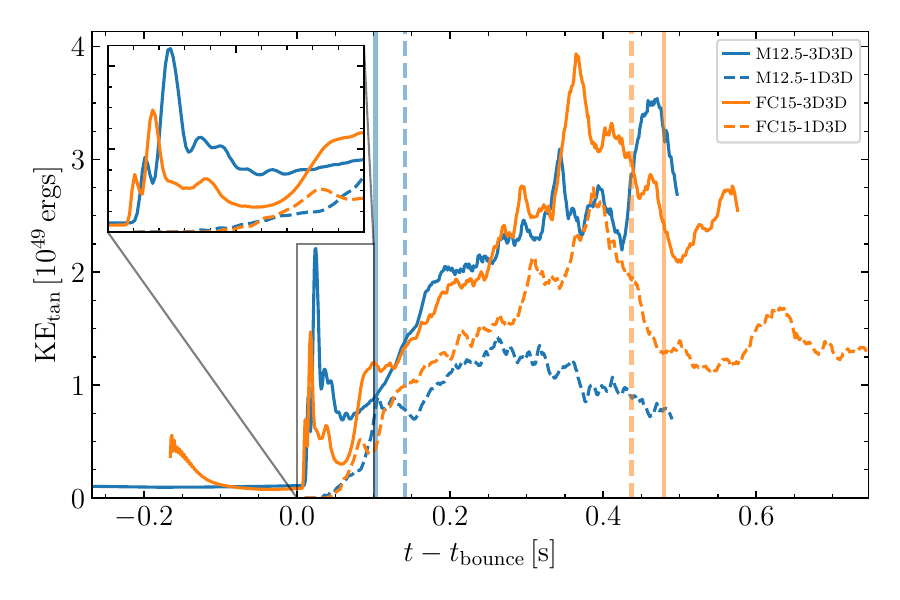}
    \caption{Integrated tangential kinetic energy (for $\rho<10^{11}$ g cm$^{-3}$) as a function of time for the four models. This is used as a proxy to measure the importance of turbulence in the models. The faded vertical lines show when the Si/O interface mean radial location is equal to the mean shock radius for each model (with colors and line styles denoted in the legend). In both cases, the \oned model has substantially less turbulence. This difference is more prominent shortly after bounce (see inset) which significantly affects the production of gravitational waves at these early times (see Fig~\ref{fig:GW}). Additionally, the turbulence around the Si/O interface can be seen in the \ddd models before bounce.}
    \label{fig:KE_tan} 
\end{figure}

In Fig.\,\ref{fig:rSiO} we illustrate, for each of the four models studied here, the maximum and minimum shock radius, the accretion rate, and the location of the Si/O (or Fe/Si, in models FC15-3D3D and FC15-1D3D) interface as a function of time after bounce (in seconds).\footnote{Historically, this transition has been associated with a silicon/oxygen boundary (see also \citealt{fryer1999, ott2018_rel}). However, we found that the density drop corresponding to a compositional interface, especially if fragmented, could also correspond to iron/silicon in the FC15-3D3D and FC15-1D3D. We refer to the Si/O interface, and more broadly, the silicon-group and oxygen-group compositional boundary, interchangeably (see also \citealt{2021ApJ...916L...5V}).} Accretion of a sharp interface across a stalled shock leads to a sharp drop in ram pressure, which may rejuvenate the stalled shock into a successful explosion \citep{fryer1999,vartanyan2018a,ott2018_rel}. Indeed, this is the case for all models except for the FC15-3D3D, which begins to explode before accreting even the Fe/Si, let alone the Si/O, interface. In both FC15 models, shock revival precedes accretion of the Si/O interface, whose subsequent accretion further accelerates shock expansion. Even for the non-exploding model M12.5-1D3D, accretion of the Si/O interface coincides with a small bump in shock radii, albeit insufficient for explosion. The tangential kinetic energy reacts to the accretion of the Si/O interface in a similar manner (see Fig.~\ref{fig:KE_tan}).

\begin{figure*}
    \includegraphics[width=0.49\linewidth]{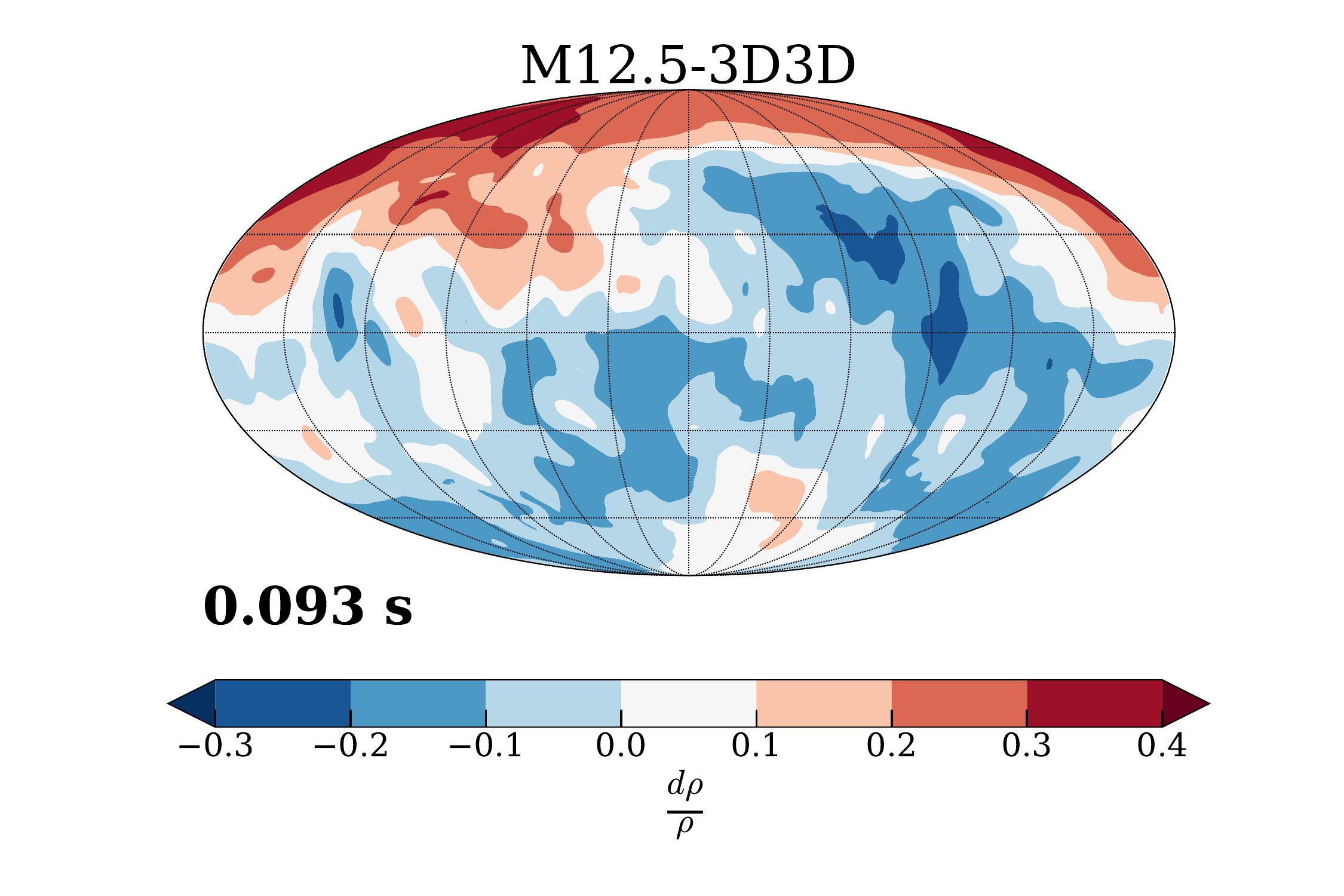}
    \includegraphics[width=0.49\linewidth]{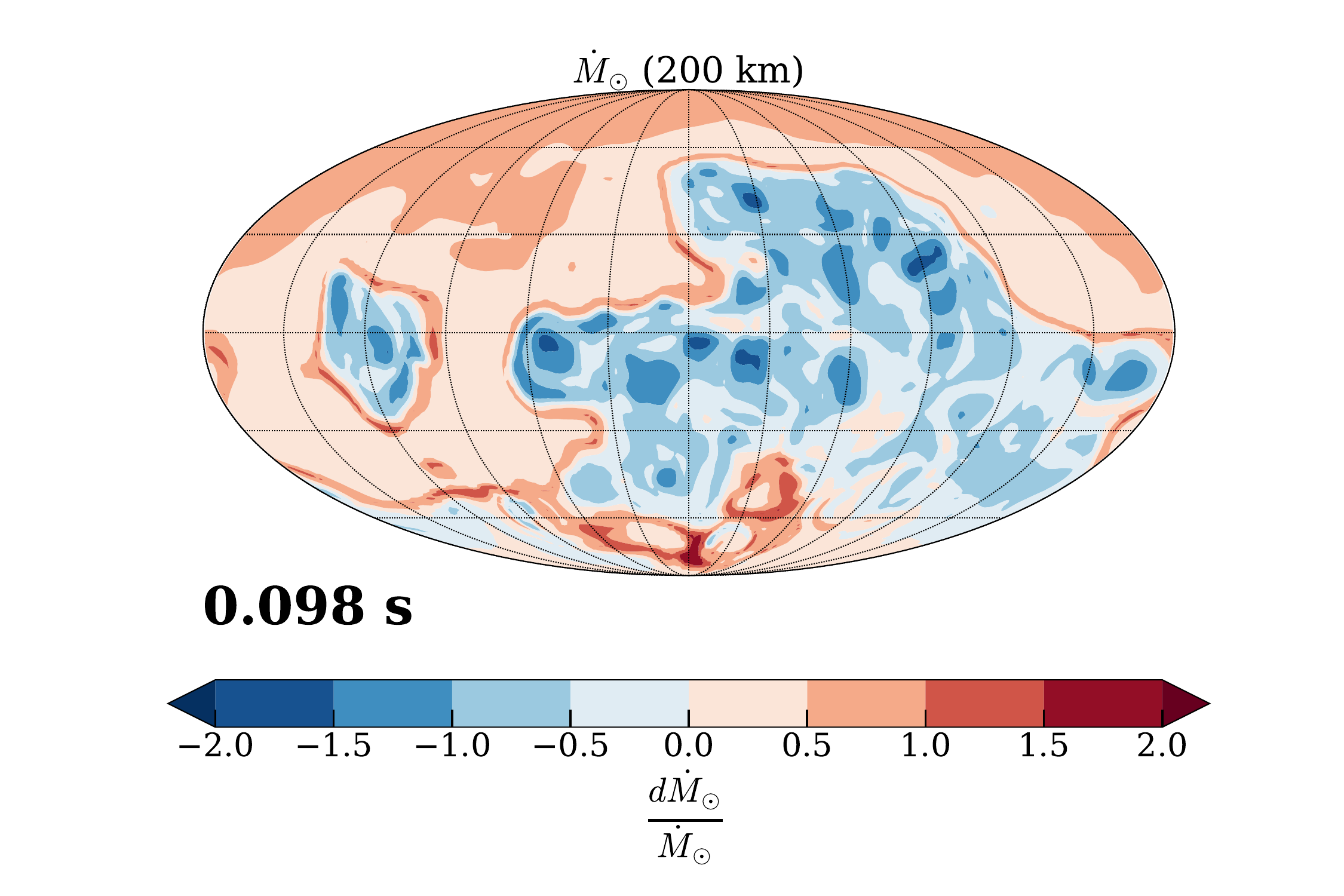}
    \includegraphics[width=0.49\linewidth]{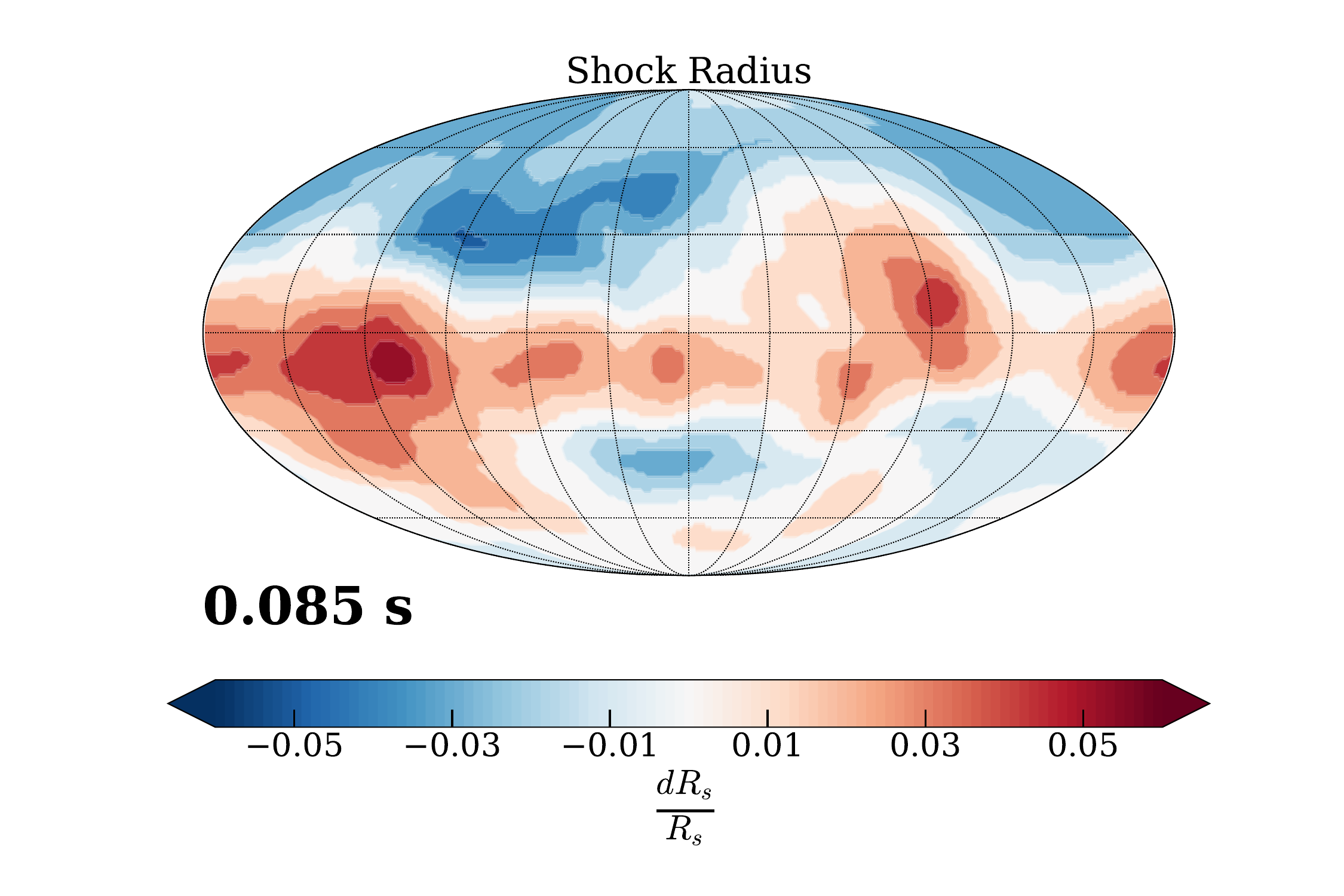}
    \includegraphics[width=0.49\linewidth]{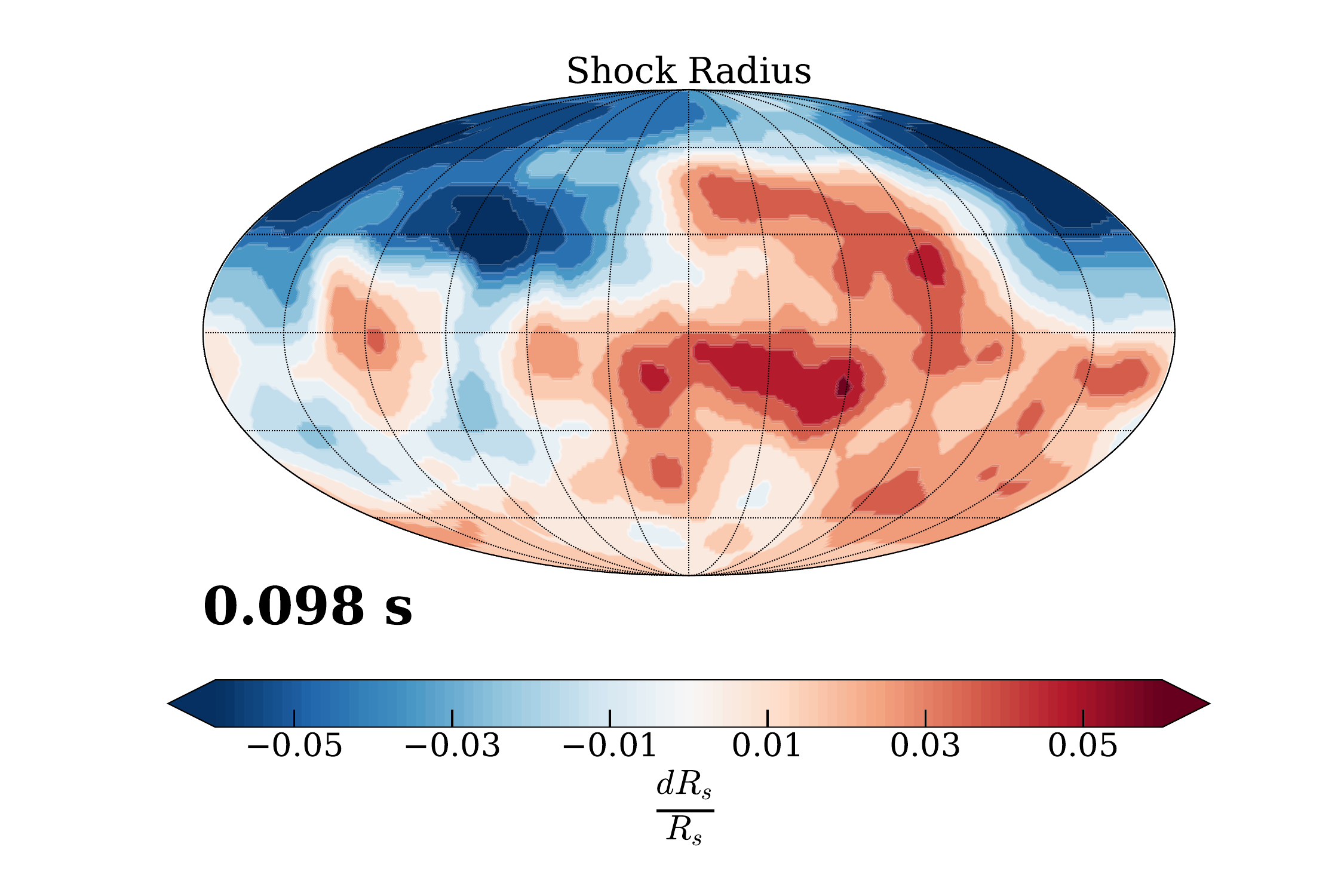}
    \caption{Mollweide projection for model M12.5-3D3D of the density (\textbf{top left}) at the Si/O interface at 93 ms post bounce, accretion rate at 200 km at 85 and 98 ms post bounce (\textbf{top right}), and shock radius (\textbf{bottom}) at 85 ms and 98 ms post bounce, before and after accretion of the Si/O interface. Note how the morphology of the density profile around the Si/O interface sets the stage for the geometry of the accretion and the emergent shock surface (which is strongly anti-correlated with the accretion rate, as seen here). This is best illustrated in the bottom panels, where the developing shock surface just prior to Si/O interface accretion (left) is very different than just after (right), with the Si/O interface molding the emergent shock. Strong shock expansion occurs in the direction of lowest accretion ram pressure and lower density.}
    \label{fig:moll}
\end{figure*}

\begin{figure*}
    \begin{minipage}{.49\linewidth}
    \centering M12.5-3D3D
    \end{minipage}
    \hfill
    \begin{minipage}{.49\linewidth}
    \centering M12.5-1D3D
    \end{minipage}
    \\
    \includegraphics[width=.49\linewidth]{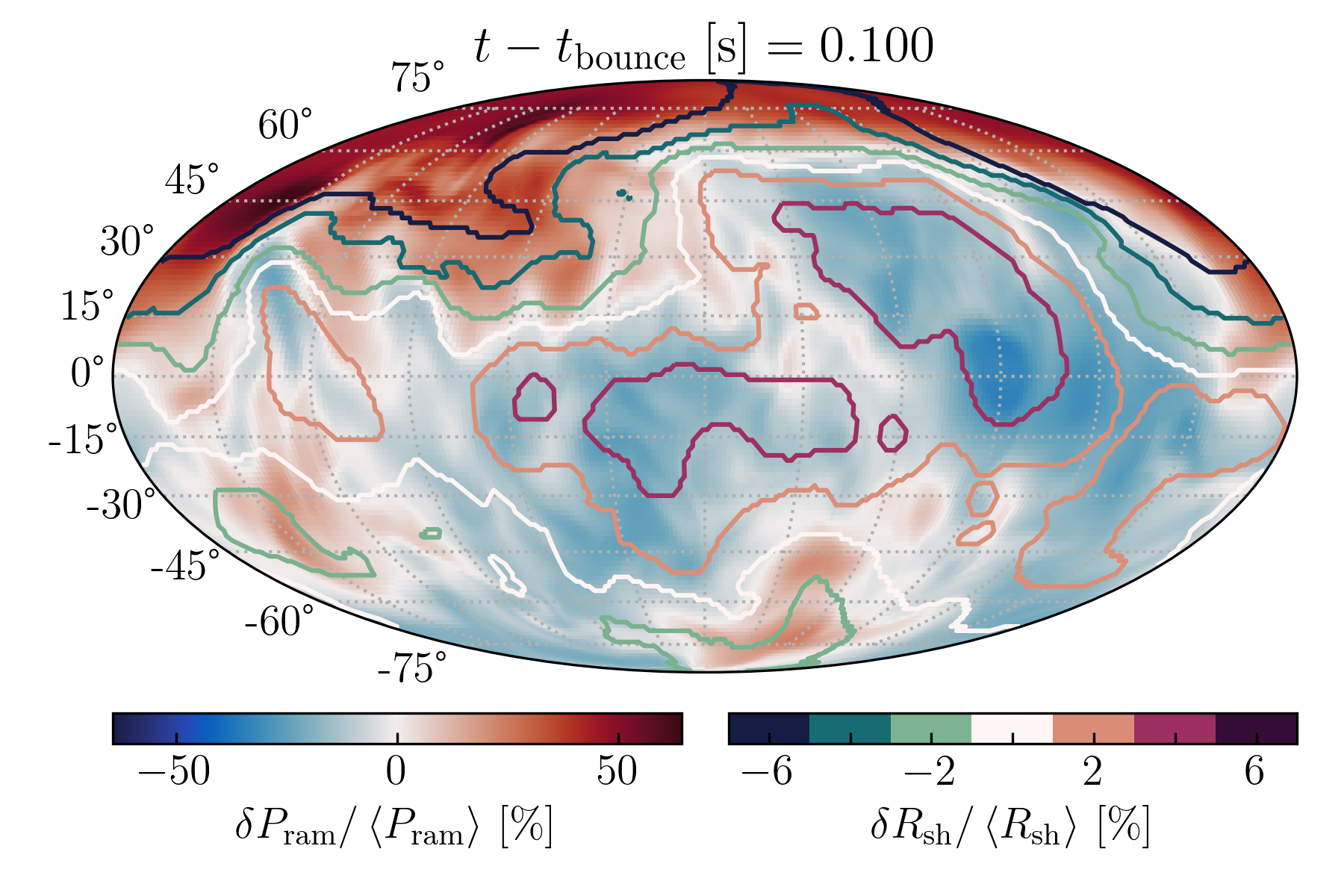}\hfill
    \includegraphics[width=.49\linewidth]{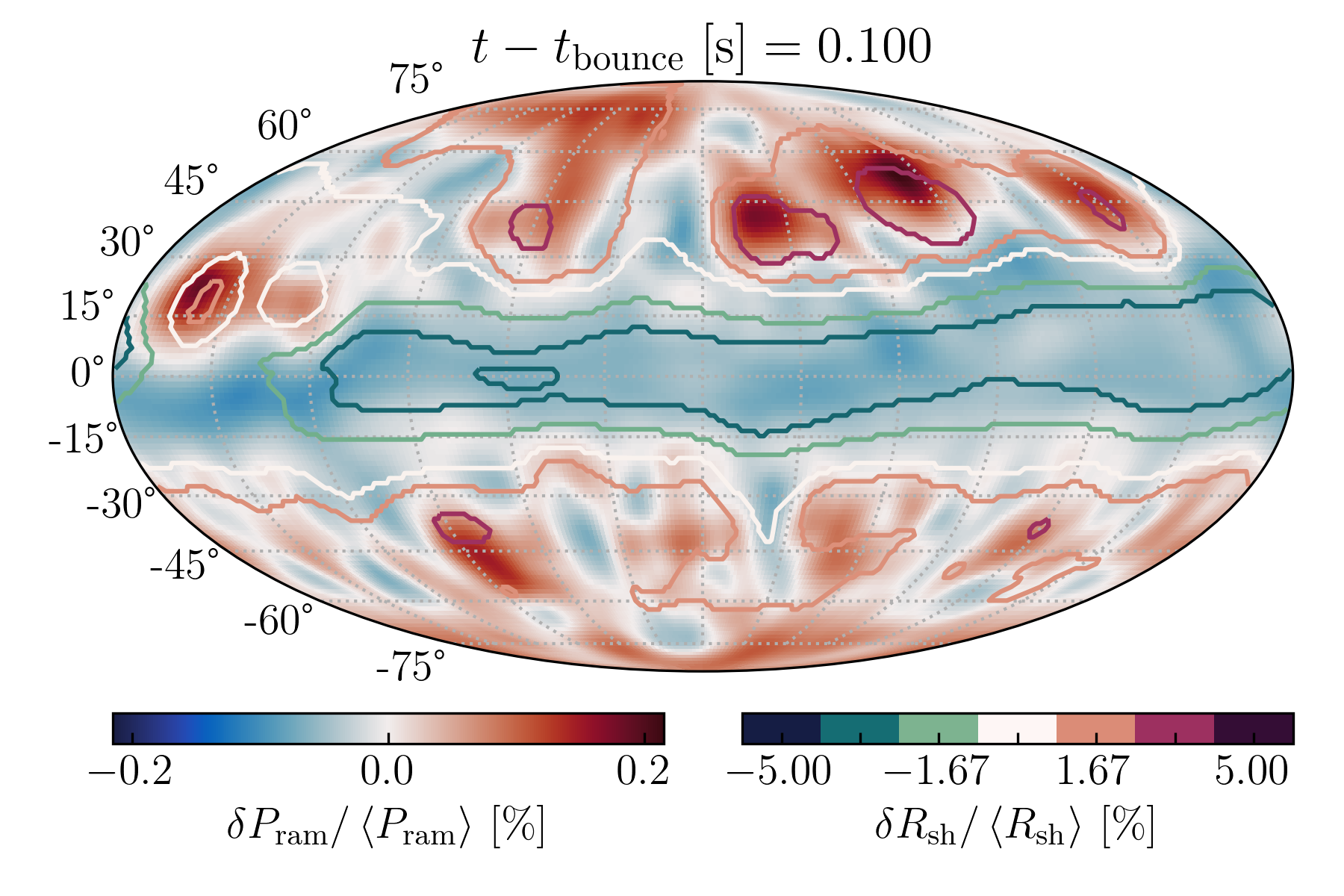}\\
    \includegraphics[width=.49\linewidth]{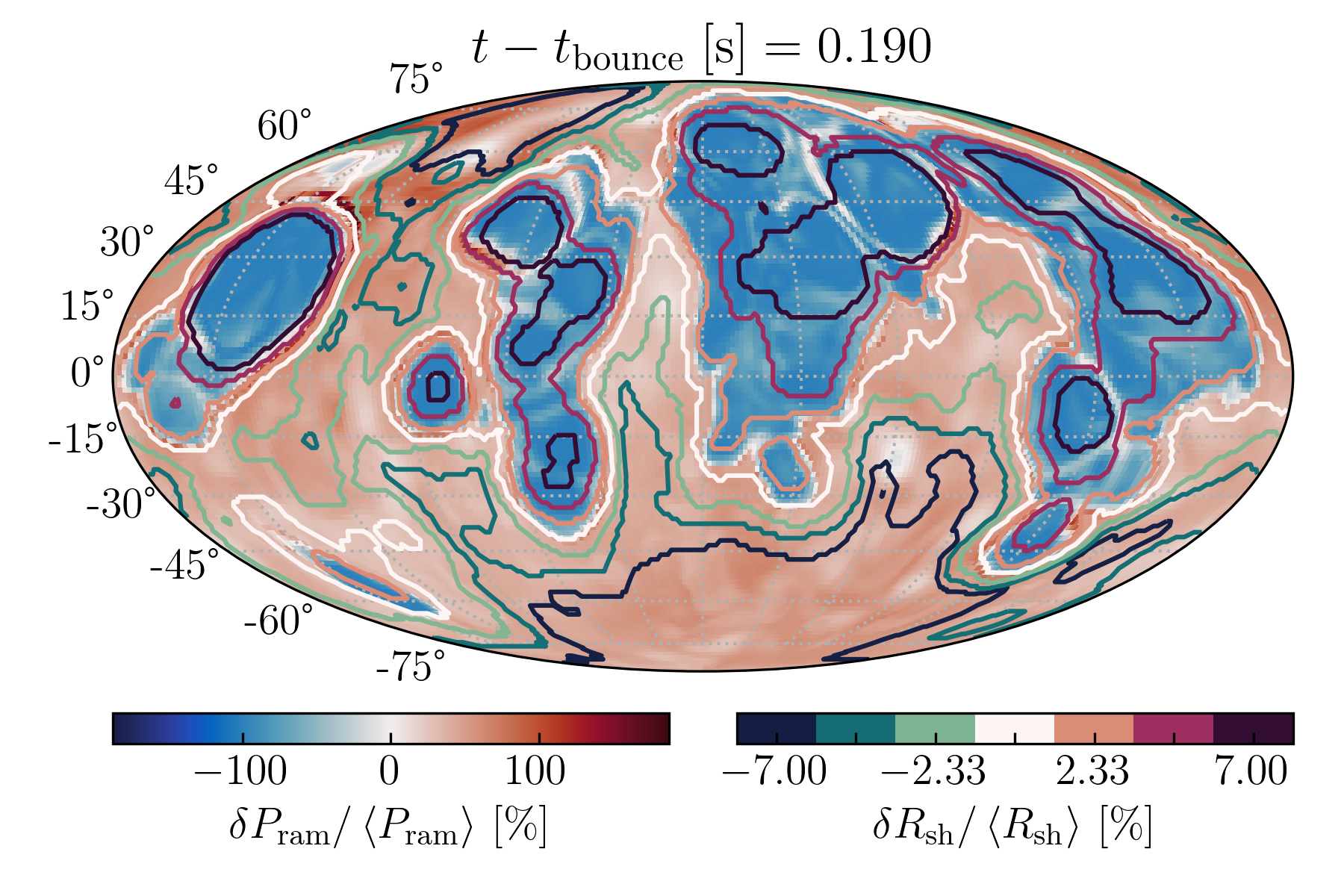}\hfill
    \includegraphics[width=.49\linewidth]{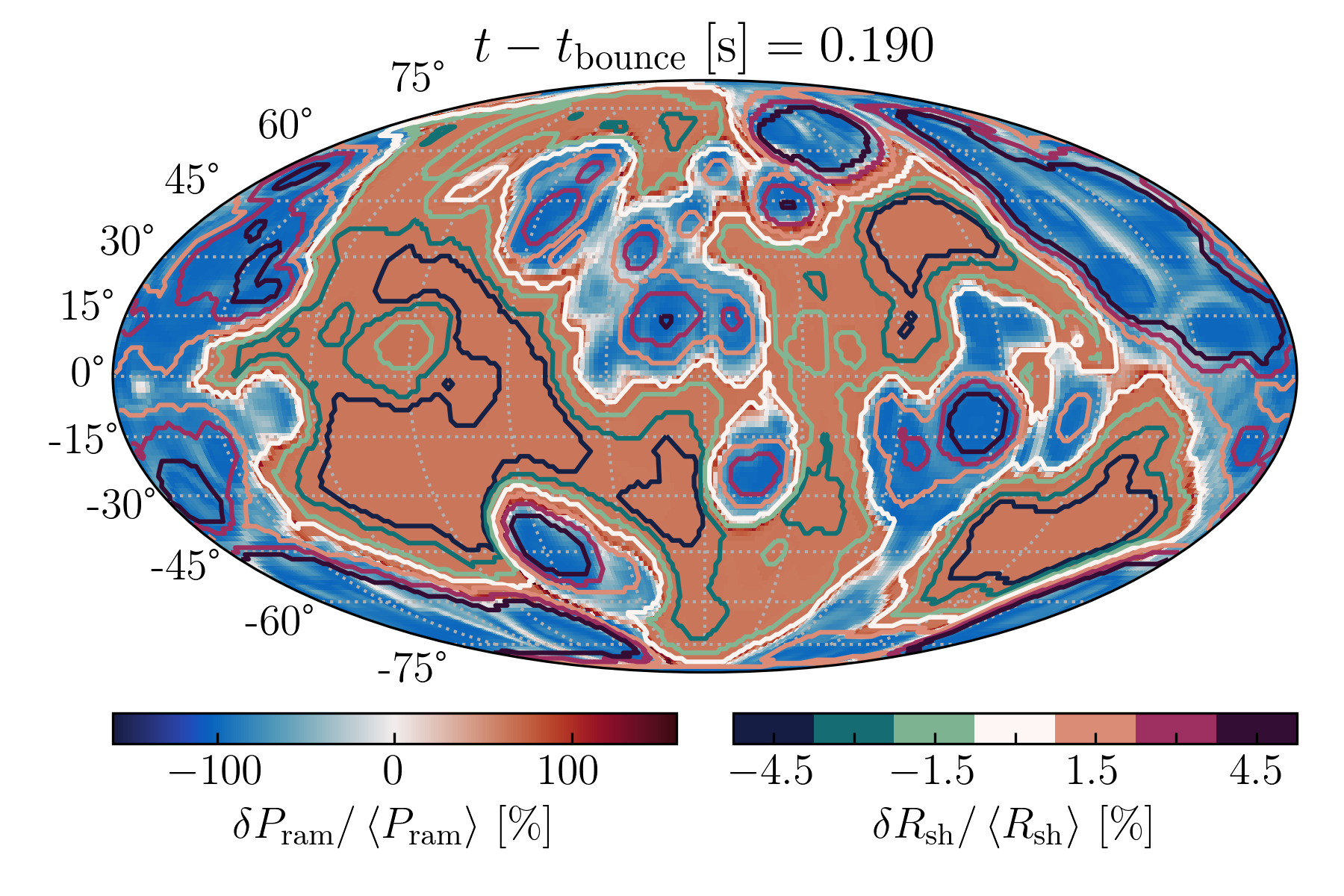}\\
    \includegraphics[width=.49\linewidth]{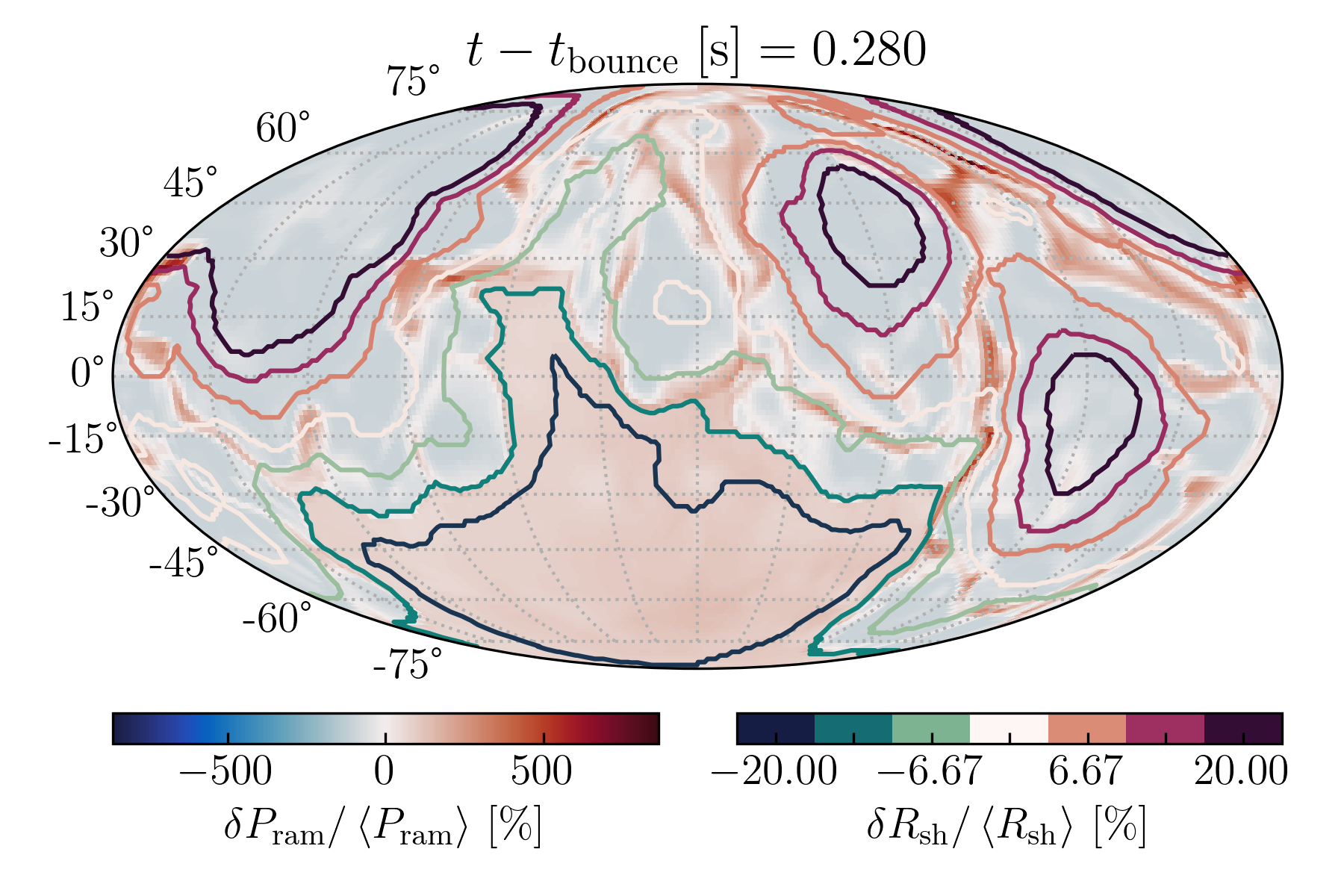}\hfill
    \includegraphics[width=.49\linewidth]{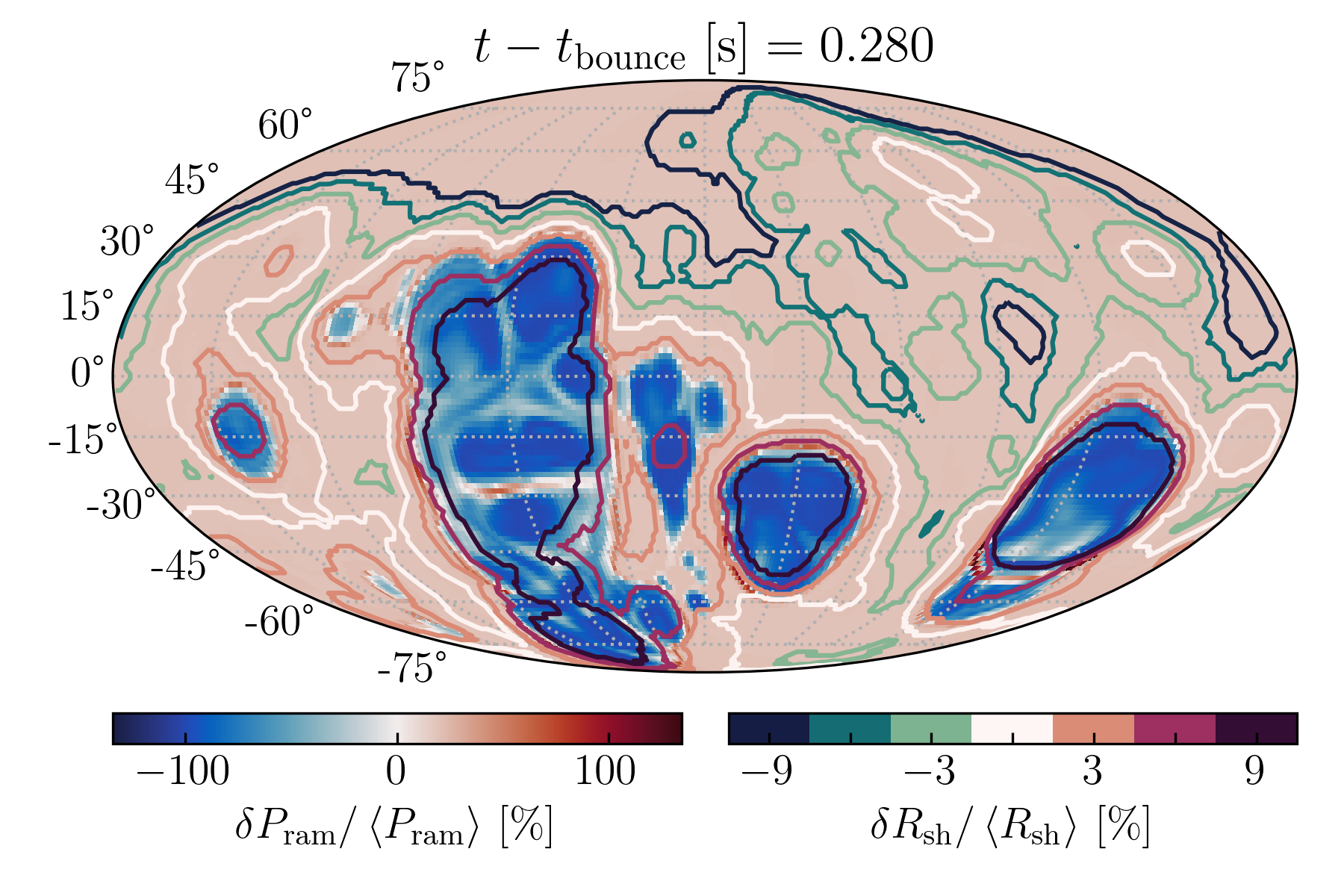}
    \caption{Time series (top to bottom) of ram pressure variations at a fixed radius with overlain contours of shock radius deviations for the two \muller models. For the \ddd model (left) the radius at which the ram pressure is plotted is 260 km, while for the \oned model (right) the radius is chosen to be 220 km. These radii correspond to the mean shock radius at 200 ms after bounce for each model. For both models the shock radius and ram pressure are anti-correlated, indicating that the shock takes the path of least resistance. Plotting accretion rate instead of ram pressure gives qualitatively similar results.}
    \label{fig:muller_pram-rshock}
\end{figure*}

\begin{figure*}
    \begin{minipage}{.49\linewidth}
    \centering FC15-3D3D
    \end{minipage}
    \hfill
    \begin{minipage}{.49\linewidth}
    \centering FC15-1D3D
    \end{minipage}
    \\
    \includegraphics[width=.49\linewidth]{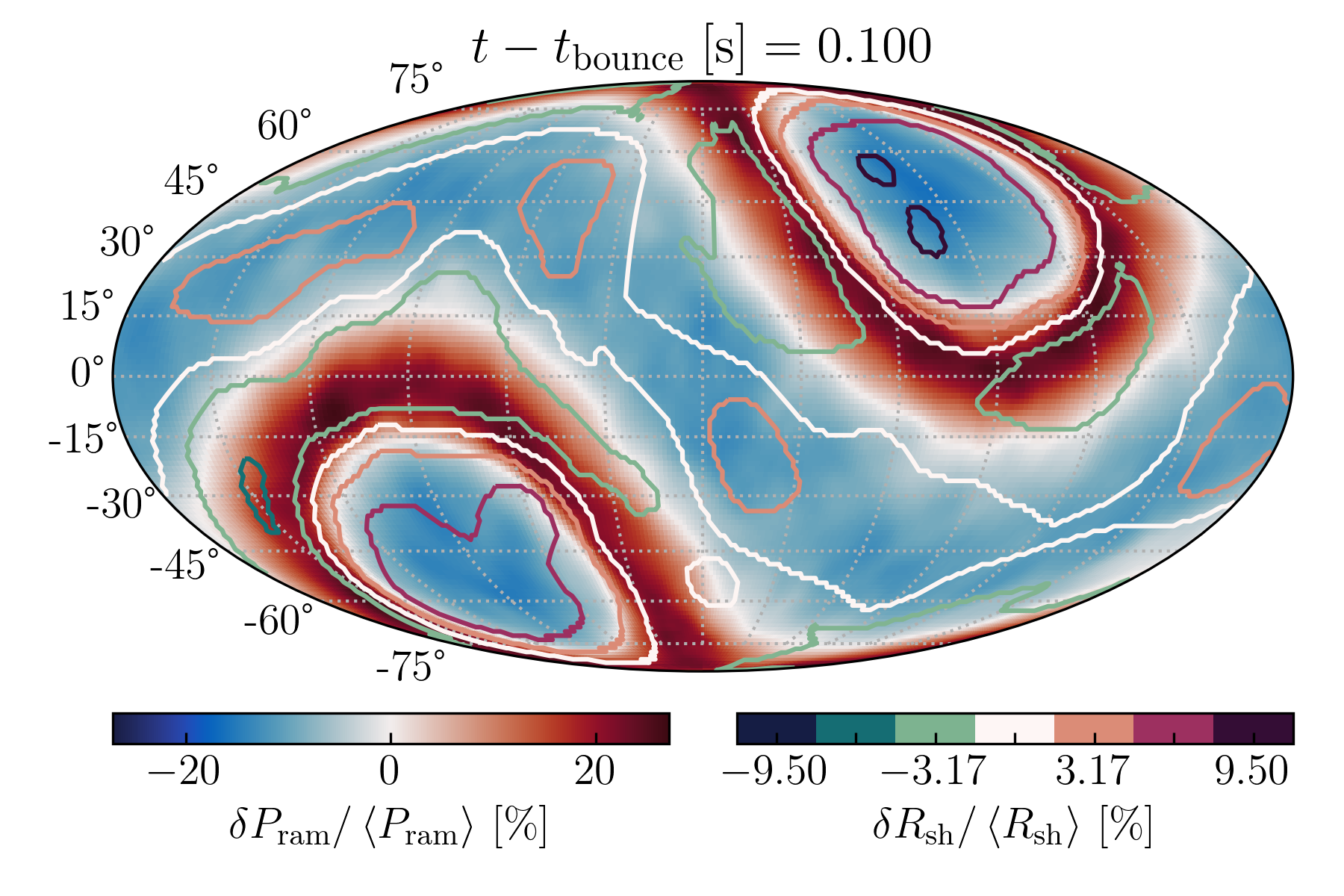}\hfill
    \includegraphics[width=.49\linewidth]{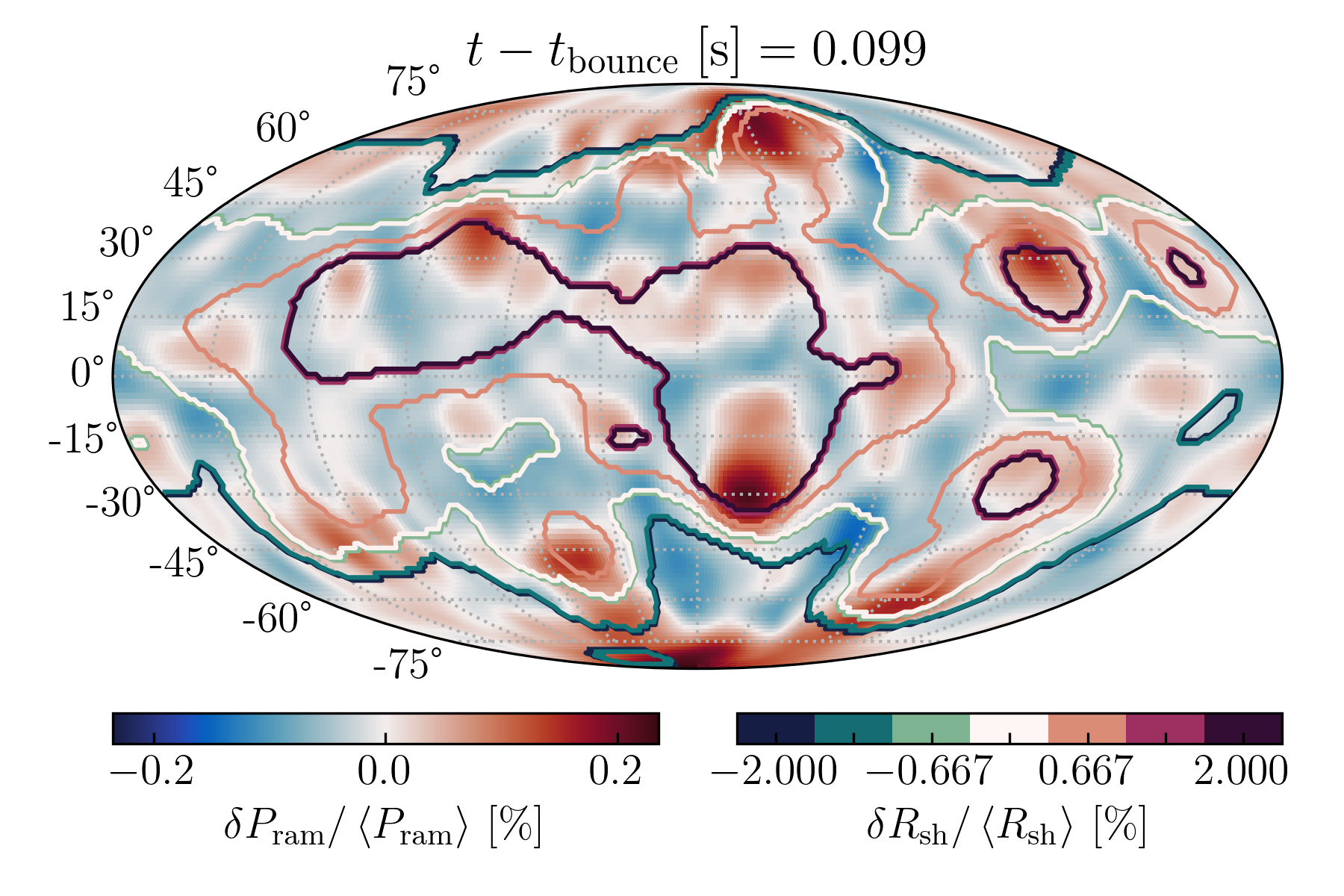}\\
    \includegraphics[width=.49\linewidth]{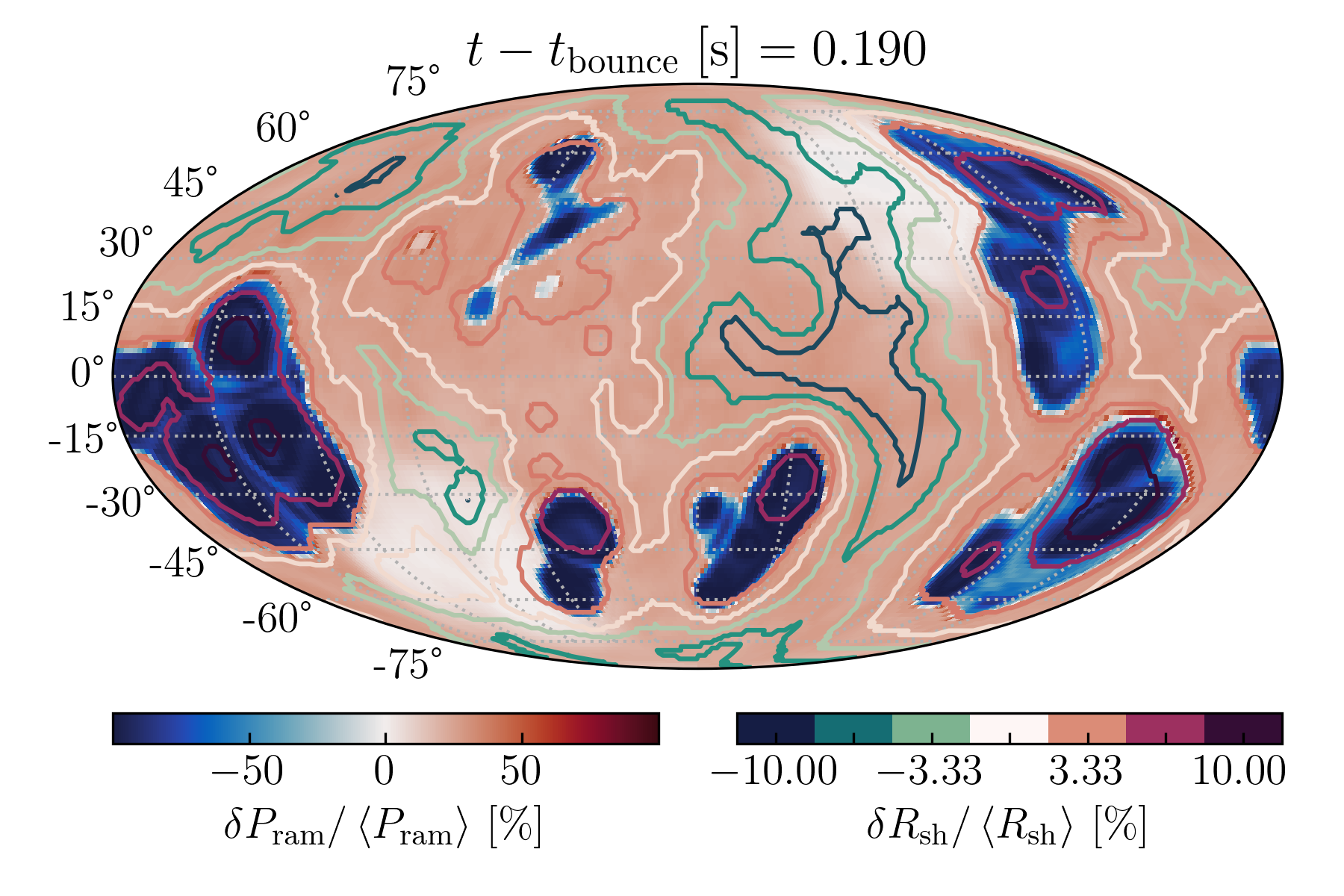}\hfill
    \includegraphics[width=.49\linewidth]{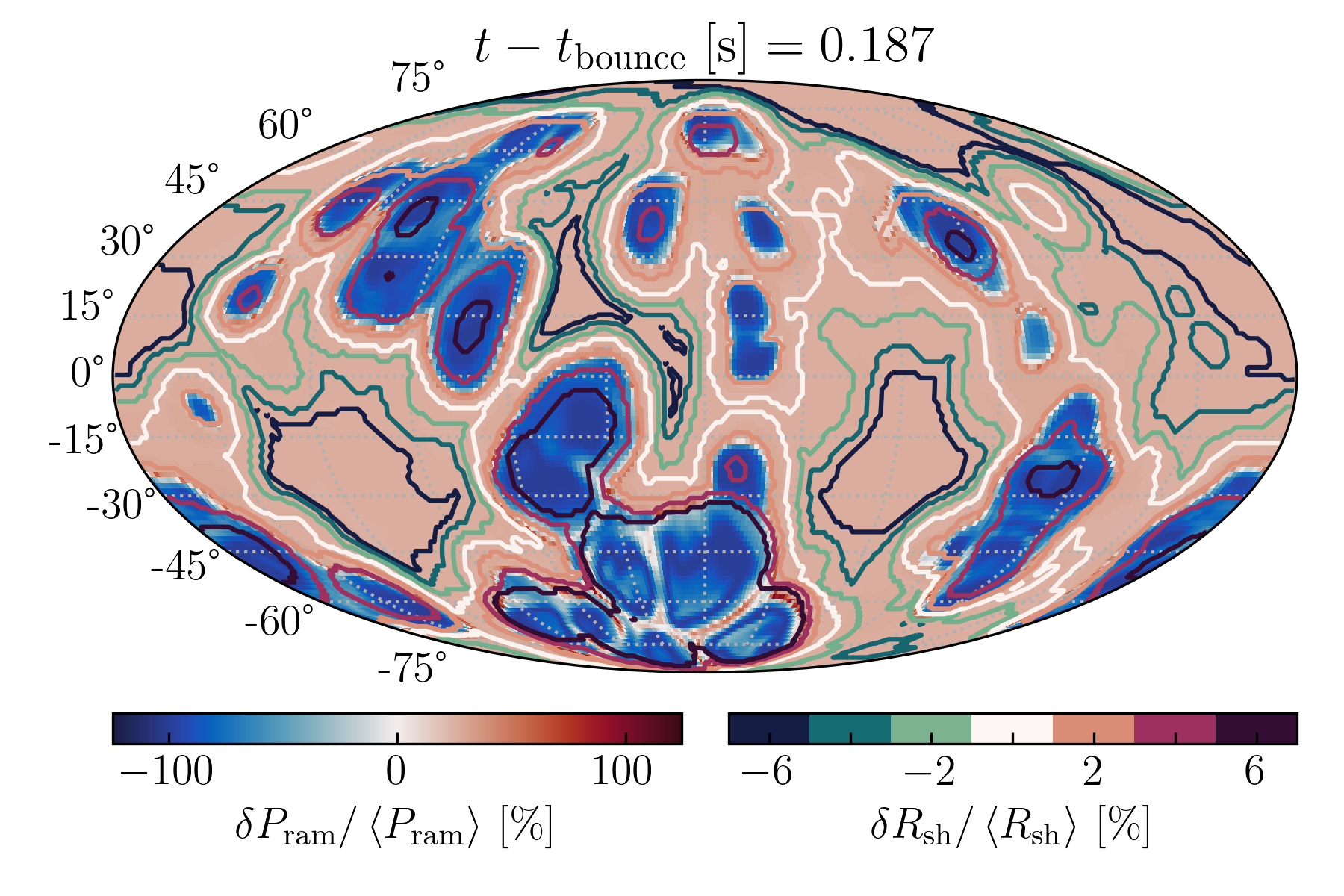}\\
    \includegraphics[width=.49\linewidth]{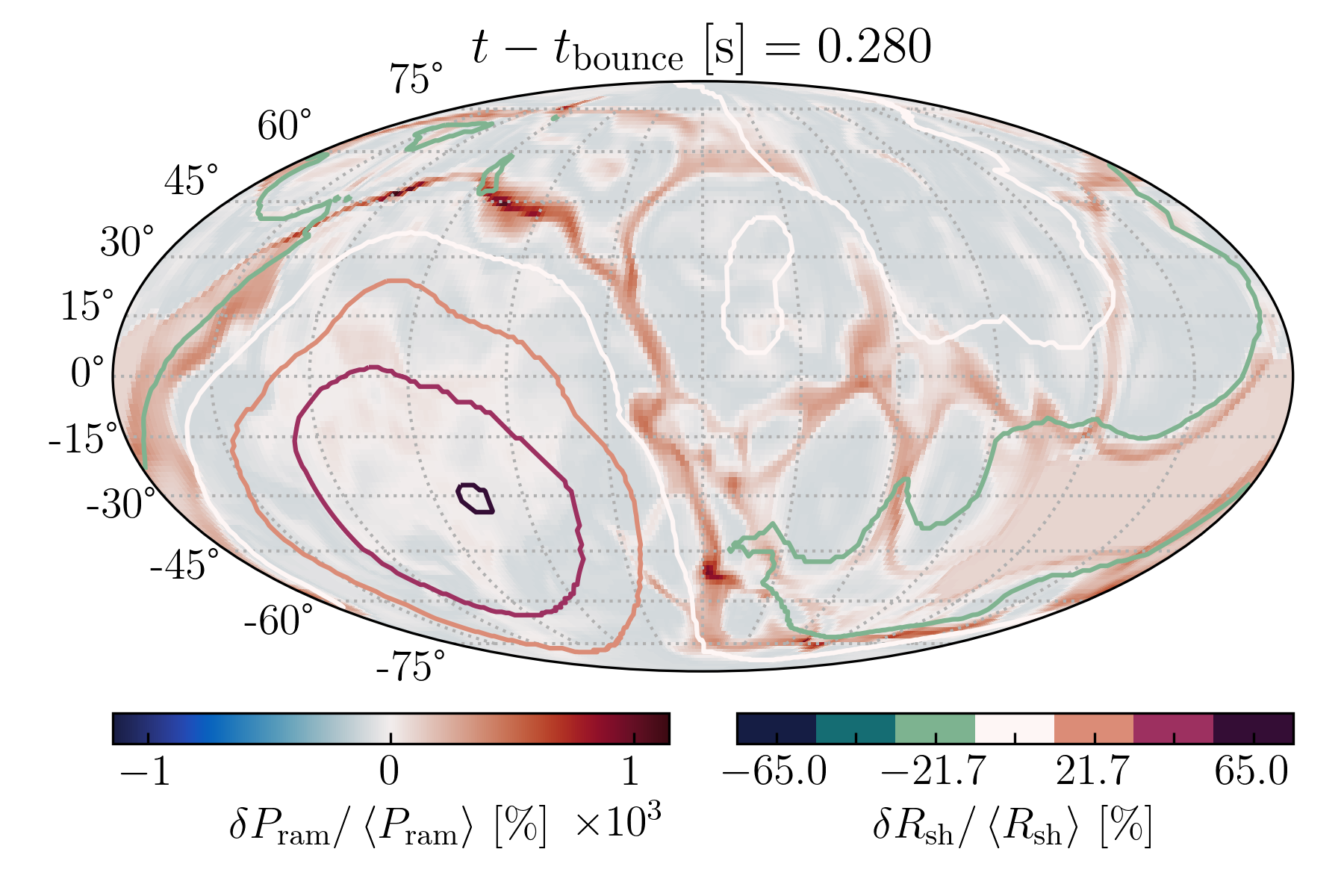}\hfill
    \includegraphics[width=.49\linewidth]{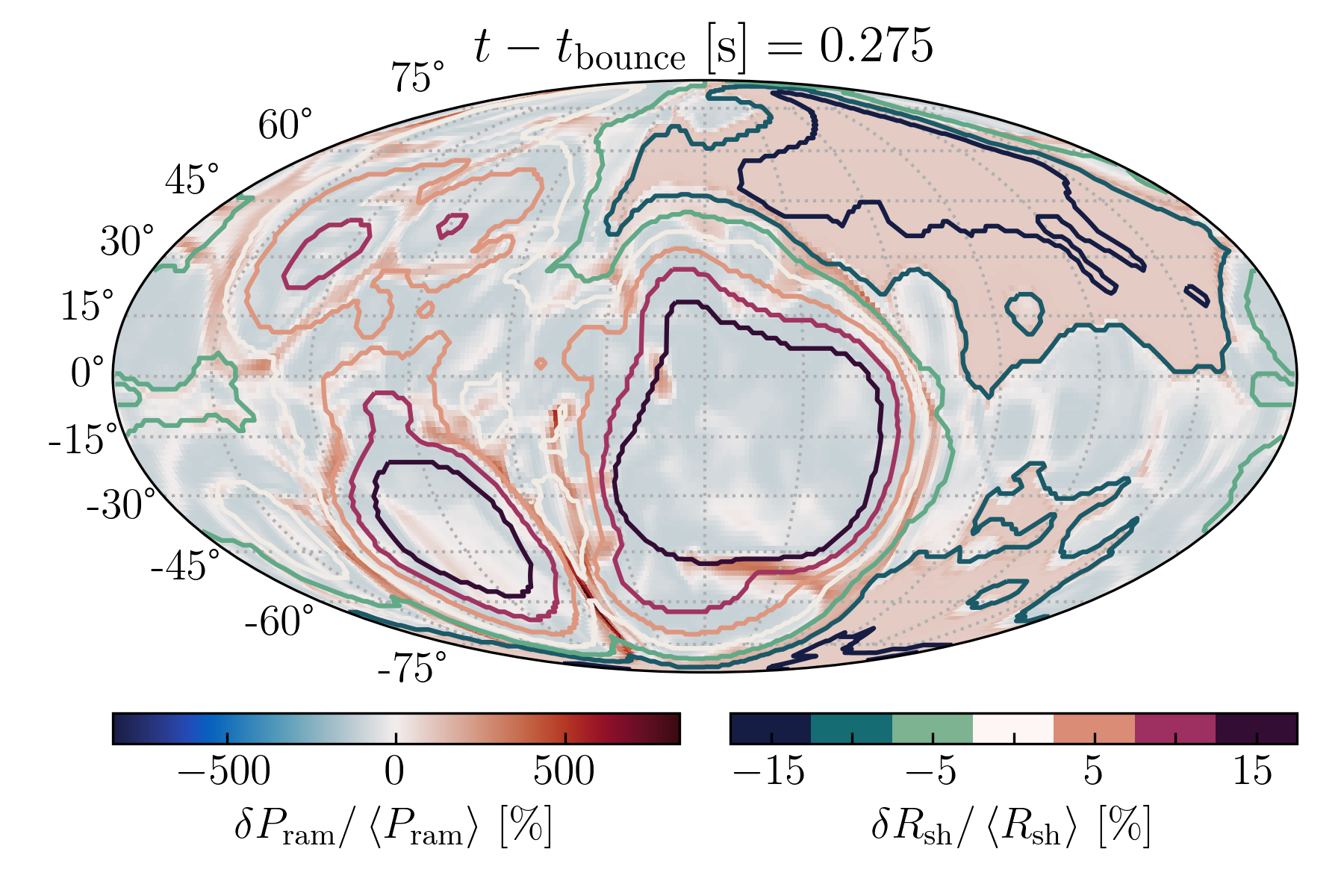}
    \caption{Same as Fig.~\ref{fig:muller_pram-rshock}, but for the two \fc models.
    For the \ddd model (left) the radius at which the ram pressure is plotted is 300 km, while for the \oned model the radius is chosen to be 250 km. As before, these radii correspond to the mean shock radius at 200 ms after bounce for each model.
    }
    \label{fig:couch_pram-rshock}
\end{figure*}

In Fig.\,\ref{fig:moll}, we depict Mollweide projections of the normalized density variations (top), accretion rate variations (middle), and shock surface variations (bottom) for model M12.5-3D3D as a case study of the role of progenitor profile on explosion outcome. We illustrate the density variations at the Si/O interface located at $\sim$350 km at 93 ms post bounce, roughly 10 ms before explosion, the accretion rate (at 200 km), and the shock surface just interior. Regions with high density have a higher accretion rate that suppresses shock expansion. The Si/O interface regulates the accretion rate and modulates the geometric development of the shock 150 km interior. Ultimately, the shock expands in the regions of lower density determined by the morphology of the matter exterior, here encompassing the Si/O interface. Notice how even small pockets of higher/lower density manifest as regions of increased/decreased accretion rate and decreased/increased shock radius expansion, respectively. These correlations continue to be imprinted on the shock radius and ram pressure to at least 100 ms post bounce (compare Fig.~\ref{fig:moll} to the upper-left panel of Fig.~\ref{fig:muller_pram-rshock}). From here on out the shock takes the path of least resistance, signified by the anti-correlation of ram pressure and shock radius shown in Fig.~\ref{fig:muller_pram-rshock} for the two M12.5 models and in Fig.~\ref{fig:couch_pram-rshock} for the two FC15 models.

\begin{figure*}
    \centering
    \includegraphics[width=0.47\textwidth]{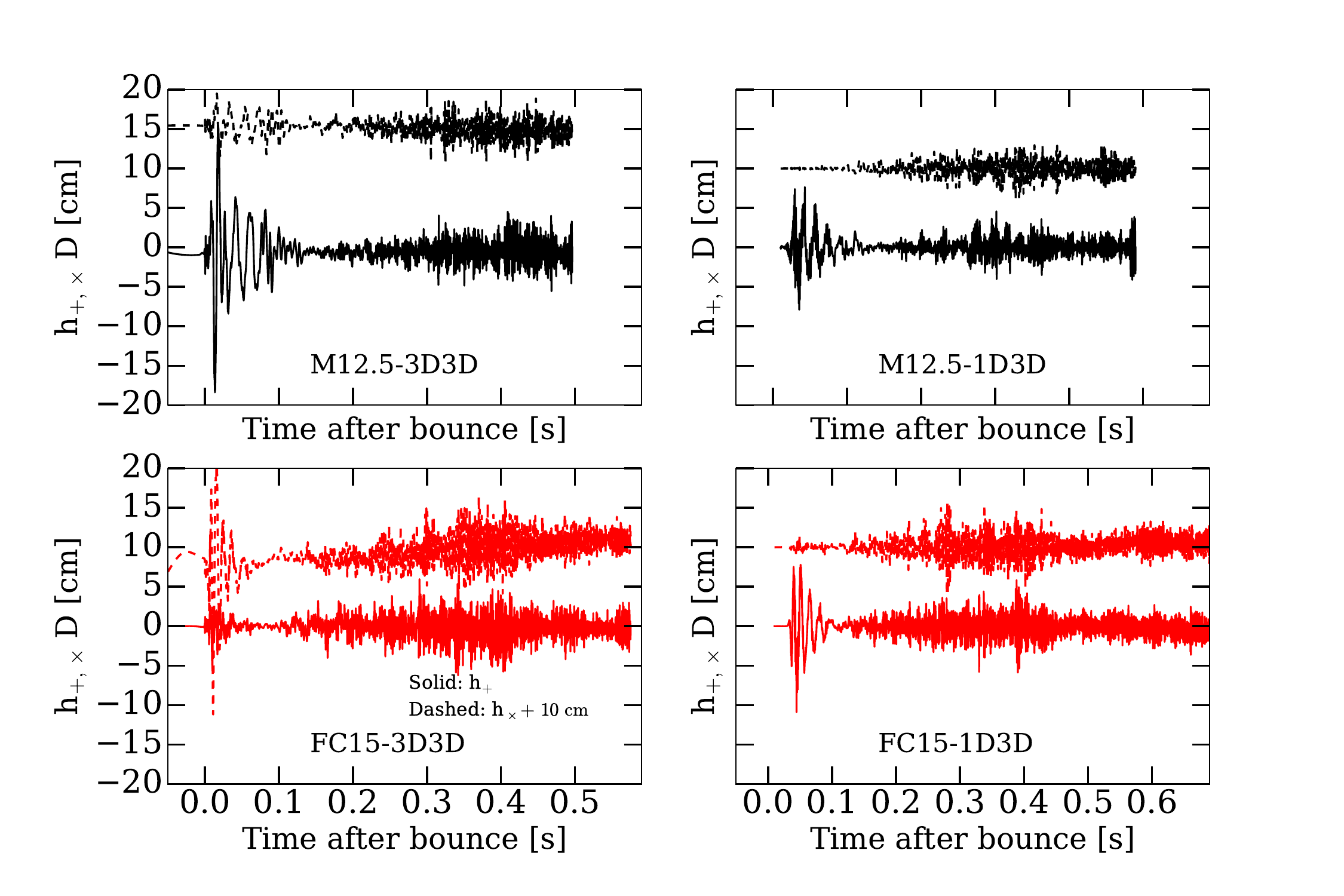}
    \includegraphics[width=0.47\textwidth]{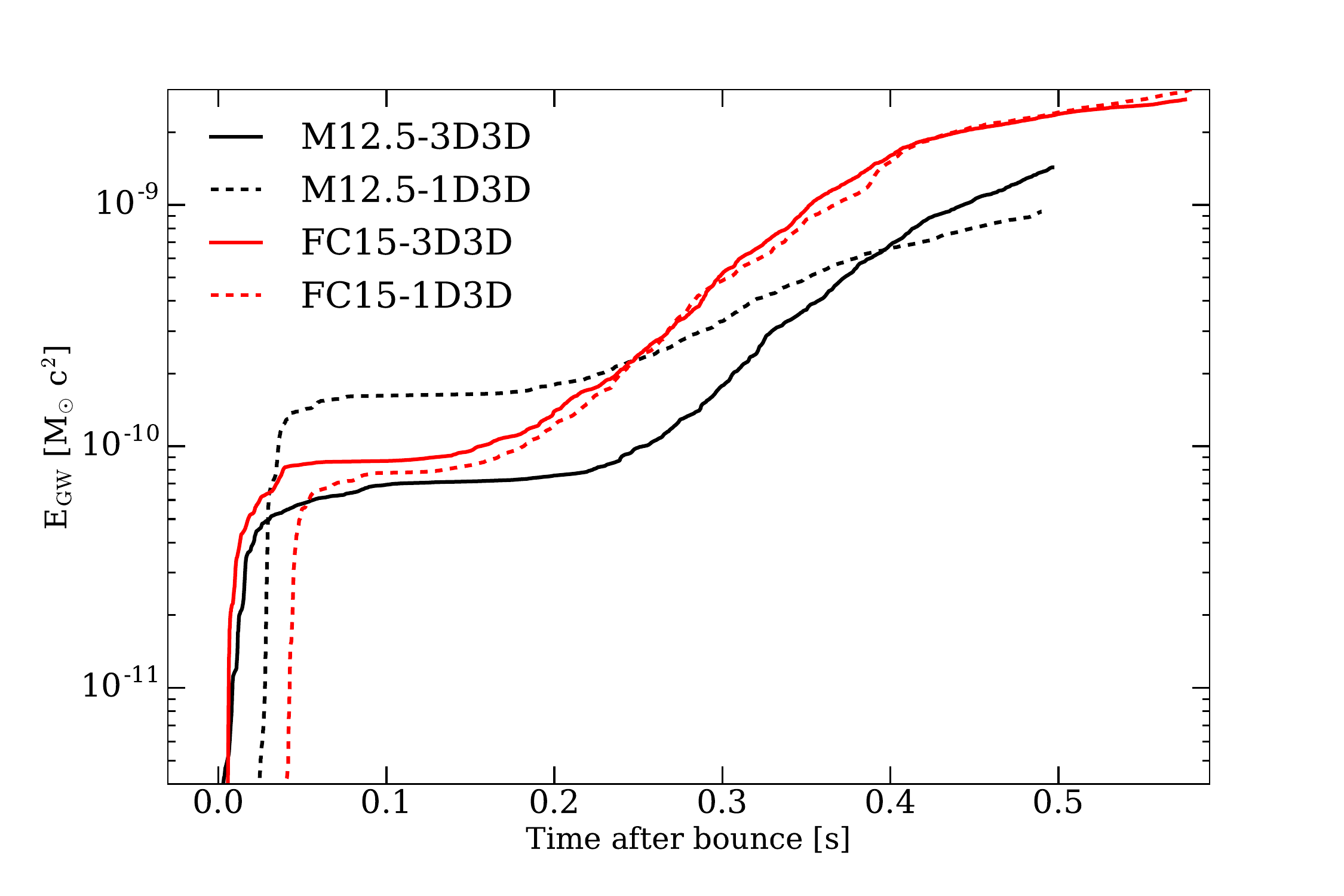}
    \caption{We illustrate the gravitational wave strain (in cm, \textbf{left}) and the gravitational wave energy (\textbf{right}) as a function of time after bounce (in s) for the models studied here. Note that the 1D MESA FC15 model (black dashed) shows a delayed growth of its gravitational wave energy compared with the fully 3D models by $\sim$25 ms. Similarly, the M12.5-1D3D model shows a rise in gravitational wave energy delayed by $\sim$40 ms compared to the 3d progenitor. Both fully 3D models rapidly generate gravitational wave energy a few ms after bounce, and have nearly plateaued while their 1d progenitor counterparts have just started generating appreciable gravitational wave energy. These earlier rise times in the fully 3D models are manifestations of the earlier onset of turbulence in the fully 3D models compared to the spherically-symmetric progenitors, generating quadrupolar deformations much closer to bounce.}
    \label{fig:GW} 
\end{figure*}

In Fig.\,\ref{fig:GW}, we look at the gravitational wave data for our models. In the left panel we plot the strains for both $+$ and $\times$ polarizations. 
The 3D3D model evinces nonzero strain before the onset of inner PNS convection $\sim$30  ms post bounce, and even pre-bounce, while the 1D3D models exhibit exactly zero strain until they are restarted in 3D 10 ms post bounce. Afterwards it still takes the 1D3D models an additional $\sim$10 ms for the strain to become noticeable.
The nonzero strain reveals early quadrupolar asymmetries in the 3D3D model present before bounce. In the right panel, we show the gravitational wave energies for all the models. 
The models with 3D progenitors rapidly generate gravitational wave energy a few ms after bounce, as their intrinsic seed turbulence enables them to undergo prompt convection much closer to bounce. 
This prompt convection is responsible for generating the significant quadrupolar deformations which produce the strong gravitational wave signals we see.
By the time the the models with 1D progenitors start rapidly generating gravitational wave energy (a few tens of ms after bounce), the corresponding 3D progenitor models have nearly plateaued. 
This $\sim$25-40 ms difference is the direct result of using progenitor models with realistic turbulence\footnote{
While 10 ms of this can be accounted for by the fact that the 1D progenitor models are started in 3D 10 ms after bounce, these delays are $\sim$2.5-4 times this value.},
and has potentially observable ramifications.
Given the significantly different accretion history and the differences in the onset of prompt convection, the expectation is that the initial plateau of the gravitational wave energy is stochastic in nature, which is likely why the M12.5-1D3D model initial plateaus to a higher level compare to its 3D progenitor counterpart.

\subsection{Effects of Turbulence}

As we have alluded to earlier in this paper, the disparity in the turbulent nature of the \oned and \ddd models is significant. 
Turbulence is inherently a 3D phenomena, and the \oned have zero initial turbulence, which subsequently plays a significant role in their evolution.
While the importance of turbulence in aiding the successful explosion of a CCSN is not new  (see also \citealt{nagakura2019b} on the effects of resolution on turbulence),
here we show how the initial, physically-consistent, turbulence of the \ddd models aids their resulting explosions by directly comparing them to their \oned counterparts.

\begin{figure*}
    \centering
    \includegraphics[width=.49\linewidth]{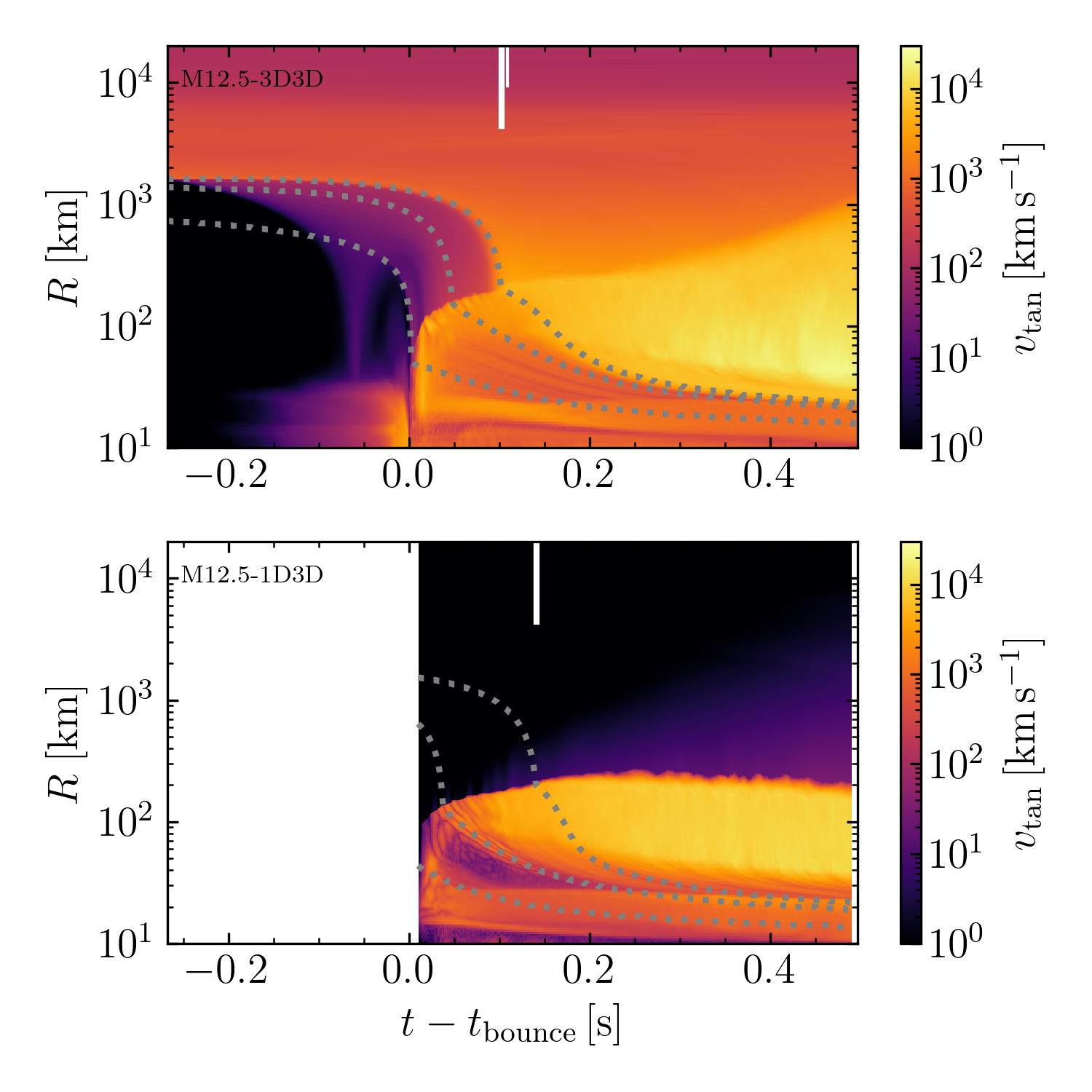}
    \includegraphics[width=.49\linewidth]{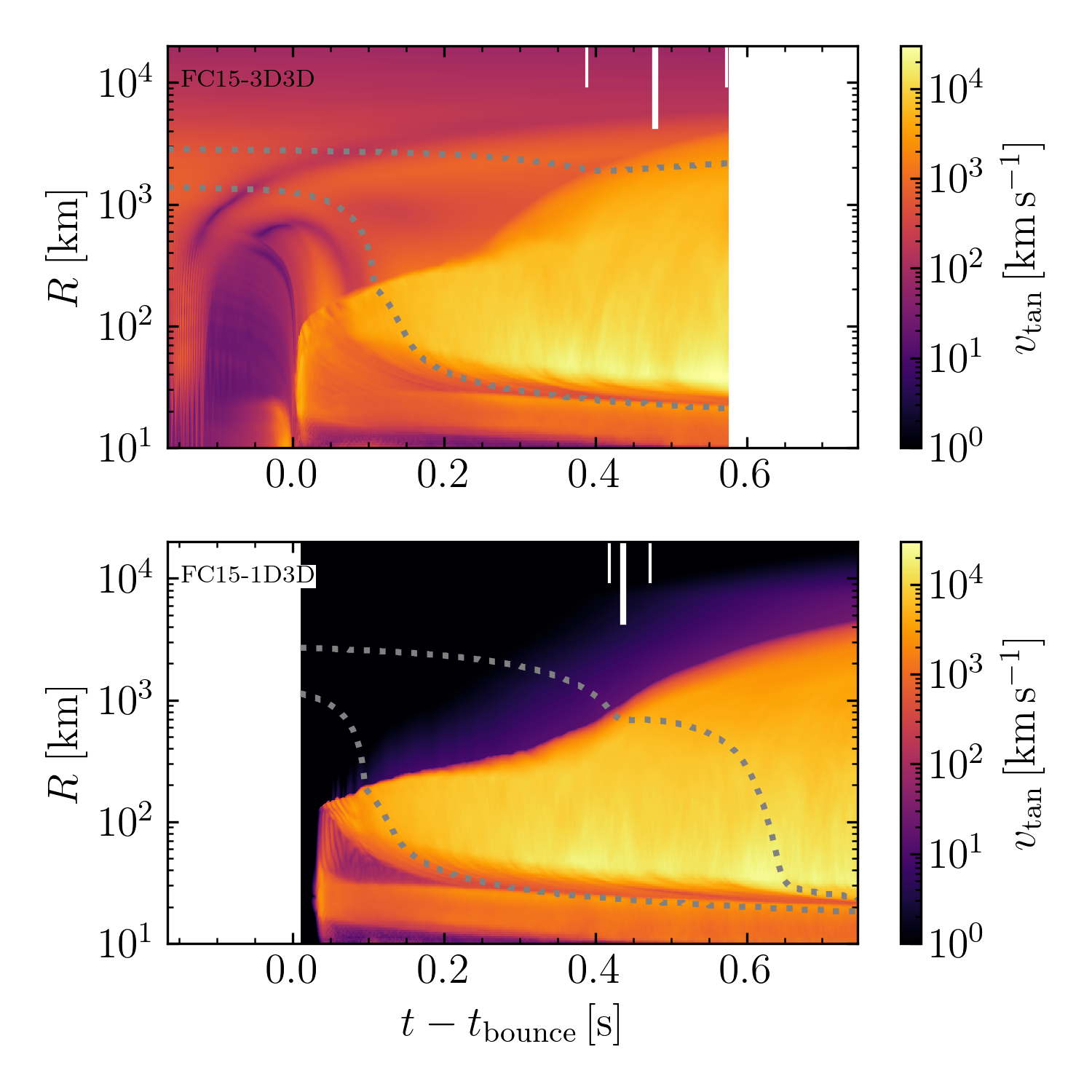}
    \caption{Space-time diagram of tangential velocity, $\left<v_{\rm tan}\right>_{\Omega}(t,r)=\left<\sqrt{v_\theta^2+v_\phi^2}\right>_{\Omega}$ (a proxy for turbulence; see also Eqn.~\ref{eqn:shell-mean}) for simulations with the 3D progenitor (top) and 1D progenitor (bottom): \muller models (left) and \fc (right). The dashed gray lines, from outermost to innermost, correspond to the mass coordinates of the Si/O, Fe/Si, and (only for the M12.5 models) Fe radiative/convective interfaces of the initial progenitor, respectively. The shock radius ($R_{\rm sh}$) is roughly where $v_{\rm tan}$ falls below $\sim 2\times10^3$ km/s and is noticeable in all the models. In all four models, $R_{\rm sh}$ is greater than 200 km by 100 ms post bounce. 
    For the \oned \muller model, $R_{\rm sh}$ can be seen to reverse direction at $\sim$300 ms post bounce, indicating a failed explosion, while all other models successfully explode. 
    One of the important features of this figure is the low-velocity (purple) region present in the \oned models beneath the shock up to $\sim$0.2 s after bounce. This feature is absent in the \ddd models, signifying the impact of the seed turbulence.
    This deficit in turbulence (compared to the \ddd case) is larger in the \muller models and is still significant at late time. This may explain why the \oned \muller model fails to explode here.
    }
    \label{fig:ST-vtan}
\end{figure*}

\begin{figure*}
    \centering
    \includegraphics[width=.49\linewidth]{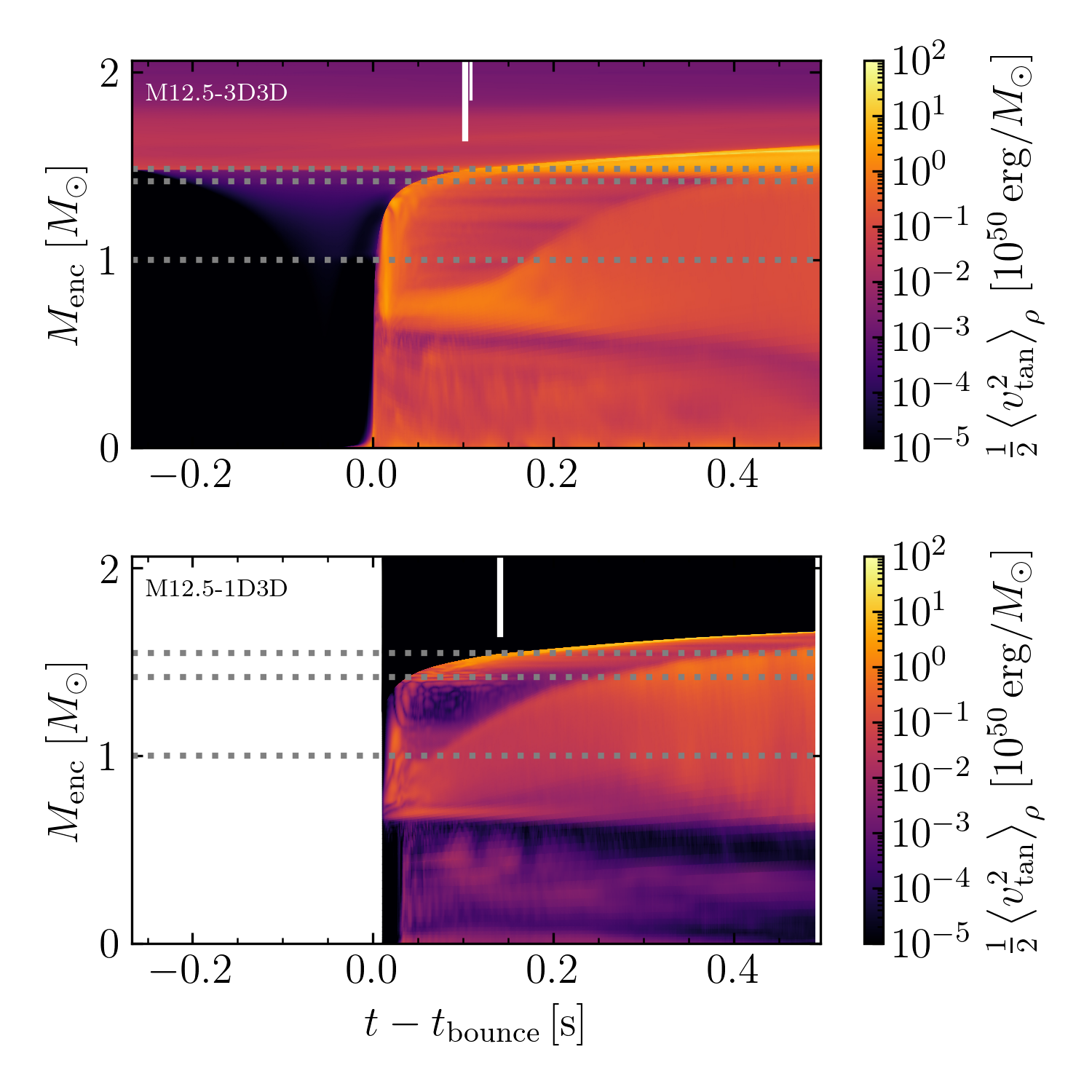}
    \includegraphics[width=.49\linewidth]{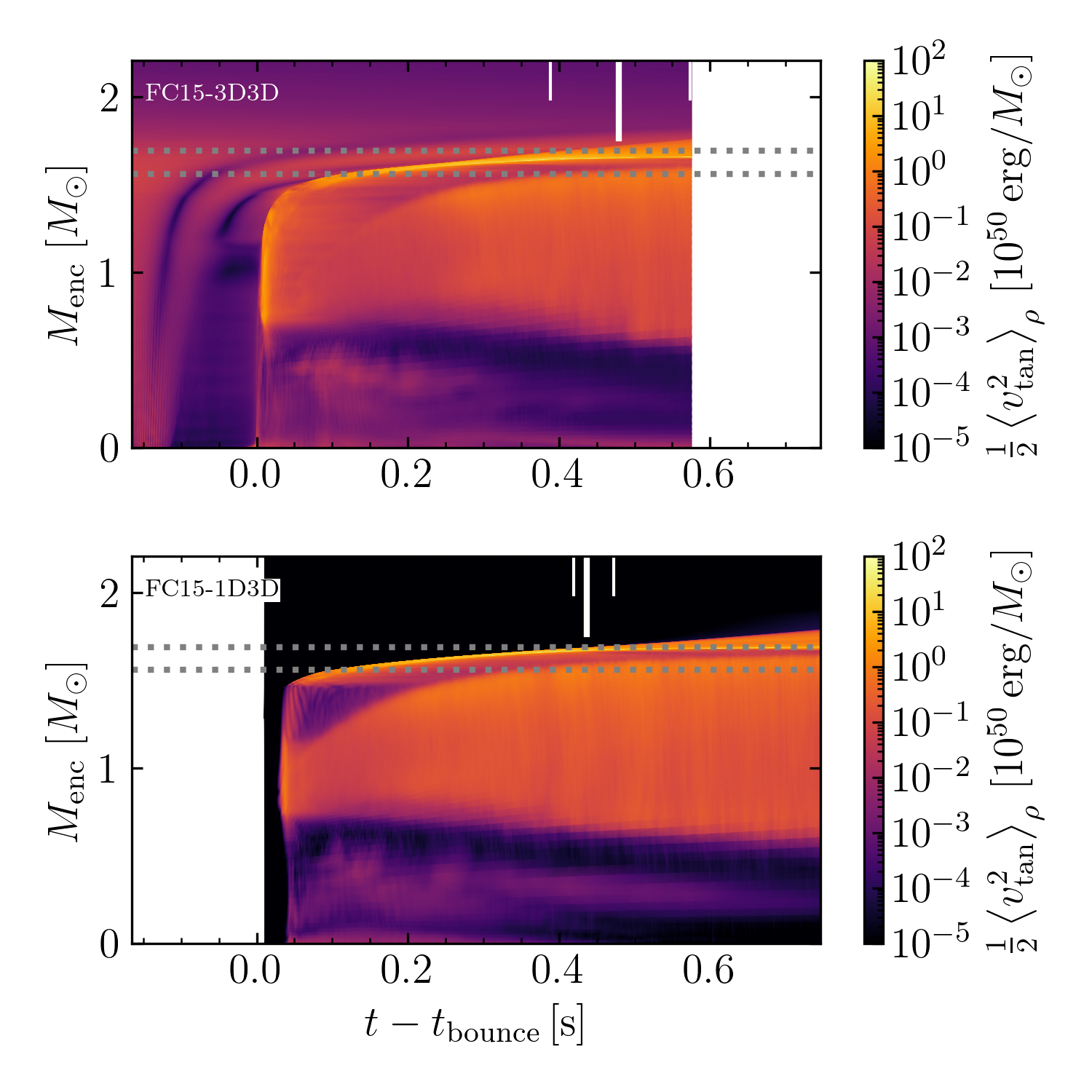}
    \caption{Mass-time diagram of specific tangential kinetic energy $\frac{1}{2}\left<v_{\rm tan}^2\right>_{\rho}(t,r)$ (a proxy for turbulence) for simulations with the 3D progenitor (top) and 1D progenitor (bottom): \muller models (left) and Couch (right). The horizontal dashed gray lines, from outermost to innermost, correspond to the mass coordinates of the Si/O, Fe/Si, and (only for the M12.5 models) Fe radiative/convective interfaces of the initial progenitor, respectively. All four models have a region of reduced turbulence below the shock and above the region of PNS convection (which extends up to $\sim1\,M_{\sun}$ shortly after bounce). 
    For both the \muller and \fc cases, it is clear that this deficit in turbulence is much more significant in the \oned models compared to the \ddd model, as we point out in Fig.~\ref{fig:ST-vtan} by noting the purple region just below the shock in \oned models at early times. 
    This extra turbulence in the \ddd models likely helps foster the explosion in these cases, explaining why these models explode more easily compared to their \oned counterparts.
    }
    \label{fig:MT-vtan}
\end{figure*}

The shell burning occurring just above the Si/O interface during the final stages of stellar evolution sets the stage for the convective instability to generate the initial turbulence found in the 3D models we obtained from \citet{muller_lowmass} and \citet{fields2020}.
Due to the intrinsic connection between the Si/O interface and turbulence, the accretion of the Si/O interface through the shock can significantly modify the evolution of the explosion (or lack thereof).
We find that when the Si/O interface is accreted before explosion, the turbulence between the mass coordinate of the initial Si/O interface and the shock radius is significantly enhanced (see Figs.~\ref{fig:ST-vtan} and \ref{fig:MT-vtan}).
For the same amount of input energy, turbulent stresses provide more support than thermal pressure, allowing this increase in turbulence to foster the explosion \citep{bhf1995,burrows2018}.

\begin{figure*}
    \includegraphics[width=\linewidth]{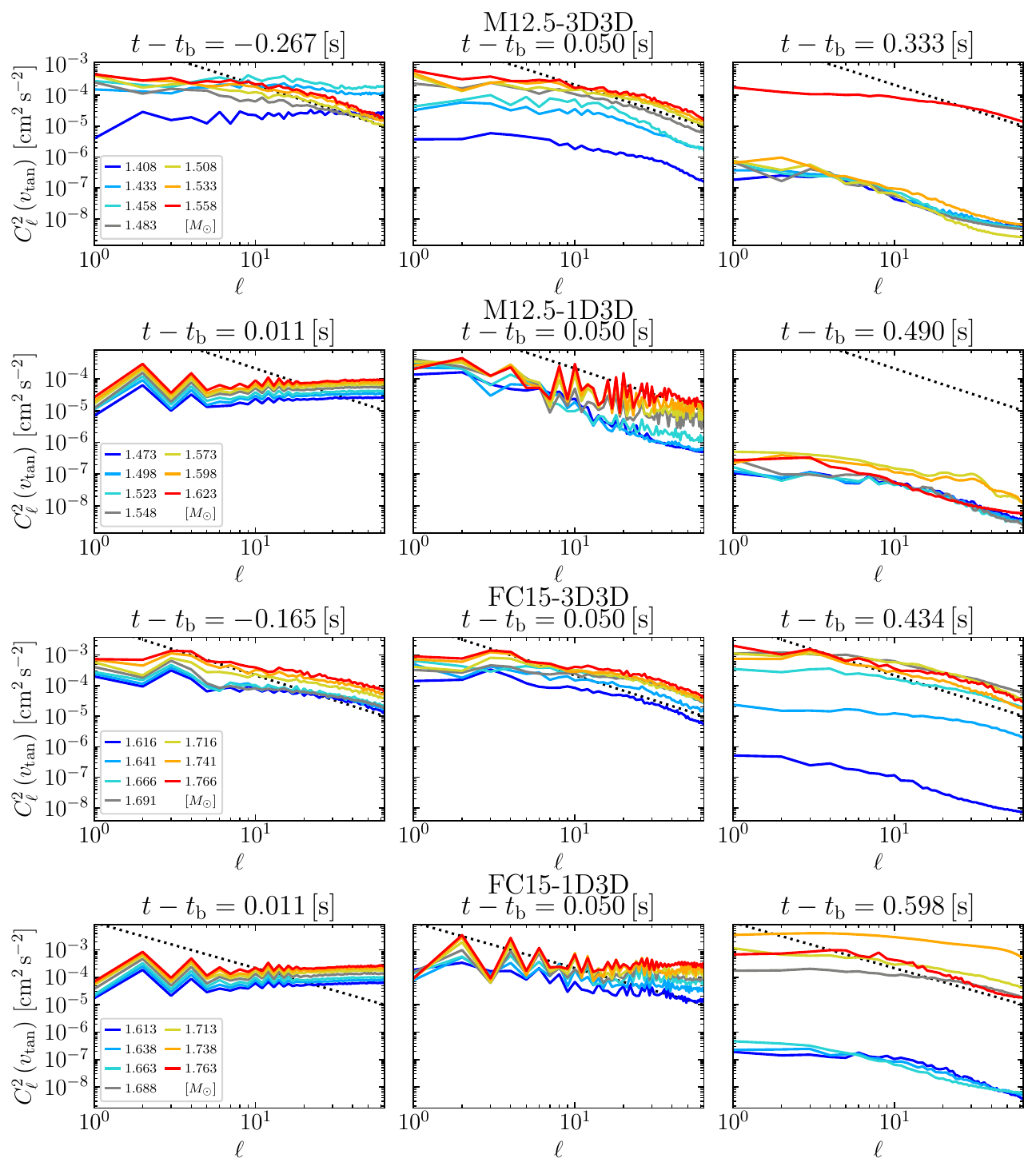}
    \vspace*{-2.5em}
    \caption{Power spectra of the tangential velocity, $C_\ell^2\left(v_{\rm tan}\right)$ (see Eqn.~\ref{eqn:cl}), as a function of model (top to bottom), time (left to right) and mass bin (color). For each model we choose seven adjacent mass bins of width $\delta m=0.025\,M_{\sun}$ with the middle mass bin (gray) centered on the mass coordinate of the initial Si/O interface, making blue lines below this interface and yellow/red lines above it. The numeric labels in the legends are the center of the mass bin in solar masses; these bins are held fixed for each row/model. The dotted black line is $\propto \ell^{-5/3}$, indicating Kolmogorov-like turbulence. There are 4 main features of this figure:
    \emph{1)} In the \oned models, the initial power spectra are flat for all shown mass bins, demonstrating a lack of turbulence. With the exception of the blue lines in the M12.5-1D3D model, this continues to be the case out to at least 50 ms post bounce.
    \emph{2)} Conversely, the \ddd models have at least some bins that turn down at large $\ell$ (small angular scales) to a power-law slope resembling $-5/3$ indicating the existence of turbulence. For the \muller case, we see that the mass bins below the Si/O interface (blue lines) lack turbulence while at and above this interface turbulence is present. The \fc \ddd model exhibits turbulence initially at all shown mass bins (both above and below the Si/O interface). 
    \emph{3)} At later times, it appears that turbulence exists in all shown mass bins for all models. 
    \emph{4)} As time progresses some mass bins dramatically lose power at all scales, indicating that they have been accreted onto the PNS.}
    \label{fig:power_spectra}
\end{figure*}

We can also track the turbulence directly by examining the turbulent power spectra of the tangential velocity, $C_\ell^2\left(v_{\rm tan}\right)$, defined as
\begin{subequations}
\begin{align}
     \label{eqn:cl}
     C_{\ell m}\left(f\right)&\equiv\int f\, Y_{\ell m}\, {\rm d} \Omega\\
     C_\ell^2\left(f\right)&\equiv\sum_{m=-\ell}^\ell\left|C_{\ell m}\left(f\right)\right|^2,
\end{align}
\end{subequations}
where $Y_{\ell m}$ are the spherical harmonics.
In Fig.~\ref{fig:power_spectra}, we follow the evolution of this power spectrum near the Si/O interface for all 4 models. 
The models with 3D progenitors exhibit turbulence before bounce, whereas the other models do not develop it until significantly after bounce (over 50 ms). 
Furthermore, the M12.5-3D3D initial progenitor model clearly shows disparate power spectra above and below the Si/O interface, with material below the interface having extremely flat spectra. 
Seed perturbations enable a more rapid growth of turbulence in the models with 3D progenitors, enabling them to explode more easily.

\begin{figure}
    \includegraphics[width=\linewidth]{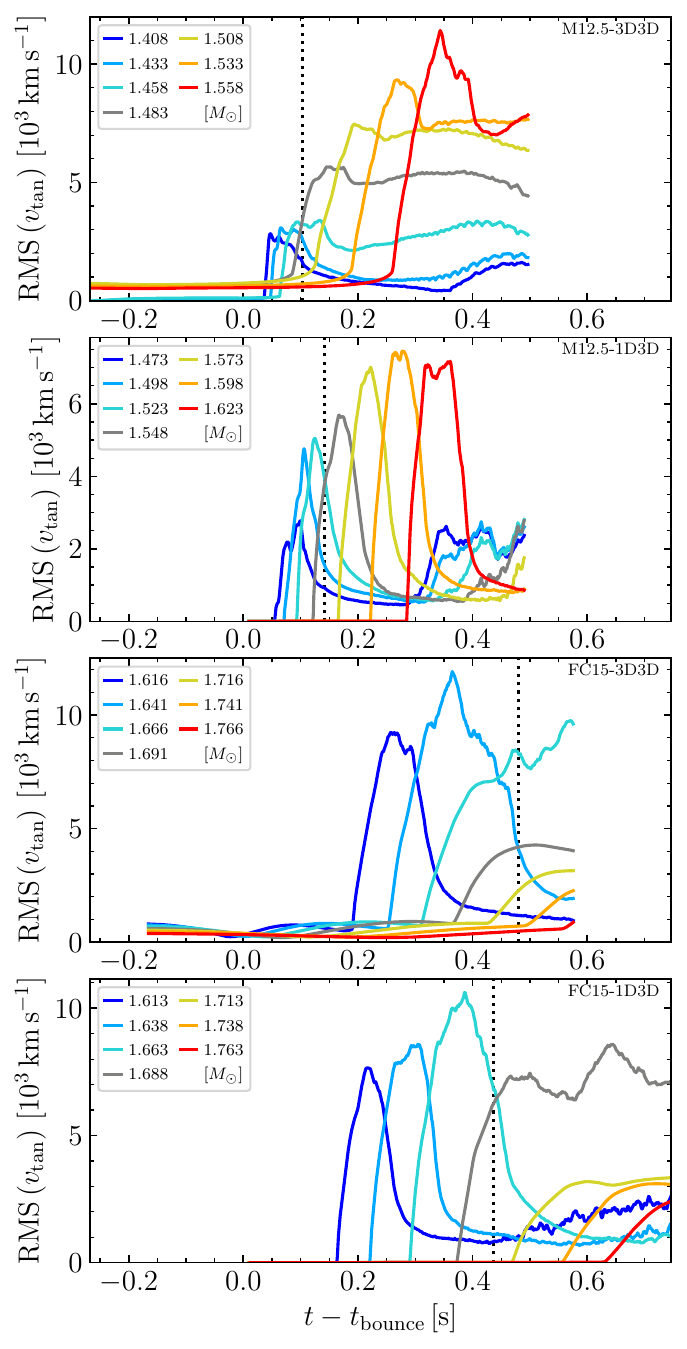}
    \vspace*{-2em}
    \caption{RMS $v_{\rm tan}$ ($\sqrt{\left<v_{\rm tan}^2\right>_\rho}$) as a function of time, using the same mass bins as Fig.~\ref{fig:power_spectra}. As before, the gray line corresponds to the Si/O interface. Here, the vertical dotted black line can be thought of as when the Si/O interface is accreted; this line corresponds to the time when the mean shock radius is equal to the radius of the mass coordinate corresponding to the initial Si/O interface. For both \ddd models, we can see that $v_{\rm tan}>0$ initially, while the \oned models do not develop noticeable tangential velocities until the shock has interacted with a given mass bin. We also note that for the \muller \ddd model the initial region with $v_{\rm tan}>0$ does not extend below the Si/O interface, while for the analogous \fc \ddd model it does. This appears to significantly affect the evolution of turbulence as the shock migrates through the model, where the mass bins at and above the Si/O interface in the \muller \ddd model receive a significant amplification of $v_{\rm tan}$. Alternatively, in the \fc \ddd model the shock has already become an explosion and is propagating outwards by the time it meets the Si/O interface, providing more of a gentle ``nudge" to those mass bins. This illustrates that the Si/O interface plays a more important role in the evolution of the \muller model.
    }
    \label{fig:rms_vtan_t}
\end{figure}

We also followed the evolution of RMS $v_{\rm tan}$ for several mass coordinates (the same ones shown in Fig.~\ref{fig:power_spectra}) surrounding and including the Si/O interface in Fig.~\ref{fig:rms_vtan_t}.
For model M12.5-3D3D, the layers at and above the Si/O interface show significant enhancement of the RMS of $v_{\rm tan}$ after passing through the shock, compared to the layers below the Si/O interface
which do not experience significant initial turbulence, signifying the potential importance of the seed turbulence present in the 3D progenitors. 
Further solidifying this point is the fact that the M12.5-3D3D model explodes, while the corresponding \oned model fails to do so.

\subsection{LESA}

\begin{figure}
    \centering
    \includegraphics[width=0.47\textwidth]{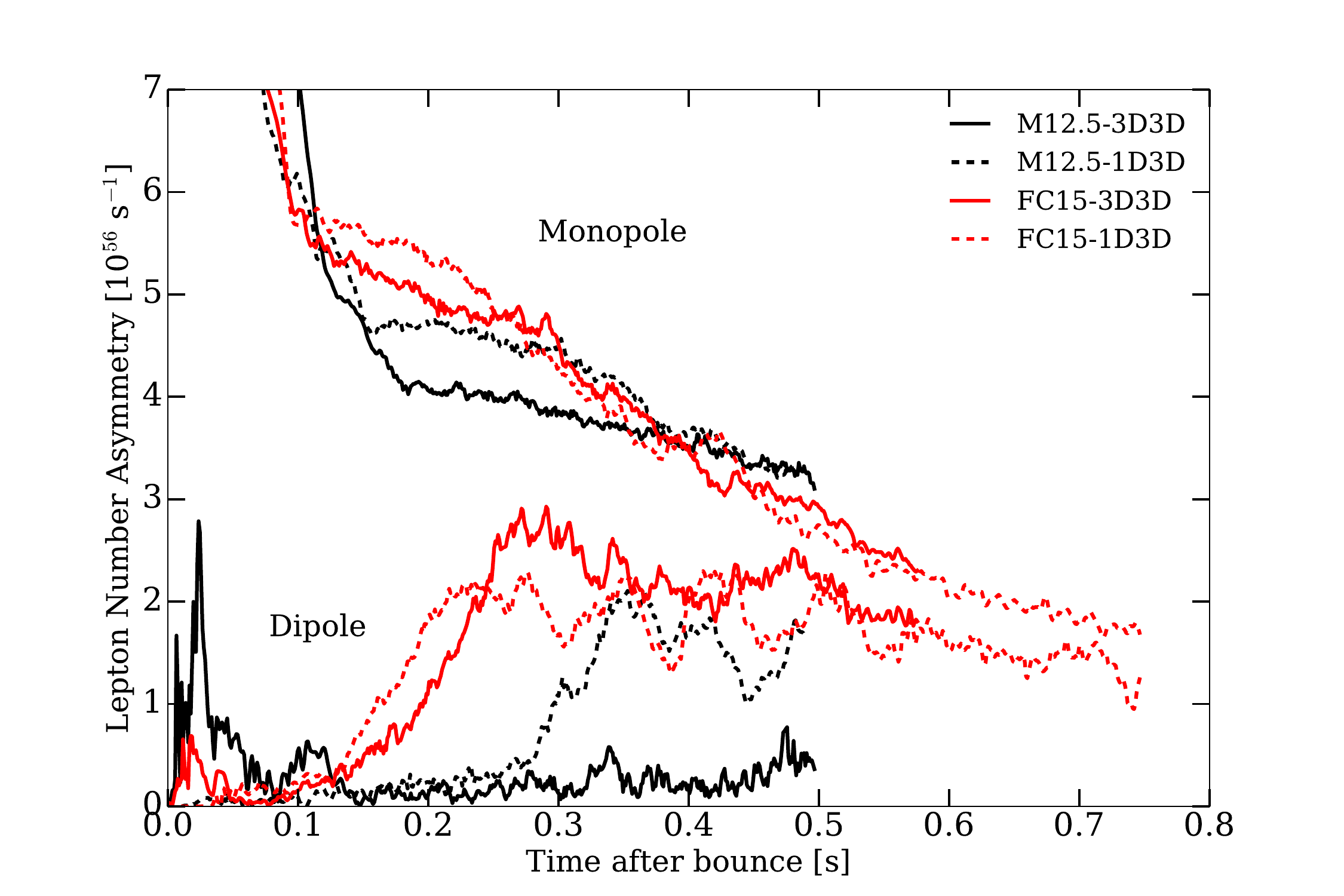}
    \caption{The monopole and dipole components of the LESA (at 500 km) as a function of time after bounce (in seconds). Note the presence of a strong dipole $\sim$30 ms after bounce in the 3D3D models absent in the 1D3D models. All models except M12.5-3D3D develop a strong LESA within the first $\sim$500 ms post bounce except model M12.5-3D3D. Note the strong dipole component shortly post bounce in the 3D3D models absent in the 1D3D models.}
    \label{fig:LESA}
\end{figure}

We comment here briefly on the development of the LESA, the lepton-emission self-sustained asymmetry, first identified in \cite{tamborra2014} and seen in recent 3D studies (\citealt{oconnor_couch2018b,glas2018,vartanyan2019,bollig2021}, but see also \citealt{2020ApJ...896..102K}). In Fig.\,\ref{fig:LESA}, we illustrate the monopole and dipole components of the lepton asymmetry as a function of time after bounce at 500 km. All models except for M12.5-3D3D illustrate the development of the LESA to varying degrees except for M12.5-3D3D. This low-mass model explodes relatively earlier, reversing accretion, which may be necessary to power the LESA mechanism. Furthermore, models FC15-3D3D and FC15-1D3D, which ran for an additional $\sim$250 ms beyond the M12.5 models, show stronger evidence of LESA in terms of the lepton asymmetry dipole magnitude than model M12.5-1D3D. This suggests that there may be a stronger LESA at later times ($\sim$800 ms post bounce), as the lepton asymmetry monopole magnitude continues to fall.

\section{Conclusions}\label{sec:conc}

Only in the last five years have three-dimensional simulations fruitfully modeled the final stages of the stellar evolution of massive stars up to core collapse. We presented here a total of four 3D core-collapse supernovae simulations encompassing two stellar models, a 12.5- and a 15-M${_\odot}$ progenitor, each evolved to core collapse in spherically-symmetric 1D and in 3D. We found that models evolved in 3D to core collapse, which better encapsulates the turbulent convective environments in the later stages of stellar evolution, preferentially exploded.

All of our models exploded, save for the model M12.5-1D3D. We found, most notably for model M12.5-3D3D, that the density profile around the Si/O interface exterior to the shock surface tracks the infalling accretion rate prior to shock revival, and this in turn determines in part the geometry of shock expansion. 
In other words, the morphology of the Si/O interface, and the turbulence around it, sets the early explosion morphology. 
The turbulence present in the 3D initial progenitor models, around and interior to the Si/O interface, serve as seed turbulence which amplifies on collapse.
This facilitates the onset of robust turbulence behind the shock, thereby enhancing shock expansion and promoting explosion. 
We found that, for model FC15-3D3D, shock revival precedes accretion of the Si/O interface, in contrast to model M12.5-3D3D, where accretion of the Si/O interface precipitates shock revival. 

Although neither of the exploding models reach large positive explosion energies within the first $\sim$600 ms simulated here, the 3D3D models manifest significant explosion energy growth rates of several tenths of a Bethe per second. We reiterate that multi-second post-bounce 3D simulations of core-collapse supernovae are essential to witness the late-time behavior of exploding massive stars, including their asymptotic explosion energy.

In addition, we provided corresponding neutrino and gravitational wave data, which encode information about the explosion timing and geometry.
We identify the presence of a LESA in all models, but it is weakest in model M12.5-3D3D, whose early explosion and low-mass may reverse accretion possibly responsible for maintaining the LESA. 
We also see that, generally, the LESA manifests more strongly at later times, after $\sim$500 ms, as the lepton-number asymmetry monopole declines and the dipole begins to dominate.
In particular, we also note that the 3D3D models have an earlier gravitational wave signal at, and even before, core bounce.
The latter effect is a consequence of turbulence in 3D pre-collapse progenitors, with possible observational ramifications.

\section*{Acknowledgements}
The authors acknowledge Bernhard M{\"u}ller, Carl Fields, and Sean Couch for providing us with the progenitor models studied here and for valuable correspondence throughout. We are also grateful to Chris White, David Radice, and Hiroki Nagakura for helpful discussions. We acknowledge support from the U.S. Department of Energy Office of Science and the Office
of Advanced Scientific Computing Research via the Scientific Discovery
through Advanced Computing (SciDAC4) program and Grant DE-SC0018297
(subaward 00009650) and support from the U.S. NSF under Grants AST-1714267
and PHY-1804048 (the latter via the Max-Planck/Princeton Center (MPPC) for Plasma Physics).
A generous award of computer time was provided
by the INCITE program. That research used resources of the
Argonne Leadership Computing Facility, which is a DOE Office of Science
User Facility supported under Contract DE-AC02-06CH11357. We are also grateful for our computational resources through the Texas Advanced Computing Center (TACC) at The University of Texas at Austin via Frontera Large-Scale Community Partnerships under grant SC0018297 as well as the Leadership Resource Allocation under grant number 1804048. In addition, this overall research
project was part of the Blue Waters sustained-petascale computing project,
which is supported by the National Science Foundation (awards OCI-0725070
and ACI-1238993) and the state of Illinois. Blue Waters was a joint effort
of the University of Illinois at Urbana-Champaign and its National Center
for Supercomputing Applications. This general project was also part of
the ``Three-Dimensional Simulations of Core-Collapse Supernovae" PRAC
allocation support by the National Science Foundation (under award \#OAC-1809073).
Moreover, we acknowledge access under the local award \#TG-AST170045
to the resource Stampede2 in the Extreme Science and Engineering Discovery
Environment (XSEDE), which is supported by National Science Foundation grant
number ACI-1548562. Finally, the authors employed computational resources provided by the TIGRESS high
performance computer center at Princeton University, which is jointly
supported by the Princeton Institute for Computational Science and
Engineering (PICSciE) and the Princeton University Office of Information
Technology, and acknowledge our continuing allocation at the National
Energy Research Scientific Computing Center (NERSC), which is
supported by the Office of Science of the US Department of Energy
(DOE) under contract DE-AC03-76SF00098.

\section*{Data Availability}

The data underlying this article will be shared on reasonable request to the corresponding author.

\clearpage
\bibliographystyle{mnras}
\bibliography{References}

\ifMNRAS
\bsp	
\label{lastpage}
\else
\fi

\end{document}